\documentclass[5p,times]{elsarticle}

\usepackage{amssymb}
\usepackage{amsmath}

\usepackage{algorithm}
\usepackage[caption=false,labelfont=sf,textfont=sf]{subfig}
\usepackage{graphicx}

\usepackage{hyperref}
\usepackage{tabularx,booktabs}
\usepackage{cases}
\usepackage[capitalize]{cleveref}
\usepackage{multirow}
\usepackage{hhline}
\usepackage{siunitx}

\journal{Future Generation Computer Systems}

\begin{document}

\begin{frontmatter}



\title{A reliability- and latency-driven task allocation framework for workflow applications in the edge-hub-cloud continuum}


\author[label1,label2]{Andreas Kouloumpris\corref{cor1}}
\ead{kouloumpris.andreas@ucy.ac.cy}
\cortext[cor1]{Corresponding author.}

\author[label2]{Georgios L. Stavrinides}

\author[label1,label2]{Maria K. Michael}

\author[label1,label2]{Theocharis Theocharides}

\affiliation[label1]{organization={Department of Electrical and Computer Engineering, University of Cyprus},
            addressline={1 Panepistimiou Avenue, Aglantzia, P.O. Box 20537}, 
            city={Nicosia},
            postcode={1678}, 
            country={Cyprus}}

\affiliation[label2]{organization={KIOS Research and Innovation Center of Excellence, University of Cyprus},
            addressline={1 Panepistimiou Avenue, Aglantzia, P.O. Box 20537}, 
            city={Nicosia},
            postcode={1678}, 
            country={Cyprus}}

\begin{abstract}

A growing number of critical workflow applications leverage a streamlined edge-hub-cloud architecture, which diverges from the conventional edge computing paradigm. An edge device, in collaboration with a hub device and a cloud server, often suffices for their reliable and efficient execution. 
However, task allocation in this streamlined architecture is challenging due to device limitations and diverse operating conditions. 
Given the inherent criticality of such workflow applications, where reliability and latency are vital yet conflicting objectives, an exact task allocation approach is typically required to ensure optimal solutions. 
As no existing method holistically addresses these issues, we propose an exact multi-objective task allocation framework to jointly optimize the overall reliability and latency of a workflow application in the specific edge-hub-cloud architecture. We present a comprehensive binary integer linear programming formulation that considers the relative importance of each objective. It incorporates time redundancy techniques, while accounting for crucial constraints often overlooked in related studies.
We evaluate our approach using a relevant real-world workflow application, as well as synthetic workflows varying in structure, size, and criticality.
In the real-world application, our method achieved average improvements of 84.19\% in reliability and 49.81\% in latency over baseline strategies, across relevant objective trade-offs.
Overall, the experimental results demonstrate the effectiveness and scalability of our approach across diverse workflow applications for the considered system architecture, highlighting its practicality with runtimes averaging between 0.03 and 50.94 seconds across all examined workflows.

\end{abstract}




\end{frontmatter}

\sloppy


\section{Introduction}
\label{sec:intro}

The technological advancements in edge computing and the Internet of Things (IoT) are spearheading the emergence of critical applications that require real-time response and reliable execution near the network edge \cite{Liu2025}.
Such applications typically comprise several component tasks with precedence relationships and are commonly referred to as \emph{workflows} \cite{Abdi2025}. 
A prominent example is the use of an unmanned aerial vehicle (UAV) equipped with sensing and computational capabilities for the autonomous inspection of critical infrastructures (e.g., power transmission towers and lines) or for search-and-rescue missions in areas that are difficult to access from the ground \cite{Savva2021, Vladan2021}. Another example is the use of implantable biomedical edge devices for remote patient monitoring and support, such as smart pacemakers \cite{Zheng2021, Wang2017, Agarwal2022, Alam2017}. 
These use-cases \cite{Savva2021, Vladan2021, Zheng2021, Wang2017, Agarwal2022, Alam2017} commonly involve pre-programmed operations and leverage a streamlined architecture based on the \emph{edge-hub-cloud} paradigm, where the edge device (e.g., UAV or smart pacemaker) interacts with a hub device (e.g., a laptop or smartphone), which in turn communicates with a cloud server to process certain tasks in a more reliable and timely manner. 
In such safety-critical scenarios, reliability and latency are of paramount importance. Consequently, their joint optimization for workflow task allocation in the edge-hub-cloud continuum, subject to device- and application-specific constraints, is essential.

The edge-hub-cloud paradigm deviates from the conventional edge computing model, which typically comprises edge devices in the bottom layer, edge servers in the middle layer, and cloud servers in the top layer. In contrast, an edge-hub-cloud architecture employs edge and hub devices that may be battery-operated in the bottom layer, and cloud servers in the top layer. While a hub device may not be as powerful as an edge server, it is strategically positioned closer to the edge devices, facilitating their coordination and communication with the remote cloud servers more effectively. 
This streamlined architecture has become increasingly prevalent in real-world domains such as UAV-enabled infrastructure inspection and biomedical monitoring \cite{Savva2021, Vladan2021, Zheng2021, Wang2017, Agarwal2022, Alam2017}, where a single edge device, a hub device, and a cloud server often suffice for reliable and real-time execution.
 
However, the reliability, processing power, memory, storage, and energy capacity of the devices, as well as the capacity of the communication channels, typically decline when moving from the cloud towards the edge of the network \cite{Kolomvatsos2019}.
These issues are further amplified by the diverse characteristics of the operating environment \cite{Rasouli2025}.
Specifically, the reliability of a device is not only dependent on technological factors, such as integrated circuit wear-out and aging, which may cause permanent errors, but it also depends on various environmental factors, such as temperature, humidity, electromagnetic interference, and cosmic rays, which may cause transient errors (e.g., soft errors).
Moreover, the proliferation of hardware errors, either permanent or transient, is well known to be workload dependent \cite{Mo2023}. 
Thus, an edge device, due to its lower manufacturing cost, higher integration, and the environmental conditions it operates in, is typically more vulnerable to reliability threats than a hub device, which in turn is often less reliable than a cloud server.

A widely used approach to increase the reliability of an application is time redundancy \cite{Ottavi2014}.
According to this method, a task is replicated and executed multiple times (two for error detection, three for error mitigation), and the outputs of all task executions are compared (dual execution) or subjected to a majority vote (triple execution).
The number of executions for a specific task typically depends on the expected reliability of the task when executed on a particular device, as well as the criticality of the task or application.
However, the additional overhead incurred by such techniques may impair the overall performance, especially in architectures with limited resources like the one considered in this work.  
Hence, maximizing the reliability and minimizing the latency of an application in the specific edge-hub-cloud environment are conflicting goals.

To achieve an optimized balance between reliability and latency based on the relative importance of each objective, task allocation plays a crucial role, as it determines which device should execute each task of the workflow. This is distinct from task scheduling, which focuses on the execution order and timing of tasks on a given device.
In its general form, the task allocation problem is NP-hard \cite{Khalil2018}.
The inherent criticality of the targeted applications necessitates the optimal mapping of tasks to system devices.
This can be attained by utilizing an exact optimization method, namely binary integer linear programming (BILP), which yields a true Pareto-optimal solution for a given combination of objective weights \cite{Grodzevich2006}.
In contrast to heuristics, which typically provide only approximate or near-Pareto-optimal solutions and thus may fail to satisfy all constraints, BILP can optimally capture the interplay between multiple constraints and objectives, enabling meaningful design-time trade-off exploration.
Given the pre-programmed nature of the considered applications, BILP can be executed offline to accommodate its computational demands, provided that a solution can be obtained within a reasonable time.

Although the applications that motivate this work utilize a streamlined edge-hub-cloud architecture, the allocation of their tasks entails all the aforementioned challenges that must be addressed.
To this end, we propose an exact, design-time (i.e., offline) multi-objective BILP-based optimization framework to allocate the tasks of a workflow application on an edge device, a hub device, and a cloud server. Our approach jointly optimizes the overall reliability and latency of the application, based on the desired trade-off between the two objectives.
It takes into account the criticality of the application and multiple constraints, including the computational and energy limitations of the devices, the varied bandwidth of the communication links, and inter-device connectivity. 
The main contributions of this work, which builds upon our preliminary research in \cite{Kouloumpris2019b, Kouloumpris2020}, are summarized as follows:
\begin{enumerate}
    \item We present a comprehensive formulation for modeling the considered problem as a multi-objective BILP model. The formulation is facilitated by a two-step task graph transformation technique, which encapsulates the application and system models, as well as the adopted energy and reliability models, while integrating time redundancy methods such as dual and triple task execution.
    
    \item Our formulation jointly optimizes reliability and latency while holistically considering constraints often ignored in related research efforts, such as the limited memory, storage, and energy capacities of the devices, the reliability requirements of the tasks, as well as the computational and communication latency and energy required to execute each task.
    
    \item We evaluate our approach using a real-world workflow application for the autonomous UAV-based aerial inspection of power transmission towers and lines. 
    
    \item To further validate and investigate the scalability and applicability of our method to workflows of different structures, sizes, and criticality levels, we use a suite of diverse synthetic workflows that we generated for this purpose.
\end{enumerate}

To our knowledge, no existing method considers the same architecture, addresses the same objectives and constraints, and employs an exact multi-objective task allocation technique as our approach.
The rest of the paper is organized as follows. \cref{sec:related} provides an overview of related literature. \cref{sec:framework} describes the proposed optimization framework. \cref{sec:evaluation} presents the experimental setup and the evaluation results, whereas \cref{sec:conclusions} concludes the paper.

\section{Related work}
\label{sec:related}

\subsection{State-of-the-art task allocation methods for workflow applications}
\label{subsec:comparisonWithSOTA}

Several studies address the allocation of workflow tasks in distributed environments, with objectives targeting reliability \cite{Fang2025}, latency \cite{Jiang2025, Nastic2021, Li2026, Santos2023, Marchese2023}, or the joint optimization of both \cite{Wang2011, Tang2022, Khurana2023, Asghari2023, Wang2018, Huang2020, Taghinezhad2023, Taghinezhad2024, Saeedi2020, Peng2022, DeMaio2020, Khaleel2023, Ramezani2021, Fan2021}.
While some of these works also involve scheduling, our discussion is focused on task allocation.

\subsubsection{Works optimizing either reliability or latency}
Fang et al. \cite{Fang2025} propose a multi-objective evolutionary algorithm for task allocation in multi-cloud systems that addresses cost, energy, and reliability, but not latency. 
Moreover, memory, storage, and communication energy constraints are not considered. 
In contrast, other works target latency as the optimization objective \cite{Jiang2025, Nastic2021, Li2026, Santos2023, Marchese2023}. 
Jiang et al. \cite{Jiang2025}, Nastic et al. \cite{Nastic2021}, and Li et al. \cite{Li2026} investigate task allocation in heterogeneous multi-tier systems, but do not incorporate reliability into their objectives and constraints, while \cite{Nastic2021} and \cite{Li2026} also overlook the energy aspects of the problem. 
On the other hand, Santos et al. \cite{Santos2023} and Marchese and Tomarchio \cite{Marchese2023} address container-based orchestration where reliability is included in the constraints, but the energy and storage requirements of the tasks are not taken into account.
Although these studies \cite{Fang2025, Jiang2025, Nastic2021, Li2026, Santos2023, Marchese2023} investigate multi-tier environments, none of them examine the edge-hub-cloud architecture considered in this work.
Among them, only Santos et al. \cite{Santos2023} employ an exact approach, formulating the problem as a mixed integer linear programming model.

\subsubsection{Works jointly optimizing reliability and latency}
A large body of research has focused on the joint optimization of reliability and latency.
Specifically, Wang et al. \cite{Wang2011}, Tang \cite{Tang2022}, and Khurana et al. \cite{Khurana2023} propose task placement heuristics for computational resources of varying reliability. 
Asghari-Alaie et al. \cite{Asghari2023} present a hybrid task allocation approach that leverages heuristics to explore the solution space, based on failure rates derived from system logs. 
Wang et al. \cite{Wang2018} propose a heuristic method based on iterative task duplication. 
Although these works \cite{Wang2011, Tang2022, Khurana2023, Asghari2023, Wang2018} generally address both computational and communication latency, they overlook the memory, storage, and energy requirements of the tasks.

On the other hand, Huang et al. \cite{Huang2020} propose a reliability- and latency-aware task allocation heuristic that also considers the energy consumption of the system. 
Similarly, Taghinezhad-Niar and Taheri \cite{Taghinezhad2023, Taghinezhad2024} present a task placement framework that utilizes task duplication and schedule gaps to improve reliability and latency while reducing the energy consumption of computational resources. 
In the same context, Saeedi et al. \cite{Saeedi2020} develop an enhanced many-objective particle swarm optimization approach, incorporating greedy heuristics for initial solution generation. 
Peng et al. \cite{Peng2022} introduce heuristics based on the reliability-to-performance ratio of the application, under energy consumption constraints.
However, none of these works \cite{Huang2020, Taghinezhad2023, Taghinezhad2024, Saeedi2020, Peng2022} consider the energy required for inter-task communication, nor the memory and storage required by each task.

Efforts that specifically target multi-tier environments include \cite{DeMaio2020} and \cite{Khaleel2023}. In \cite{DeMaio2020}, De Maio et al. propose a genetic algorithm-based task allocation metaheuristic, aimed at finding the best trade-off between latency and reliability, while considering the storage requirements of the tasks. Similarly, a critical path-based task placement heuristic that takes into account the memory needs of the tasks is presented by Khaleel in \cite{Khaleel2023}. Although both of these studies investigate multi-tier systems, they do not consider the specific architecture examined in this work, nor the energy demands of the application.
While none of the multi-objective research efforts \cite{Wang2011, Tang2022, Khurana2023, Asghari2023, Wang2018, Huang2020, Taghinezhad2023, Taghinezhad2024, Saeedi2020, Peng2022, DeMaio2020, Khaleel2023} utilize an exact method for solving the specific problem, an exact integer non-linear programming approach is employed by Ramezani et al. in \cite{Ramezani2021}. The proposed method involves the segmentation of the application task graph and the optimization of each segment separately to achieve a balance between reliability and latency. 
Although an exact method is employed, the inter-task communication latency, a multi-tier environment, and the storage and energy requirements of the application are not taken into account. 
The study by Fan et al. \cite{Fan2021}, which is most closely related to our work, investigates the joint optimization of reliability and latency in a multi-tier system. However, similar to the majority of prior works, their approach neither incorporates energy constraints nor considers the specific edge-hub-cloud architecture, and it does not utilize an exact optimization method.

\subsubsection{Summary of research gaps}
Overall, the presented state-of-the-art task allocation strategies for workflow applications \cite{Fang2025, Jiang2025, Nastic2021, Li2026, Santos2023, Marchese2023, Wang2011, Tang2022, Khurana2023, Asghari2023, Wang2018, Huang2020, Taghinezhad2023, Taghinezhad2024, Saeedi2020, Peng2022, DeMaio2020, Khaleel2023, Ramezani2021, Fan2021} do not comprehensively address the joint optimization of reliability and latency under all the constraints considered in our work (i.e., memory, storage, computational and communication latency and energy, and reliability constraints), nor do they examine the specific edge-hub-cloud architecture. Moreover, in contrast to our approach, the vast majority of these efforts do not use an exact method to solve the task allocation problem. 
With respect to these aspects, \cref{table:comparison} summarizes the qualitative comparison of our work with the state of the art.

\begin{table*}[t]
\centering
\caption{Qualitative comparison with related state-of-the-art works on task allocation for workflow applications.}
\label{table:comparison}
\resizebox{0.69\textwidth}{!}{
    \begin{tabular}{lccccccccccc} 
        \toprule
        \multirow{3}{*}{Ref.}  & \multicolumn{2}{c}{Objectives} & \multicolumn{7}{c}{Constraints} & \multirow{2}{*}{Multi-Tier} & \multirow{2}{*}{Exact}\\
        \cmidrule(lr){2-3}
        \cmidrule(lr){4-10}
                                    & \multirow{2}{*}{Reliability} & \multirow{2}{*}{Latency} & \multirow{2}{*}{Memory} & \multirow{2}{*}{Storage} & Comp.      & Comp.      & Comm.      & Comm.      & \multirow{2}{*}{Reliability} & \multirow{2}{*}{Environ.}          & \multirow{2}{*}{Method}\\
                                    &            &            &                         &                          & Latency    & Energy     & Latency    & Energy     &                              &                         &\\  
        \midrule  
        \cite{Fang2025} & \checkmark & - & - & - & \checkmark & \checkmark & \checkmark & - & \checkmark & \checkmark & -\\
        \cite{Jiang2025} & - & \checkmark & - & - & \checkmark & \checkmark & \checkmark & \checkmark & - & \checkmark & -\\
        \cite{Nastic2021} & - & \checkmark & \checkmark & - & \checkmark & - & \checkmark & - & - & \checkmark & -\\
        \cite{Li2026} & - & \checkmark & \checkmark & - & \checkmark & - & \checkmark & - & - & \checkmark & -\\
        \cite{Santos2023} & - & \checkmark & \checkmark & - & \checkmark & - & \checkmark & - & \checkmark & \checkmark & \checkmark\\
        \cite{Marchese2023} & - & \checkmark & \checkmark & - & \checkmark & - & \checkmark & - & \checkmark& \checkmark & - \\

        \cite{Wang2011}             & \checkmark & \checkmark & -                       & -                        & \checkmark & -          & -          & -          & \checkmark                   & -             & -\\
        \cite{Tang2022}             & \checkmark & \checkmark & -                       & -                        & \checkmark & -          & \checkmark & -          & \checkmark                   & -             & -\\
        \cite{Khurana2023}          & \checkmark & \checkmark & -                       & -                        & \checkmark & -          & \checkmark & -          & \checkmark                   & -             & -\\
        \cite{Asghari2023}          & \checkmark & \checkmark & -                       & -                        & \checkmark & -          & \checkmark & -          & \checkmark                   & -             & -\\
        \cite{Wang2018}             & \checkmark & \checkmark & -                       & -                        & \checkmark & -          & \checkmark & -          & \checkmark                   & -             & -\\
        \cite{Huang2020}            & \checkmark & \checkmark & -                       & -                        & \checkmark & \checkmark & \checkmark & -          & \checkmark                   & -             & -\\
        \cite{Taghinezhad2023}      & \checkmark & \checkmark & -                       & -                        & \checkmark & \checkmark & \checkmark & -          & \checkmark                   & -             & -\\
        \cite{Taghinezhad2024}      & \checkmark & \checkmark & -                       & -                        & \checkmark & \checkmark & \checkmark & -          & \checkmark                   & \checkmark    & -\\ 
        \cite{Saeedi2020}           & \checkmark & \checkmark & -                       & -                        & \checkmark & \checkmark & \checkmark & -          & \checkmark                   & -             & -\\
        \cite{Peng2022}             & \checkmark & \checkmark & -                       & -                        & \checkmark & \checkmark & \checkmark & -          & \checkmark                   & -             & -\\
        \cite{DeMaio2020}           & \checkmark & \checkmark & -                       & \checkmark               & \checkmark & -          & \checkmark & -          & \checkmark                   & \checkmark    & -\\
        \cite{Khaleel2023}          & \checkmark & \checkmark & \checkmark              & -                        & \checkmark & -          & \checkmark & -          & \checkmark                   & \checkmark    & -\\
        \cite{Ramezani2021}         & \checkmark & \checkmark & \checkmark              & -                        & \checkmark & -          & -          & -          & \checkmark                   & -             & \checkmark\\
        \cite{Fan2021} & \checkmark & \checkmark & \checkmark & \checkmark & \checkmark & - & \checkmark & - & \checkmark & \checkmark & -\\
        
        Ours                   & \checkmark & \checkmark & \checkmark              & \checkmark               & \checkmark & \checkmark & \checkmark & \checkmark & \checkmark                   & \checkmark    & \checkmark\\
        \bottomrule
    \end{tabular}
}
\end{table*}

\subsection{Comparison with our preliminary research}
\label{subsec:comparisonWithPreliminaryReserach}
In our preliminary studies \cite{Kouloumpris2019b, Kouloumpris2020}, we explored task allocation in an edge-hub-cloud environment, applying time redundancy techniques to gain initial insights into the resulting overall latency \cite{Kouloumpris2019b} and reliability \cite{Kouloumpris2020}, respectively. However, these initial studies were limited to single objectives and lacked a comprehensive optimization approach, motivating the current multi-objective problem. 
In another prior study \cite{Kouloumpris2024}, we considered either latency or energy optimization without incorporating reliability, using a simplified single-step task graph extension to facilitate the problem formulation.
Moreover, although in \cite{Kouloumpris2024b} we investigated both task allocation and scheduling in an edge-hub-cloud architecture, we solely focused on latency optimization without considering reliability. In contrast, this work advances beyond these preliminary studies by formulating a holistic multi-objective task allocation framework that jointly optimizes reliability and latency. Additionally, we propose an enhanced two-step task graph transformation technique incorporating advanced models for both energy and reliability, thereby providing a novel and comprehensive solution to the problem under consideration.

\section{Proposed multi-objective optimization framework}
\label{sec:framework}

\subsection{Framework overview}
We formulate the examined task allocation problem as a multi-objective BILP model.  
BILP is an exact method suitable for modeling and solving problems of this nature and size at design-time \cite{Grodzevich2006}.
Our formulation is facilitated by a two-step task graph transformation approach.
In the first step, we transform the initial task graph (TG) of the workflow application into an intermediate edge-hub-cloud graph (EG). EG represents the possible allocations of the tasks on the considered devices, encapsulating both the application and system models. It also facilitates the definition of the energy and reliability models. 
In the second step, we transform EG into the final reliability-aware edge-hub-cloud graph (REG), taking into account the reliability requirements of each task, by integrating dual and triple task execution techniques. 
Finally, we use the resulting REG to formulate the considered problem. The problem is solved according to the defined trade-off between the two competing objectives, reliability and latency.
Intuitively, the transformation from TG to EG and subsequently to REG makes all feasible task allocations and reliability-aware execution configurations explicit, enabling the BILP solver to select exactly one candidate allocation and time redundancy technique (when applicable) per task, while jointly optimizing reliability and latency.

\subsection{Modeling assumptions}
The proposed framework targets pre-programmed safety-critical workflow applications executed in a streamlined edge-hub-cloud architecture, where the workflow structure is static and known a priori.
In line with the majority of related works in the field \cite{DeMaio2020, Fan2021, Fang2025, Jiang2025, Peng2022, Tang2022, Khaleel2023, Asghari2023, Khurana2023, Wang2018}, we adopt an offline (design-time) optimization approach where the application, system, and network parameters are obtained through profiling or real-world measurements and treated as fixed inputs. 
This static approach is typical and appropriate in the targeted applications \cite{Savva2021, Vladan2021, Zheng2021, Wang2017, Agarwal2022, Alam2017}, where deterministic guarantees on reliability and latency are required. In contrast, dynamic exact methods are computationally infeasible, while heuristic online methods cannot guarantee optimal solutions, nor provide the same level of predictability.


\subsection{Step 1 of TG transformation -- intermediate EG}
\label{subsec:step1Transformation}

The TG of a workflow application that comprises a set of tasks $\mathcal{T} = \{1, \allowbreak 2, \allowbreak \dots, \allowbreak |\mathcal{T}|\}$, is typically represented by a directed acyclic graph $G=( \mathcal{N, \mathcal{A}})$ \cite{Stavrinides2019}.  $\mathcal{N}=\{ N_{1}, \allowbreak N_{2}, \allowbreak \dots, \allowbreak N_{|\mathcal{T}|} \}$ is the set of its nodes, whereas $\mathcal{A} = \{ A_{i \rightarrow j} \, | \, N_{i}, \allowbreak N_{j} \in \mathcal{N}, \, N_i \neq N_j, \, N_{i} \rightarrow N_{j}  \}$ is the set of its arcs.
A node $N_{i} \in \mathcal{N}$ corresponds to a task $i \in \mathcal{T}$ of the application. An arc $A_{i \rightarrow j} \in \mathcal{A}$ between a node $N_{i}$ (corresponding to a parent task $i$) and a node $N_{j}$ (corresponding to a child task $j$), denotes the communication and precedence relationship between the two tasks.

In the first step of the task graph transformation, the initial TG $G$ is transformed into the intermediate EG $\dot{G}=(\dot{\mathcal{N}}, \dot{\mathcal{A}})$, based on the main principles of our approach in \cite{Kouloumpris2024}.
Specifically, each node $N_{i} \in \mathcal{N}$ in $G$  is transformed into a composite node (i.e., a set of nodes) $\dot{N}_{i} \in \dot{\mathcal{N}}$ in $\dot{G}$, such that:
\begin{equation}
\label{eq:nodeTransformationEG}
 \dot{\phi}_{\mathrm{node}} (N_{i}) = \dot{N}_{i} = \{ N_{ik} \, | \, k \in \mathcal{Q}_{i} \subseteq \mathcal{U} \},
\end{equation}
where $\dot{\phi}_{\mathrm{node}}$ is the function that transforms the node $N_{i}$ into the set $\dot{N}_{i}$ as shown above, and $\mathcal{U}$ is the set of devices in the examined edge-hub-cloud environment.
$k$ is a device in the subset of devices $\mathcal{Q}_{i} \subseteq \mathcal{U}$ where task $i$ can be allocated.
If task $i$ can be allocated on any device in $\mathcal{U}$, then $\mathcal{Q}_{i} = \mathcal{U}$, otherwise $\mathcal{Q}_{i} \subset \mathcal{U}$.
Hence, an individual node $N_{ik} \in \dot{G}$ denotes the possible allocation of task $i$ on device $k$.

Each arc $A_{i \rightarrow j} \in \mathcal{A}$ in $G$ is transformed into a composite arc (i.e., a set of arcs) $\dot{A}_{i \rightarrow j} \in \dot{\mathcal{A}}$ in $\dot{G}$, such that:
\begin{equation}
\label{eq:arcTransformationEG}
 \begin{split}
     \dot{\phi}_{\mathrm{arc}} (A_{i \rightarrow j}) = \dot{A}_{i \rightarrow j}  =  \{ & A_{ik \rightarrow jl}  = N_{ik} \rightarrow N_{jl}\\ 
     &  \, | \, k \in \mathcal{Q}_{i} \subseteq \mathcal{U}, \,  l \in \mathcal{Q}_{j} \subseteq \mathcal{U} \},
 \end{split}
\end{equation}
where $\dot{\phi}_{\mathrm{arc}}$ is the function that transforms the arc $A_{i \rightarrow j}$ into the set $\dot{A}_{i \rightarrow j}$ as shown above.
An individual arc $A_{ik \rightarrow jl} \in \dot{G}$ represents the communication and precedence relationship between the nodes it connects, $N_{ik}$ and $N_{jl}$.
It is noted that if the communication between devices $k$ and $l$ cannot be performed directly (e.g., between an edge device and a cloud server in the examined architecture), it is performed through an intermediate device (e.g., a hub device), incurring additional communication latency and energy costs. If both tasks $i$ and $j$ are on the same device (i.e., $k = l$), the involved communication cost is considered negligible compared to the case where $k \neq l$.

\subsubsection{EG node parameters}
\label{subsubsec:EGNodeParams}

A node $N_{ik} \in \dot{G}$ has the following parameters:
\begin{itemize}
    \item $M_{i}$: main memory required by task $i$.
    
    \item $S_{i}$: storage required by task $i$.
    
    \item $D_{i}$: size of output data generated by task $i$.
    
    \item $L_{ik}$: execution time of task $i$ on device $k$.
   
    \item $P_{ik}$: power required to execute task $i$ on device $k$. 
    
    \item $E_{ik}$: energy required to execute task $i$ on device $k$.

    \item $V_{ik}$: vulnerability factor of task $i$ on device $k$ (i.e., probability that the execution of task $i$ on device $k$ fails).

    \item $R_{ik}$: reliability of task $i$ on device $k$ (i.e., probability that task $i$ is executed on device $k$ without any failures).
     
    \item $\epsilon_{ik}$: execution mode of task $i$ when allocated on device $k$, indicating whether task $i$ requires single execution (SE) on device $k$, or dual execution (DE) or triple execution (TE), where the replicated executions may be performed on the same or on different devices to increase reliability.
 
    \item $\nu_{i}$: number of child tasks of task $i$ (i.e., number of immediate successor nodes of node $N_{i} \in G$).  
\end{itemize}

Parameters $M_{i}$, $S_{i}$, $D_{i}$, and $\nu_{i}$ are  device-independent, whereas  $L_{ik}$, $P_{ik}$, $E_{ik}$, $V_{ik}$, $R_{ik}$, and $\epsilon_{ik}$ are device-dependent.
$M_{i}$, $S_{i}$, $D_{i}$, $L_{ik}$, and $P_{ik}$ can be defined through performance profiling and power monitoring tools, as explained in \cref{sec:evaluation}. 
$E_{ik}$ is given by \eqref{eq:compEnergy}, based on the energy model described in \cref{subsec:energyModel}.
$V_{ik}$ can be profiled via well-known approaches based on fault injection techniques and/or architecturally correct execution (ACE) analysis \cite{Mukherjee2003, Maniatakos2014}.
$R_{ik}$ and $\epsilon_{ik}$ are defined by \eqref{eq:reliability} and \eqref{eq:execMode}, respectively, based on the reliability model presented in \cref{subsec:reliabilityModel}.
Finally, $\nu_{i}$ is a structural characteristic of $G$ and thus easily derived.

\subsubsection{EG arc parameters}
\label{subsubsec:EGArcParams}

An arc $A_{ik \rightarrow jl} \in \dot{G}$ has the following parameters:
\begin{itemize}
     \item $\zeta_{ik \rightarrow jl}^{o}$: binary parameter indicating whether $A_{ik \rightarrow jl}$ involves indirect communication between devices $k$ and $l$, through intermediate device $o$:
    \begin{equation}
    \label{eq:zeta}
         \zeta_{ik \rightarrow jl}^{o} =
         \begin{cases}
            1, & \text{if devices $k$ and $l$ cannot communicate}\\
               & \text{directly and communication is}\\
               & \text{performed via device $o$},\\
            0, & \text{otherwise}.
         \end{cases}
    \end{equation}
    
    \item $CL_{ik \rightarrow jl}$: time required to transfer the output data $D_{i}$ of task $i$ (on device $k$) to task $j$ (on device $l$): 
    \begin{equation}
    \label{eq:commLatency}
         CL_{ik \rightarrow jl} =  
         \begin{cases}
            D_{i} / \sigma_{kl}, & \text{if $\zeta_{ik \rightarrow jl}^{o}=0$},\\
            & k \neq l,\\
            D_{i}  \left( 1 / \sigma_{ko} + 1 / \sigma_{ol} \right), & \text{if $\zeta_{ik \rightarrow jl}^{o}=1$},\\
            0, & \text{if $k = l$},
         \end{cases}
    \end{equation}
    where $\sigma_{kl}$, $\sigma_{ko}$, and $\sigma_{ol}$ denote the bandwidth of the communication channels between the respective devices.
    
    \item $CE_{ik \rightarrow jl}$: energy required to transfer the output data $D_{i}$ of task $i$ (on device $k$) to task $j$ (on device $l$). It is given by \eqref{eq:commEnergy}, based on the energy model in \cref{subsec:energyModel}.
\end{itemize}

\begin{figure*}[!t]
    \centering
    \includegraphics[width=0.93\textwidth]{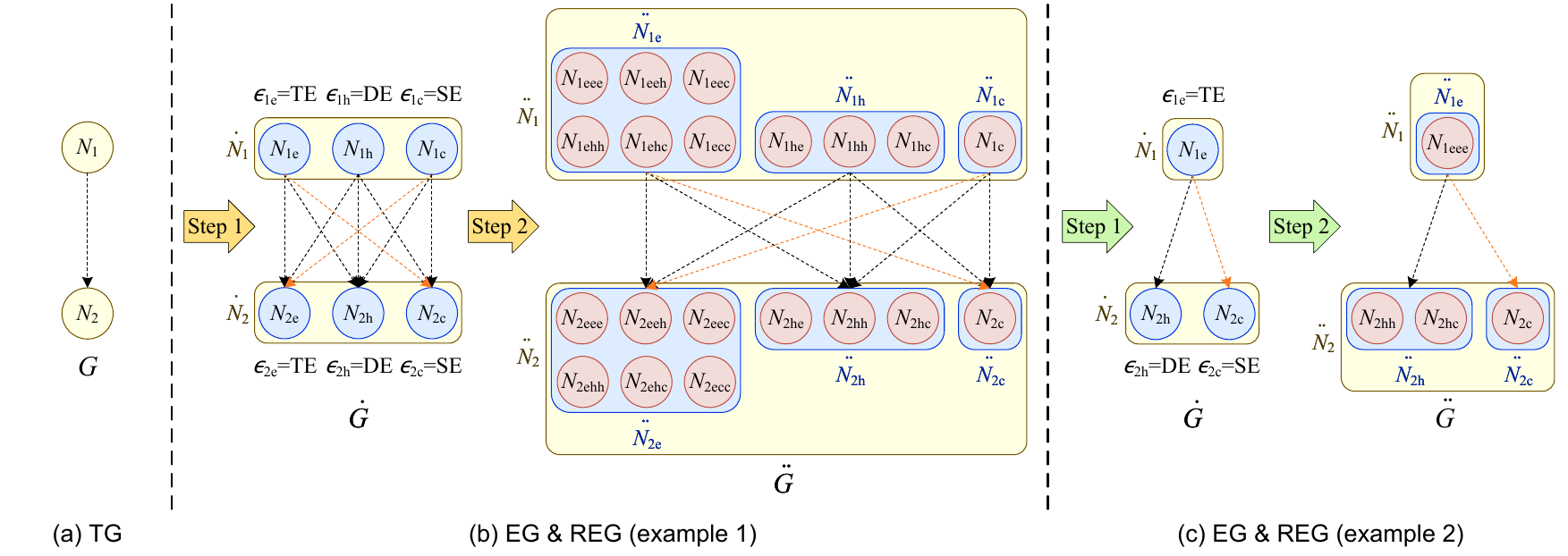}
     \caption{Examples of transforming an application's initial TG $G$ into its corresponding intermediate EG $\dot{G}$ and final REG $\ddot{G}$.}
    \label{fig:transformation}
\end{figure*}

\subsubsection{Transformation examples (EG)}
\label{subsubsec:examplesStep1}
\cref{fig:transformation} illustrates two examples of our two-step transformation technique. 
The initial TG $G$ of an application with two tasks is shown in \cref{fig:transformation}a.
The considered edge-hub-cloud architecture comprises a single edge device ($\mathrm{e}$), a single hub device ($\mathrm{h}$), and a single cloud server ($\mathrm{c}$), i.e., $\mathcal{U} = \{ \mathrm{e}, \mathrm{h}, \mathrm{c}\}$.
In step 1 of the first example (\cref{fig:transformation}b), $G$ is transformed into $\dot{G}$ considering that both tasks can be allocated on any device, i.e., $\mathcal{Q}_1 = \mathcal{Q}_2 = \mathcal{U} = \{ \mathrm{e}, \allowbreak \mathrm{h}, \allowbreak \mathrm{c} \}$. 
In step 1 of the second example (\cref{fig:transformation}c), $G$ is transformed into $\dot{G}$ considering that task 1 requires fixed allocation on device $\mathrm{e}$ ($\mathcal{Q}_1 = \{ \mathrm{e} \}$), and task 2 on device $\mathrm{h}$ or $\mathrm{c}$ ($\mathcal{Q}_2 = \{ \mathrm{h}, \allowbreak \mathrm{c} \}$).
In both examples, we assume that a task requires triple ($\mathrm{TE}$), dual ($\mathrm{DE}$), or single ($\mathrm{SE}$) execution if it is allocated on device $\mathrm{e}$, $\mathrm{h}$, or $\mathrm{c}$, respectively.
The arcs involving devices $\mathrm{e}$ and $\mathrm{c}$, which cannot communicate directly, are depicted in orange. Step 2 of each example is described in \cref{subsubsec:examplesStep2}.

\subsection{Energy model}
\label{subsec:energyModel}

The computational energy $E_{ik}$ required to execute task $i$ on device $k$ is given by \cite{Saeedi2020, Taghinezhad2024}:
\begin{equation}
\label{eq:compEnergy}
    E_{ik} = L_{ik} \,  P_{ik}.
\end{equation}
The communication energy $CE_{ik \rightarrow jl}$ required to transfer the output data $D_{i}$ of task $i$, allocated on device $k$, to task $j$, allocated on device $l$, is defined as \cite{Wang2017}:
\begin{equation}
\label{eq:commEnergy}
    CE_{ik \rightarrow jl} = 
    \begin{cases}
        D_{i}  \left(\tau_{kl} + \rho_{kl} \right), & \text{if $\zeta_{ik \rightarrow jl}^{o}=0$},\\ 
         & k \neq l,\\
        D_{i}  \left(\tau_{ko} + \rho_{ko} + \tau_{ol} + \rho_{ol}  \right), & \text{if $\zeta_{ik \rightarrow jl}^{o}=1$},\\
        0, & \text{if $k = l$},
    \end{cases}
\end{equation}
where $\tau_{kl}$, $\tau_{ko}$, and $\tau_{ol}$ ($\rho_{kl}$, $\rho_{ko}$, and $\rho_{ol}$) denote the energy required to transmit (receive) a unit of data over the respective communication channels.
This energy model builds upon established formulations widely adopted in related studies \cite{Wang2017, Saeedi2020, Taghinezhad2024}. 
It captures both the computational and communication energy required for workflow execution in the examined edge-hub-cloud architecture, while remaining independent of low-level architectural details for broader applicability.

\subsection{Reliability model}
\label{subsec:reliabilityModel}

We consider a criticality level $\mathit{\Lambda} \in \{1, \allowbreak 2, \allowbreak \dots, \allowbreak \mathit{\Lambda}_{\mathrm{max}} \}$ for the examined application.
A higher $\mathit{\Lambda}$ denotes a higher criticality and thus a lower vulnerability tolerance ($\mathit{\Lambda}_{\mathrm{max}}$ is the highest criticality level).
We employ two thresholds to represent the vulnerability tolerance of the application: (a) a vulnerability threshold for \emph{dual task execution}, which is inversely proportional to $\mathit{\Lambda}$:
\begin{equation}
\label{eq:vtDE}
    VT_{\mathrm{DE}} = \frac{\kappa} {\mathit{\Lambda}},
\end{equation}
and (b) a vulnerability threshold for \emph{triple task execution}:
\begin{equation}
\label{eq:vtTE}
    VT_{\mathrm{TE}} = \lambda \, VT_{\mathrm{DE}}, \, \lambda > 1,
\end{equation}
where $\kappa$ and $\lambda$ are adjustment coefficients. 
$VT_{\mathrm{DE}}$ and $VT_{\mathrm{TE}}$ are used in \eqref{eq:execMode} to determine if a task $i$ of the application, when allocated on a device $k$, requires (based on its vulnerability factor $V_{ik}$) duplication or triplication, respectively, to increase its reliability \cite{Ottavi2014}.
As $V_{ik}$ is the probability the execution of task $i$ on device $k$ fails, we consider that $0 < V_{ik} < 1$.
Hence, the reliability of task $i$ on device $k$ is given by \cite{Xie2020, Taghinezhad2023}:
\begin{equation}
\label{eq:reliability}
    R_{ik} = 1 - V_{ik}.
\end{equation}

The execution mode $\epsilon_{ik}$ of task $i$ when allocated on device $k$ depends on $V_{ik}$, $VT_{\mathrm{DE}}$, and $VT_{\mathrm{TE}}$, as follows: 
\begin{equation}
\label{eq:execMode}
    \epsilon_{ik} = 
    \begin{cases}
        \mathrm{SE}, &  \text{if $V_{ik} < VT_{\mathrm{DE}}$},\\

        \mathrm{DE}, &  \text{if $VT_{\mathrm{DE}} \leq V_{ik} < VT_{\mathrm{TE}}$},\\

        \mathrm{TE}, &  \text{if $VT_{\mathrm{TE}} \leq V_{ik}$}.
    \end{cases}
\end{equation}
If $\epsilon_{ik}=\mathrm{SE}$, task $i$ is executed only once on device $k$, as its reliability requirements are met. Otherwise,  if $\epsilon_{ik}=\mathrm{DE}$ or $\epsilon_{ik}=\mathrm{TE}$, task $i$ is replicated and executed twice or three times, respectively, either sequentially on the same device or in parallel on different devices, to increase its reliability. 
If $\epsilon_{ik}=\mathrm{DE}$, the outputs of both executions of task $i$ are compared (for error detection), whereas if $\epsilon_{ik}=\mathrm{TE}$, the outputs of all three executions are subjected to a majority vote (for error mitigation).
It is noted that in this work we consider transient errors.
We use $\epsilon_{ik}$ to transform $\dot{G}$ into $\ddot{G}$, as described in \cref{subsec:step2Transformation}.
The adopted reliability model is based on widely used dual and triple execution techniques for detecting and mitigating transient errors \cite{Ottavi2014, Xie2020, Taghinezhad2023}. By linking application criticality to vulnerability thresholds, it enforces stricter reliability requirements at higher criticality levels, effectively capturing the dependencies between reliability and task allocation in the examined edge-hub-cloud architecture.

\subsection{Step 2 of TG transformation -- final REG}
\label{subsec:step2Transformation}

In the second step of the task graph transformation, the intermediate EG $\dot{G}$ is transformed into the final REG  $\ddot{G} = ( \ddot{\mathcal{N}}, \ddot{\mathcal{A}} )$, 
by considering for each node $N_{ik} \in \dot{G}$ the task replication technique determined by \eqref{eq:execMode}. 
Specifically, each composite node $\dot{N}_{i} \in \dot{\mathcal{N}}$ in $\dot{G}$ is transformed into another composite node $\ddot{N}_{i} \in \ddot{\mathcal{N}}$ in $\ddot{G}$, such that:
\begin{equation}
\label{eq:nodeTransformationREG}
    \ddot{\phi}_{\mathrm{node}} (\dot{N}_{i}) = \ddot{N}_{i} = \{ \ddot{N}_{ik} \, | \, \ddot{N}_{ik} = \ddot{\psi} (N_{ik}), \, N_{ik} \in \dot{N}_{i} \},
\end{equation}
where $\ddot{\phi}_{\mathrm{node}}$ is the function that transforms the set $\dot{N}_{i}$ into the final set $\ddot{N}_{i}$ as shown above. 
$\ddot{\psi}$ is the function that transforms each individual node $N_{ik} \in \dot{N}_{i}$ into the set $\ddot{N}_{ik} \in \ddot{N}_{i}$, so that:
\begin{equation}
\label{eq:candidateNodes}
   \ddot{N}_{ik} = 
    \begin{cases}
    \left\{ N_{ik} = \{ r_{ik}^1 \} \right\}, &  \text{if $\epsilon_{ik} = \mathrm{SE}$},\\
   
    \left\{ N_{ikl} = \{ r_{ik}^1, r_{il}^2 \} \, | \, l \in \mathcal{Q}_{i} \right\}, &  \text{if $\epsilon_{ik} = \mathrm{DE}$},\\
     
    \left\{ N_{iklm} = \{ r_{ik}^1, r_{il}^2, r_{im}^3 \} \, | \, l, m \in \mathcal{Q}_{i} \right\},
     &  \text{if $\epsilon_{ik} = \mathrm{TE}$,}
    \end{cases}
\end{equation}
where $r_{in}^z$ represents the allocation of primary task $i$ or one of its replicas on device $n$. 
Specifically, if $z = 1$, it denotes \emph{primary} task $i$, allocated on its initially assigned (as in $\dot{G}$) device $k$ (\emph{primary device}). 
If $z = 2$ or $z = 3$, it denotes a \emph{replica} of task $i$, allocated on device $l \in \mathcal{Q}_i$ or $m \in \mathcal{Q}_i$, respectively ($l$ and/or $m$ may be the same as the primary device $k$).

Consequently, if $\epsilon_{ik} = \mathrm{SE}$, $N_{ik} \in \dot{G}$ remains as one node $N_{ik} \in \ddot{G}$, representing the allocation of primary task $i$ on primary device $k$.
On the other hand, if $\epsilon_{ik} = \mathrm{DE}$, $N_{ik} \in \dot{G}$ is transformed into a set of nodes $\ddot{N}_{ik} \in \ddot{G}$.
Each individual node $N_{ikl} \in \ddot{N}_{ik}$ represents the duplication of task $i$, so that primary task $r_{ik}^1$ is allocated on primary device $k$ and its replica $r_{il}^2$ is allocated on device $l$.
If $l \neq k$, $r_{il}^2$ sends its output back to device $k$.
The outputs of $r_{ik}^1$ and $r_{il}^2$ are compared on device $k$ (for error detection).
Similarly, if $\epsilon_{ik} = \mathrm{TE}$, $N_{ik} \in \dot{G}$ is transformed into a set of nodes $\ddot{N}_{ik} \in \ddot{G}$. 
Each individual node $N_{iklm} \in \ddot{N}_{ik}$ represents the triplication of task $i$, so that $r_{ik}^1$ is allocated on device $k$, whereas $r_{il}^2$ and $r_{im}^3$ are allocated on devices $l$ and $m$, respectively.
If $l \neq k$ and/or $m \neq k$, $r_{il}^2$ and/or $r_{im}^3$, respectively, send their outputs back to device $k$. A majority vote is taken on device $k$ on the outputs of $r_{ik}^1$, $r_{il}^2$, and $r_{im}^3$ (for error mitigation).

For notational convenience, an individual node $N_{ik}$, $N_{ikl}$, or $N_{iklm}$ in $\ddot{G}$ ($k, l, m \in \mathcal{Q}_i$) is denoted in abstracted form as $N_{i\hat{k}}$, and is referred to as \emph{candidate node}.  
The possible cases of the non-abstracted form of $N_{i\hat{k}}$, with respect to the devices where its task replicas may be allocated, are defined as:
\begin{align}
N_{i\hat{k}} &= N_{ik},   && \text{if}\ \epsilon_{ik}=\mathrm{SE}, \label{eq:case1}\\
             &= N_{ikk},  && \text{if}\ \epsilon_{ik}=\mathrm{DE},\, k = l, \label{eq:case2}\\
             &= N_{ikl},  && \text{if}\ \epsilon_{ik}=\mathrm{DE},\, k \neq l, \label{eq:case3}\\
             &= N_{ikkk}, && \text{if}\ \epsilon_{ik}=\mathrm{TE},\, k = l = m, \label{eq:case4}\\
             &= N_{ikkm}, && \text{if}\ \epsilon_{ik}=\mathrm{TE},\, k = l \neq m, \label{eq:case5}\\
             &= N_{ikll}, && \text{if}\ \epsilon_{ik}=\mathrm{TE},\, k \neq l = m, \label{eq:case6}\\
             &= N_{iklm}, && \text{if}\ \epsilon_{ik}=\mathrm{TE},\, k \neq l \neq m. \label{eq:case7}
\end{align}
These cases are referenced in \eqref{eq:totalLatency}--\eqref{eq:totalVulnerability} to maintain brevity.

For an arc $A_{ik \rightarrow jl} \in \dot{G}$, the communication between any two candidate nodes $N_{i\hat{k}} \in \ddot{N}_{ik}$ and $N_{j\hat{l}} \in \ddot{N}_{jl}$ in $\ddot{G}$ is performed over the communication channel between primary devices $k$ and $l$, regardless of where the task replicas in $N_{i\hat{k}}$ and $N_{j\hat{l}}$ are allocated.
Therefore, each composite arc $\dot{A}_{i \rightarrow j} \in \dot{\mathcal{A}}$ in $\dot{G}$ is transformed into another composite arc $\ddot{A}_{i \rightarrow j} \in \ddot{\mathcal{A}}$ in $\ddot{G}$, such that:
\begin{equation}
\begin{split}
    \ddot{\phi}_{\mathrm{arc}} (\dot{A}_{i \rightarrow j}) = \ddot{A}_{i \rightarrow j} = \{ & A_{ik \rightarrow jl} = \ddot{N}_{ik} \rightarrow \ddot{N}_{jl} \\ 
    & \, | \, A_{ik \rightarrow jl} \in \dot{A}_{i \rightarrow j}  \},
\end{split}
\end{equation}
where $\ddot{\phi}_{\mathrm{arc}}$ is the function that transforms the set $\dot{A}_{i \rightarrow j}$ into the final set $\ddot{A}_{i \rightarrow j}$ as shown above. 
Hence, an individual arc $A_{ik \rightarrow jl}$ between nodes $N_{ik}$ and $N_{jl}$ in $\dot{G}$ is retained as arc $A_{ik \rightarrow jl}$ between corresponding sets $\ddot{N}_{ik}$ and $\ddot{N}_{jl}$ in $\ddot{G}$.

\subsubsection{REG candidate node parameters}
\label{subsubsec:REGNodeParams}

A candidate node $N_{i\hat{k}} \in \ddot{G}$ inherits $\epsilon_{ik}$ and $\nu_{i}$ from $N_{ik} \in \dot{G}$. 
Each primary task or replica $r_{in}^z \in N_{i\hat{k}}$ inherits $M_{i}$, $S_{i}$, $D_{i}$, $L_{in}$, $P_{in}$, $E_{in}$, $V_{in}$, and $R_{in}$ from the respective node $N_{in} \in \dot{G}$.
Additionally, $N_{i\hat{k}}$ has the following aggregate parameters:
\begin{itemize} 
    \item $L_{i\hat{k}}$:
    total time required to execute primary task $i$ and its replicas on their allocated devices, including the time required to transfer the input data $D_j$ from primary device $k$ to the devices of the replicas, return the output $D_i$ of each replica back to device $k$, and perform the comparison/voting on device $k$.
    It is defined based on the different forms of $N_{i\hat{k}}$ in \eqref{eq:case1}--\eqref{eq:case7}:
    \begin{equation}
    \label{eq:totalLatency}
     L_{i\hat{k}} =
     \begin{cases}
        \hspace{-1pt} L_{ik}, &  \hspace{-5pt} \text{if \eqref{eq:case1}},\\

        \hspace{-1pt} 2L_{ik} + \beta_{k}^{\mathrm{DE}}, & \hspace{-5pt} \text{if \eqref{eq:case2}},\\
        
        \hspace{-1pt} \max \{ L_{ik}, \frac{D_j}{\sigma_{kl}} + L_{il} + \frac{D_i}{\sigma_{lk}} \} + \beta_{k}^{\mathrm{DE}}, & \hspace{-5pt} \text{if \eqref{eq:case3}},\\

        \hspace{-1pt} 3L_{ik} + \beta_{k}^{\mathrm{TE}}, & \hspace{-5pt} \text{if \eqref{eq:case4}},\\
            
        \hspace{-1pt} \max \{ 2L_{ik}, \frac{D_j}{\sigma_{km}} + L_{im} + \frac{D_i}{\sigma_{mk}} \} + \beta_{k}^{\mathrm{TE}}, & \hspace{-5pt} \text{if \eqref{eq:case5}},\\

        \hspace{-1pt} \max \{ L_{ik}, \frac{D_j}{\sigma_{kl}} + 2 ( L_{il} + \frac{D_i}{\sigma_{lk}} ) \} + \beta_{k}^{\mathrm{TE}}, & \hspace{-5pt} \text{if \eqref{eq:case6}},\\
        
        \hspace{-1pt} \max \{ L_{ik}, \frac{D_j}{\sigma_{kl}} + L_{il} + \frac{D_i}{\sigma_{lk}}, & \\ 
        \hspace{42pt} \frac{D_j}{\sigma_{km}} + L_{im} + \frac{D_i}{\sigma_{mk}} \} \hspace{-2pt} + \beta_{k}^{\mathrm{TE}}, & \hspace{-5pt} \text{if \eqref{eq:case7}},
        \end{cases}
    \end{equation}
    where $\beta_{k}^{\mathrm{DE}}$ and $\beta_{k}^{\mathrm{TE}}$ denote the time required to perform the comparison and voting, respectively, on device $k$.

    \item $E_{in}^{z, i\hat{k}}$:
    total energy required by device $n$ to execute $r_{in}^z$, including (a) the energy required to receive $r_{in}^z$'s input $D_j$ from primary device $k$ and transmit back its output $D_i$ (if $n \neq k$), or (b) the energy required to transmit $D_j$ to the devices of the replicas, receive $D_i$ from each replica, and execute the comparison/voting (if $n = k$).
    It is defined based on the different forms of $N_{i\hat{k}}$ in \eqref{eq:case1}--\eqref{eq:case7}, considering if $r_{in}^z$ is the primary task ($z=1$) or not ($z \in \{2,3\}$), and if $n=k$ or $n \neq k$:
    \begin{equation}
    \label{eq:totalEnergyPerReplicaDevice}
        E_{in}^{z, i\hat{k}} =
        \begin{cases}
        \hspace{-1pt} E_{ik}, & \hspace{10pt} \text{if $ z \hspace{-1pt} = \hspace{-1pt} 1 \hspace{-1pt} \land \hspace{-1pt} \eqref{eq:case1}$}\\
            \multicolumn{2}{r}{\hspace{3pt} \text{$\lor \hspace{-1pt} \left( z \hspace{-1pt} \in \hspace{-1pt} \{ 2, 3 \} \hspace{-1pt} \land \hspace{-1pt} \left( \eqref{eq:case2} \hspace{-1pt} \lor \hspace{-1pt} \eqref{eq:case4} \hspace{-1pt} \lor \hspace{-1pt} \left( \eqref{eq:case5} \hspace{-1pt} \land n \hspace{-1pt} = \hspace{-1pt} k \right) \right) \right)$},}\\

        \hspace{-1pt} E_{ik} \hspace{-1pt} + \hspace{-1pt} \delta_{k}^{\mathrm{DE}}, & \hspace{10pt} \text{if $z \hspace{-1pt} = \hspace{-1pt} 1 \hspace{-1pt} \land \hspace{-1pt} \eqref{eq:case2}$},\\

        \hspace{-1pt} E_{ik} \hspace{-1pt} + \hspace{-1pt} D_{j}\tau_{kl} \hspace{-1pt} + \hspace{-1pt} D_{i}\rho_{lk} \hspace{-1pt} + \hspace{-1pt} \delta_{k}^{\mathrm{DE}}, & \hspace{10pt} \text{if $z \hspace{-1pt} = \hspace{-1pt} 1 \hspace{-1pt} \land \hspace{-1pt} \eqref{eq:case3}$},\\

        \hspace{-1pt} E_{ik} \hspace{-1pt} + \hspace{-1pt} \delta_{k}^{\mathrm{TE}}, & \hspace{10pt} \text{if $z \hspace{-1pt} = \hspace{-1pt} 1 \hspace{-1pt} \land \hspace{-1pt} \eqref{eq:case4}$},\\

        \hspace{-1pt} E_{ik} \hspace{-1pt} + \hspace{-1pt} D_{j}\tau_{km} \hspace{-1pt} + \hspace{-1pt} D_{i}\rho_{mk} \hspace{-1pt} + \hspace{-1pt} \delta_{k}^{\mathrm{TE}}, & \hspace{10pt} \text{if $z \hspace{-1pt} = \hspace{-1pt} 1 \hspace{-1pt} \land \hspace{-1pt} \eqref{eq:case5}$},\\

        \hspace{-1pt} E_{ik} \hspace{-1pt} + \hspace{-1pt} D_{j}\tau_{kl} \hspace{-1pt} + \hspace{-1pt} 2D_{i}\rho_{lk} \hspace{-1pt} + \hspace{-1pt} \delta_{k}^{\mathrm{TE}}, & \hspace{10pt} \text{if $z \hspace{-1pt} = \hspace{-1pt} 1 \hspace{-1pt} \land \hspace{-1pt} \eqref{eq:case6}$},\\

        \hspace{-1pt} E_{ik} \hspace{-1pt} + \hspace{-1pt} D_{j}\left(\tau_{kl} \hspace{-1pt} + \hspace{-1pt} \tau_{km}\right) & \\
        \hspace{13pt} + \hspace{1pt} D_{i}\left(\rho_{lk} \hspace{-1pt} + \hspace{-1pt} \rho_{mk}\right) \hspace{-1pt} + \hspace{-1pt} \delta_{k}^{\mathrm{TE}}, & \hspace{10pt} \text{if $z \hspace{-1pt} = \hspace{-1pt} 1 \hspace{-1pt} \land \hspace{-1pt} \eqref{eq:case7}$},\\

        \hspace{-1pt} D_{j}\rho_{kl} \hspace{-1pt} + \hspace{-1pt} E_{il} \hspace{-1pt} + \hspace{-1pt} D_{i}\tau_{lk}, & \hspace{10pt} \text{if $z \hspace{-1pt} \in \hspace{-1pt} \{ 2, 3 \}$}\\
        \multicolumn{2}{r}{\text{$\land \hspace{-1pt} \left( \eqref{eq:case3} \hspace{-1pt} \lor \hspace{-1pt} \eqref{eq:case6} \hspace{-1pt} \lor \hspace{-1pt} \left( \eqref{eq:case7} \hspace{-1pt} \land n \hspace{-1pt} = \hspace{-1pt} l \right) \right)$},}\\

        \hspace{-1pt} D_{j}\rho_{km} \hspace{-1pt} + \hspace{-1pt} E_{im} \hspace{-1pt} + \hspace{-1pt} D_{i}\tau_{mk}, & \hspace{10pt} \text{if $z \hspace{-1pt} \in \hspace{-1pt} \{ 2, 3 \}$}\\
        \multicolumn{2}{r}{\text{$\land \hspace{-1pt} \left( \eqref{eq:case5} \hspace{-1pt} \lor \hspace{-1pt} \eqref{eq:case7} \right) \hspace{-1pt} \land n \hspace{-1pt} = \hspace{-1pt} m$},}
        \end{cases}
    \end{equation}
    where $\delta_{k}^{\mathrm{DE}} = \beta_{k}^{\mathrm{DE}} \, \gamma_{k}^{\mathrm{DE}}$ and $\delta_{k}^{\mathrm{TE}} = \beta_{k}^{\mathrm{TE}} \, \gamma_{k}^{\mathrm{TE}}$ denote the energy required to perform the comparison and voting, respectively, on device $k$. $\gamma_{k}^{\mathrm{DE}}$ and $\gamma_{k}^{\mathrm{TE}}$ indicate the power consumption in each case.

    \item $V_{i\hat{k}}$: total vulnerability factor of $N_{i\hat{k}}$, i.e., probability that the execution of primary task $i$ and all of its replicas fails.
    It depends on the different forms of $N_{i\hat{k}}$ in \eqref{eq:case1}--\eqref{eq:case7}, as follows \cite{Taghinezhad2023}:
    \begin{equation}
    \label{eq:totalVulnerability}
        V_{i\hat{k}}  = 
        \begin{cases}
        V_{ik},  &  \text{if \eqref{eq:case1}},\\
        \left( V_{ik} \right)^2, &  \text{if \eqref{eq:case2}},\\
        V_{ik}\,V_{il}, &  \text{if \eqref{eq:case3}},\\
        \left( V_{ik} \right)^3, & \text{if \eqref{eq:case4}},\\
        \left( V_{ik} \right)^2\,V_{im}, & \text{if \eqref{eq:case5}},\\
        V_{ik}\,\left( V_{il} \right)^2, & \text{if \eqref{eq:case6}},\\
        V_{ik}\,V_{il}\,V_{im}, & \text{if \eqref{eq:case7}}.
        \end{cases}
    \end{equation}

    \item $R_{i\hat{k}}$: total reliability of $N_{i\hat{k}}$, i.e., probability that primary task $i$ or at least one of its replicas is executed without any failures.
    It is given by \cite{Xie2020}:
    \begin{equation}
    \label{eq:totalReliability}
        R_{i\hat{k}} = 1 - V_{i\hat{k}}.
    \end{equation}
\end{itemize}

Without loss of generality, in \eqref{eq:totalLatency} and \eqref{eq:totalEnergyPerReplicaDevice} we consider that each pair of devices communicate directly.
If the communication is indirect, then the same conditions hold, but the relevant $\sigma$, $\tau$, and $\rho$ parameters are used.

\subsubsection{REG arc parameters}
\label{subsubsec:REGArcParams}
An arc $A_{ik \rightarrow jl} \in \ddot{G}$ has the same parameters $\zeta_{ik \rightarrow jl}^{o}$, $CL_{ik \rightarrow jl}$, and $CE_{ik \rightarrow jl}$ as $A_{ik \rightarrow jl} \in \dot{G}$.

\subsubsection{Transformation examples (REG)}
\label{subsubsec:examplesStep2}
Each intermediate EG $\dot{G}$ described in \cref{subsubsec:examplesStep1} is transformed into the corresponding REG $\ddot{G}$ as illustrated by step 2 in Figs. \ref{fig:transformation}b and \ref{fig:transformation}c. 
Specifically, each node $N_{ik} \in \dot{G}$ is transformed into a set $\ddot{N}_{ik} \in \ddot{G}$, based on $\epsilon_{ik}$. 
Each set $\ddot{N}_{ik} \in \ddot{G}$ comprises candidate nodes that represent the possible allocations of the primary task and its replicas in each case.
For instance, in the first example (\cref{fig:transformation}b), if task 1 is allocated on device $\mathrm{e}$ (denoted by $N_{1\mathrm{e}} \in \dot{G}$), it requires triple execution. The possible allocations of its replicas on its eligible devices ($\mathcal{Q}_1 = \mathcal{U} = \{ \mathrm{e}, \allowbreak \mathrm{h}, \allowbreak \mathrm{c} \}$) are represented by the six candidate nodes in set $\ddot{N}_{1\mathrm{e}} \in \ddot{G}$.
On the other hand, in the second example (\cref{fig:transformation}c), task 1 can only be allocated on device $\mathrm{e}$. 
Since $N_{1{\mathrm{e}}} \in \dot{G}$ requires triple execution, $\ddot{N}_{1\mathrm{e}} \in \ddot{G}$ solely consists of candidate node $N_{1\mathrm{eee}}$.
As in $\dot{G}$, the arcs involving devices $\mathrm{e}$ and $\mathrm{c}$ are shown in orange.

\subsubsection{Impact of transformation on task graph size}
\label{subsubsec:graphSize}
Evidently, the transformation of the initial TG $G$ into the final REG $\ddot{G}$ increases the size of the resulting graph. In the worst case, where no task requires fixed allocation and all nodes in the intermediate EG $\dot{G}$ require triple execution, for $u$ different devices, the number of nodes in $\ddot{G}$ increases by $u^2 (u+1)/2$ times, whereas the number of arcs increases by $u^2$ times, with respect to those in $G$. 

\subsection{Multi-objective BILP formulation}
\label{subsec:BILPFormulation}

We utilize the resulting REG $\ddot{G}$ to formulate the problem as a BILP model, as follows:

\subsubsection{Decision variables}
\label{subsubsec:decisionVariables}
We employ the following variables:
\begin{itemize}
    \item A binary variable $x_{i\hat{k}}$ corresponding to a candidate node $N_{i\hat{k}} \in \ddot{G}$, such that $x_{i\hat{k}} = 1$ if $N_{i\hat{k}}$ is selected, and $x_{i\hat{k}} = 0$ otherwise.

    \item A binary variable $x_{ik \rightarrow jl}$ corresponding to an arc $A_{ik \rightarrow jl} \in \ddot{G}$, such that $x_{ik \rightarrow jl} = 1$ if $A_{ik \rightarrow jl}$ is selected, and $x_{ik \rightarrow jl} = 0$ otherwise.

    \item An auxiliary binary variable $x_{in}^{z, i\hat{k}}$ corresponding to a primary task or a replica $r_{in}^z \in N_{i\hat{k}}$, such that $x_{in}^{z, i\hat{k}} = 1$ if $r_{in}^z$ is selected, and $x_{in}^{z, i\hat{k}} = 0$ otherwise.

    \item An auxiliary binary variable $x_{ik}$ corresponding to a set $\ddot{N}_{ik} \in \ddot{G}$, such that $x_{ik} = 1$ if $\ddot{N}_{ik}$ is selected, and $x_{ik} = 0$ otherwise.
\end{itemize}

\subsubsection{Objective 1}
\label{subsubsec:objective1}
The first objective aims to maximize the overall reliability of the application on the target system. 
The overall reliability denotes the probability that each primary task of the application, or at least one of its replicas, is executed on its allocated device without any failures. It is given by \cite{Taghinezhad2023}:
\begin{equation}
\label{eq:overallReliability}
    R(\ddot{G}) = \prod_{ N_{i\hat{k}} \in  \ddot{\mathcal{N}}}  \left( R_{i\hat{k}} \right)^{x_{i\hat{k}}}. 
\end{equation}

To formulate the objective function, we first linearize \eqref{eq:overallReliability} by taking its logarithm \cite{Kherraf2019, Ramezani2021}: 
\begin{equation}
\label{eq:logRel}
     \log \left( R(\ddot{G}) \right) = \sum_{ N_{i\hat{k}} \in  \ddot{\mathcal{N}}} \log \left( R_{i\hat{k}} \right) x_{i\hat{k}}.
\end{equation}
Thus, we define the first objective function as:
\begin{equation}
\label{eq:objective1}
     f_{\mathrm{rel}} = \sum_{ N_{i\hat{k}} \in \ddot{\mathcal{N}}}  \log \left( R_{i\hat{k}} \right) x_{i\hat{k}}.
\end{equation}

\subsubsection{Objective 2}
\label{subsubsec:objective2}
The second objective aims to minimize the overall latency of the application, which is defined by:
\begin{equation}
\label{eq:objective2}
    f_{\mathrm{lat}} = \sum_{ N_{i\hat{k}} \in  \ddot{\mathcal{N}}}  L_{i\hat{k}} \, x_{i\hat{k}} \, + \sum_{ A_{ik \rightarrow jl} \in  \ddot{\mathcal{A}}} CL_{ik \rightarrow jl} \, x_{ik \rightarrow jl}.
\end{equation}

\subsubsection{Multi-objective}
\label{subsubsec:multiObjective}
As the two objectives, reliability and latency, have different scales, we first normalize  \eqref{eq:objective1} and \eqref{eq:objective2} in the interval $[0,1]$ \cite{Grodzevich2006}:
\begin{equation}
\label{eq:normalizedObjective1}
	f^{\prime}_{\mathrm{rel}} = \frac{f_{\mathrm{rel}} - \min \left( f_{\mathrm{rel}}\right)} {\max \left( f_{\mathrm{rel}}\right)  - \min \left( f_{\mathrm{rel}}\right)},
\end{equation}
and
\begin{equation}
\label{eq:normalizedObjective2}
	f^{\prime}_{\mathrm{lat}} = \frac{f_{\mathrm{lat}} - \min \left( f_{\mathrm{lat}}\right)} {\max \left( f_{\mathrm{lat}}\right)  - \min \left( f_{\mathrm{lat}}\right)}.
\end{equation}
Since we can either minimize or maximize the combined objective, we convert the minimization of latency to a maximization problem, by considering the negation of \eqref{eq:normalizedObjective2}:
\begin{equation}
\label{eq:normalizedObjective2Negation}
	f^{\prime \prime}_{\mathrm{lat}} = - f^{\prime}_{\mathrm{lat}}.
\end{equation}

Hence, we formulate the multi-objective function as:
\begin{equation}
\label{eq:multiobjective}
    g = w_{\mathrm{rel}} f^{\prime}_{\mathrm{rel}} + w_{\mathrm{lat}} f^{\prime \prime}_{\mathrm{lat}},
\end{equation}
where $w_{\mathrm{rel}}$ and $w_{\mathrm{lat}}$ are predefined weights reflecting the relative importance of reliability and latency, respectively, in the combined objective. It is noted that $0 \leq w_{\mathrm{rel}}, \allowbreak w_{\mathrm{lat}} \leq 1$ and $w_{\mathrm{rel}} + w_{\mathrm{lat}} = 1$.
Consequently, the considered multi-objective problem is defined as:
\begin{equation}
\label{eq:problem}
    \max g
\end{equation}
subject to the constraints defined in \cref{subsubsec:constraints}.
We utilize the weighted sum approach, as the individual objective functions and the considered constraints are linear. In the context of BILP, which is an exact method, solving \eqref{eq:problem} for a given pair of objective weights yields a single Pareto-optimal solution \cite{Grodzevich2006}. 
Hence, the proposed framework does not aim to generate the entire set of Pareto-optimal solutions (Pareto front) in a single execution.
In contrast, by varying the objective weights across different runs, different Pareto-optimal solutions can be explored, reflecting alternative trade-offs between reliability and latency. The weighted sum formulation is simple, practical, and easy to implement, while allowing direct control over the relative importance of the individual objectives.

\subsubsection{Constraints}
\label{subsubsec:constraints}
The multi-objective function \eqref{eq:multiobjective} is maximized subject to the following constraints:
\begin{equation}
    \label{eq:one}
     x_{i\hat{k}} \in \{ 0, 1 \}, \, \forall \, N_{i\hat{k}} \in \ddot{\mathcal{N}},  
\end{equation}
\begin{equation}
    \label{eq:three}
     x_{ik \rightarrow jl} \in \{ 0, 1 \}, \, \forall \, A_{ik \rightarrow jl} \in \ddot{\mathcal{A}},   
\end{equation}
\begin{equation}
    \label{eq:two}
     x_{in}^{z, i\hat{k}} \in \{ 0, 1 \}, \, \forall \, r_{in}^z \in N_{i\hat{k}}, \, \forall \, N_{i\hat{k}} \in \ddot{\mathcal{N}},  
\end{equation}
\begin{equation}
    \label{eq:three2}
    x_{ik} \in \{ 0, 1 \}, \, \forall \, \ddot{N}_{ik} \in \ddot{\mathcal{N}},
\end{equation}
\begin{equation}
    \label{eq:four}
    \sum_{N_{i\hat{k}} \in \ddot{N}_{i}} x_{i\hat{k}} = 1, \, \forall \, \ddot{N}_{i} \in \ddot{\mathcal{N}},
\end{equation}
\begin{equation}
    \label{eq:four2}
    x_{ik} = \sum_{N_{i\hat{k}} \in \ddot{N}_{ik}} x_{i\hat{k}}, \, \forall \, \ddot{N}_{ik} \in \ddot{\mathcal{N}}, 
\end{equation}
\begin{equation}
    \label{eq:five}
     x_{i\hat{k}} \, |N_{i\hat{k}}| = \sum_{r_{in}^z \in N_{i\hat{k}}} x_{in}^{z, i\hat{k}}, \, \forall \, N_{i\hat{k}} \in \ddot{\mathcal{N}},  
\end{equation}
\begin{equation}
    \label{eq:six}
    \sum_{A_{ik \rightarrow jl} \in \ddot{\mathcal{A}}} x_{ik \rightarrow jl} = \nu_{i}, \, \forall \, \ddot{N}_{i} \in \ddot{\mathcal{N}},
\end{equation}
\begin{equation}
    \label{eq:seven1}
    x_{ik \rightarrow jl} \leq x_{ik}, \, \forall \, A_{ik \rightarrow jl} \in \ddot{\mathcal{A}},
\end{equation}
\begin{equation}
    \label{eq:seven2}
    x_{ik \rightarrow jl} \leq x_{jl}, \, \forall \, A_{ik \rightarrow jl} \in \ddot{\mathcal{A}},
\end{equation}
\begin{equation}
    \label{eq:seven3}
    x_{ik \rightarrow jl} \geq x_{ik} + x_{jl} - 1, \, \forall \, A_{ik \rightarrow jl} \in \ddot{\mathcal{A}},
\end{equation}
\begin{equation}
    \label{eq:eight}
    \sum_{N_{i\hat{k}} \in \ddot{\mathcal{N}}} \sum_{r_{in}^z \in N_{i\hat{k}}} M_{i} \, x_{in}^{z, i\hat{k}} \leq M_{n}^{\mathrm{bgt}}, \, \forall \, n \in \mathcal{U},
\end{equation}
\begin{equation}
    \label{eq:nine}
    \sum_{N_{i\hat{k}} \in \ddot{\mathcal{N}}} \sum_{r_{in}^z \in N_{i\hat{k}}} S_{i} \, x_{in}^{z, i\hat{k}} \leq S_{n}^{\mathrm{bgt}}, \, \forall \, n \in \mathcal{U},
\end{equation}
\begin{equation}
    \label{eq:ten}
    \begin{split}
    & \sum_{N_{i\hat{k}} \in \ddot{\mathcal{N}}} \sum_{r_{in}^z \in N_{i\hat{k}}} E_{in}^{z, i\hat{k}} \, x_{in}^{z, i\hat{k}}\\
    & + \sum_{A_{in \rightarrow jl} \in \ddot{\mathcal{A}}} D_{i} \, x_{in \rightarrow jl} \left( \tau_{nl} \left( 1 - \zeta_{in \rightarrow jl}^{o} \right)  + \tau_{no} \, \zeta_{in \rightarrow jl}^{o} \right)\\
    & + \sum_{A_{jl \rightarrow in} \in \ddot{\mathcal{A}}} D_{j} \, x_{jl \rightarrow in} \left( \rho_{ln} \left( 1 - \zeta_{jl \rightarrow in}^{o} \right)  + \rho_{on} \, \zeta_{jl \rightarrow in}^{o} \right)\\
    & + \sum_{A_{il \rightarrow jo} \in \ddot{\mathcal{A}}} D_{i} \, x_{il \rightarrow jo} \left( \rho_{ln} + \tau_{no} \right) \zeta_{il \rightarrow jo}^{n}\\
    & \leq  E_{n}^{\mathrm{bgt}}, \, \forall \, n \in \mathcal{U}, \, n \neq l \neq o.
    \end{split}
\end{equation}

Equations \eqref{eq:one}--\eqref{eq:three2} guarantee the binary nature of the decision variables.
Constraint \eqref{eq:four} ensures that only one candidate node $N_{i\hat{k}}$ in each set $\ddot{N}_{i} \in \ddot{\mathcal{N}}$ will be selected. That is, it ensures that only one of the possible allocations of each primary task $i$ and its replicas (if applicable) will be selected.
On the other hand, \eqref{eq:four2} ensures that if a candidate node $N_{i\hat{k}}$ is selected, the corresponding set $\ddot{N}_{ik}$ in which $N_{i\hat{k}}$ belongs, will be selected as well.
\Cref{eq:five} guarantees that if a candidate node $N_{i\hat{k}}$ is selected, its corresponding primary task and its replicas $r_{in}^z \in N_{i\hat{k}}$ will also be selected.

Constraint \eqref{eq:six} ensures that for each set $\ddot{N}_{i} \in \ddot{\mathcal{N}}$ (and by extension, for each selected candidate node $N_{i\hat{k}}$), the number of outgoing arcs that will be selected will be equal to the number of child tasks of task $i$.
In the same context, \eqref{eq:seven1}--\eqref{eq:seven3} guarantee that if two candidate nodes $N_{i\hat{k}} \in \ddot{N}_{ik}$ and $N_{j\hat{l}} \in \ddot{N}_{jl}$, respectively, are selected, the corresponding arc $A_{ik \rightarrow jl}$ will also be selected.
Consequently, \eqref{eq:six}--\eqref{eq:seven3} ensure that the connectivity and precedence relationships between the candidate nodes in $\ddot{G}$ -- and thus between the nodes in the initial TG $G$ -- will be preserved.

Finally, \eqref{eq:eight}--\eqref{eq:ten} guarantee that the memory, storage, and energy budgets ($M_{n}^{\mathrm{bgt}}$, $S_{n}^{\mathrm{bgt}}$, and $E_{n}^{\mathrm{bgt}}$, respectively) for each device $n \in \mathcal{U}$ will not be exceeded for the execution of the application.   
$M_{n}^{\mathrm{bgt}}$, $S_{n}^{\mathrm{bgt}}$, and $E_{n}^{\mathrm{bgt}}$ are typically defined based on the respective resources of each device.
It is noted that the energy constraint in \eqref{eq:ten} includes both the computational and communication energy consumption of device $n \in \mathcal{U}$.
Regarding the communication energy, we consider three cases: (a) device $n$ transmits data to another device $l$ (either directly or indirectly via device $o$), (b) $n$ receives data from $l$ (directly or indirectly through $o$), and (c) $n$ is utilized for the communication between $l$ and $o$. 
As the candidate nodes in $\ddot{G}$ incorporate the reliability constraints of the tasks, our formulation ensures the exploration of only reliability-feasible allocations. 
The main notations used in our approach are listed in \cref{table:notations}.

\begin{table*}[t]
\centering
\caption{Main notations.}
\label{table:notations}
\resizebox{0.77\textwidth}{!}{
    \begin{tabular}{ll} 
        \toprule
        Notation & Definition\\
        \hline
        $\mathcal{T}$ & Set of tasks of the considered workflow application.\\
        $\mathcal{U}$ & Set of computational devices in the target system.\\
        $\mathcal{Q}_{i}$ & Subset of computational devices where task $i$ can be allocated ($\mathcal{Q}_{i} \subseteq \mathcal{U}$).\\
        $i, j$ & Tasks of the application ($i, j \in \mathcal{T}$).\\
        $k, l, m, n, o$ & Computational devices in the target system ($k, l, m, n, o \in \mathcal{U}$).\\

        $G=(\mathcal{N}, \mathcal{A})$ & Initial task graph (TG) of the application, where $\mathcal{N}$ and $\mathcal{A}$ are the sets of its nodes and arcs, respectively.\\

        $\dot{G} = (\dot{\mathcal{N}}, \dot{\mathcal{A}})$ & Intermediate edge-hub-cloud graph (EG), where $\dot{\mathcal{N}}$ and $\dot{\mathcal{A}}$ are the sets of its composite nodes and arcs, respectively.\\

        $\ddot{G} = (\ddot{\mathcal{N}}, \ddot{\mathcal{A}})$ & Final reliability-aware edge-hub-cloud graph (REG), where $\ddot{\mathcal{N}}$ and $\ddot{\mathcal{A}}$ are the sets of its composite nodes and arcs, respectively.\\  
        
        $\ddot{N}_{i}$ & Composite node in $\ddot{G}$ ($\ddot{N}_{i} \in \ddot{\mathcal{N}}$), consisting of sets of nodes  $\ddot{N}_{ik}$.\\
        $\ddot{N}_{ik}$ & Set in $\ddot{N}_{i}$, consisting of candidate nodes $N_{i\hat{k}}$.\\ 
        $N_{i\hat{k}}$ & Candidate node $N_{ik}$, $N_{ikl}$, or $N_{iklm}$ in $\ddot{N}_{ik}$, corresponding to the possible allocation of primary task $i$ on device $k$, and its replicas\\
        & (if applicable) on devices $l$ and $m$.\\

        $r_{in}^z$ & Primary task $i$ (if $z=1$) or one of its replicas (if $z \in \{2, 3\}$), allocated on device $n$, corresponding to candidate node $N_{i\hat{k}}$.\\
        
        $\ddot{A}_{i \rightarrow j}$ & Composite arc in $\ddot{G}$ ($\ddot{A}_{i \rightarrow j} \in \ddot{\mathcal{A}}$), consisting of individual arcs $A_{ik \rightarrow jl}$.\\
        $A_{ik \rightarrow jl}$ & Arc in $\ddot{A}_{i \rightarrow j}$, corresponding to the communication between the sets of nodes $\ddot{N}_{ik}$ and $\ddot{N}_{jl}$.\\ 

        $\epsilon_{ik}$ & Execution mode of task $i$ when allocated on device $k$.\\
        $\nu_{i}$ & Number of child tasks of task $i$.\\
        
        $M_{i}$ & Main memory required by task $i$.\\
        $S_{i}$ & Storage required by task $i$.\\
        $D_{i}$ & Size of output data generated by task $i$.\\
        $L_{in}$ & Execution time of task $i$ on device $n$.\\ 
        $P_{in}$ & Power required to execute task $i$ on device $n$.\\
        $E_{in}$ & Energy required to execute task $i$ on device $n$.\\
        $V_{in}$ & Vulnerability factor of task $i$ on device $n$.\\
        $R_{in}$ & Reliability of task $i$ on device $n$.\\

        $L_{i\hat{k}}$ & Total latency of candidate node $N_{i\hat{k}}$.\\
        $E_{in}^{z, i\hat{k}}$ & Total energy required by device $n$ to execute primary task or replica $r_{in}^z$ of candidate node $N_{i\hat{k}}$.\\
        $V_{i\hat{k}}$ & Total vulnerability factor of candidate node $N_{i\hat{k}}$.\\
        $R_{i\hat{k}}$ & Total reliability of candidate node $N_{i\hat{k}}$.\\

        $\zeta_{ik \rightarrow jl}^{o}$ & Binary parameter indicating whether arc $A_{ik \rightarrow jl}$ involves indirect communication between devices $k$ and $l$ through device $o$.\\
        $CL_{ik \rightarrow jl}$ & Communication latency of arc $A_{ik \rightarrow jl}$.\\
        $CE_{ik \rightarrow jl}$ & Communication energy consumption of arc $A_{ik \rightarrow jl}$.\\

        $\sigma_{kl}$ & Bandwidth of the communication channel between devices $k$ and $l$.\\

        $\tau_{kl}, \rho_{kl}$ & Energy required to transmit and receive, respectively, a unit of data between devices $k$ and $l$.\\

        $\mathit{\Lambda}$ & Criticality level of the application.\\
        $VT_{\mathrm{DE}}, VT_{\mathrm{TE}}$ & Vulnerability threshold of the application for dual and triple task execution, respectively.\\

        $\beta_{k}^{\mathrm{DE}}$, $\beta_{k}^{\mathrm{TE}}$ & Time required to perform the comparison and voting, respectively, on device $k$.\vspace{1pt}\\
    
        $\gamma_{k}^{\mathrm{DE}}$, $\gamma_{k}^{\mathrm{TE}}$ & Power required to perform the comparison and voting, respectively, on device $k$.\vspace{1pt}\\
    
        $\delta_{k}^{\mathrm{DE}}$, $\delta_{k}^{\mathrm{TE}}$ & Energy required to perform the comparison and voting, respectively, on device $k$.\\

        $x_{i\hat{k}}$ & Binary decision variable corresponding to candidate node $N_{i\hat{k}} \in \ddot{G}$.\\
        $x_{ik \rightarrow jl}$ & Binary decision variable corresponding to arc $A_{ik \rightarrow jl} \in \ddot{G}$.\\
        $x_{in}^{z, i\hat{k}}$ & Auxiliary binary decision variable corresponding to primary task or replica $r_{in}^z \in N_{i\hat{k}}$.\\ 
        $x_{ik}$ & Auxiliary binary decision variable corresponding to set $\ddot{N}_{ik} \in \ddot{G}$.\\ 
       
        $w_{\mathrm{rel}}, w_{\mathrm{lat}}$ & Weights reflecting the relative importance of overall reliability and latency, respectively, in multi-objective function $g$.\\
        
        $M^{\mathrm{bgt}}_{n}, S^{\mathrm{bgt}}_{n}, E^{\mathrm{bgt}}_{n}$ & Memory, storage, and energy budgets, respectively, of device $n$.\\
        
        \bottomrule
    \end{tabular}
    }
\end{table*}

\subsubsection{Complexity}
\label{subsubsec:complexity}
The computational complexity of the proposed BILP framework depends on both the problem size and the employed solver. Well-established solvers such as Gurobi \cite{Gurobi} typically use proprietary algorithms whose implementation details are not publicly available, and thus their computational complexity cannot be derived \cite{Mo2023}. 
On the other hand, the problem size (i.e., the number of variables and constraints) depends on the size of the resulting REG $\ddot{G}$, whose growth relative to the initial TG $G$ is analyzed in \cref{subsubsec:graphSize}. 
Despite the size increase in $\ddot{G}$, our scalability analysis in \cref{para:scalability} demonstrates that the proposed framework provides optimal solutions within practical time frames, even for TGs with 1000 tasks.  
Thus, the benefits of our approach significantly outweigh the additional complexity introduced by the size increase in $\ddot{G}$.
\section{Framework evaluation}
\label{sec:evaluation}

We evaluated our framework using a real-world workflow application for the autonomous UAV-based aerial inspection of power transmission towers and lines.
Furthermore, we investigated its scalability and applicability to applications of different structures, sizes, and criticality levels, using synthetic workflows that we generated for this purpose. 
In contrast to the qualitative comparison with state-of-the-art approaches presented in \Cref{sec:related}, a meaningful quantitative comparative evaluation is not applicable, as existing techniques (whether heuristic or exact) neither consider the specific edge-hub-cloud environment, nor address all objectives and constraints of our framework. 
Furthermore, adapting exact methods (e.g., \cite{Santos2023, Ramezani2021}) to accommodate all aspects of our framework would result in formulations essentially equivalent to the one proposed in this work, which is among our key contributions, offering limited additional insight.
Nevertheless, to further assess and highlight the strengths of our method, we provide a comparative evaluation against three baseline task allocation strategies in the real-world workflow case.

\subsection{Experimental setup}
\label{subsec:setup}

\begin{table*}[t]
\centering
\caption{Computational devices \& task execution mode percentages per device for $\mathit{\Lambda} = 3$.}
\scriptsize
\resizebox{0.9\textwidth}{!}{
\begin{tabular}{cccccrrccrrrrrr}
\toprule
\multirow{2}{*}{Device\,$n$} & \multirow{2}{*}{$\mathrm{e}$/$\mathrm{h}$/$\mathrm{c}$} & \multirow{2}{*}{Processor} & $\beta_{n}^{\mathrm{DE}}$ & $\beta_{n}^{\mathrm{TE}}$ & \multicolumn{1}{c}{$\gamma_{n}^{\mathrm{DE}}$} & \multicolumn{1}{c}{$\gamma_{n}^{\mathrm{TE}}$} & $M^{\mathrm{bgt}}_{n}$ & $S^{\mathrm{bgt}}_{n}$ & \multicolumn{1}{c}{$E^{\mathrm{bgt}}_{n}$} & \multicolumn{1}{c}{$P_{n}^{\mathrm{idle}}$} & \multicolumn{1}{c}{$P_{n}^{\mathrm{max}}$} & \multicolumn{3}{c}{Exec. Mode \% ($\mathit{\Lambda} = 3$)}\\
\cline{13-15}
 & & & (\SI{}{\micro \second}) & (\SI{}{\micro \second}) & \multicolumn{1}{c}{(W)} & \multicolumn{1}{c}{(W)} & (GiB) & (GiB) & \multicolumn{1}{c}{(Wh)} & \multicolumn{1}{c}{(W)} & \multicolumn{1}{c}{(W)} & \multicolumn{1}{c}{SE} & \multicolumn{1}{c}{DE} & \multicolumn{1}{c}{TE}\vspace{1pt}\\

\hline

Raspberry\,Pi & $\mathrm{e}$ & Broadcom\,Cortex & $1.00$ & $1.50$ & $1.20$ & $1.21$ & $1$ & $16$ & $129.96$${}^{a}$ & $1.2$ & $4.5$ & $5\%$	& $20\%$	& $75\%$ \\
3\,Model\,B & & A53\,@\,1.4\,GHz & & & & & & & & & & & &\vspace{4pt}\\

Mi\,Notebook & $\mathrm{h}$ & Intel\,i5\,8250U & $0.02$ & $0.03$ & $8.02$ & $8.03$ & $8$ & $512$ & $60.00$${}^{b}$  & $8.0$ & $44.0$ & $20\%$	& $35\%$	& $45\%$ \\
Pro & & @\,1.6\,GHz & & & & & & & & & & & &\vspace{4pt}\\

HPE\,ProLiant & $\mathrm{c}$ & Intel\,Xeon\,Gold & $0.01$ & $0.02$ & $75.23$ & $75.26$ & $400$ & $10\,240$ & \multicolumn{1}{c}{-} & $75.0$ & $900.0$ & $35\%$	& $50\%$	& $15\%$ \\
DL580\,Gen10 & & 6240\,@\,2.6\,GHz & & & & & & & & & & & &\\

\bottomrule
 \multicolumn{15}{l}{\scriptsize{${}^{a}$Battery capacity of DJI Matrice 100 UAV where device $\mathrm{e}$ is attached.}}\\
 \multicolumn{15}{l}{\scriptsize{${}^{b}$Battery capacity of device $\mathrm{h}$.}}
\end{tabular}
}
\label{table:devices}
\end{table*}

While the proposed framework can accommodate multiple devices per layer in the edge-hub-cloud continuum, our evaluation focused on applications for which a single edge device, a single hub device, and a single cloud server suffice, as discussed in \cref{sec:intro}.
In our experiments, the edge device ($\mathrm{e}$), the hub device ($\mathrm{h}$), and the cloud server ($\mathrm{c}$) were based on typical real-world counterparts, as shown in \cref{table:devices}.
Parameters $M^{\mathrm{bgt}}_{n}$, $S^{\mathrm{bgt}}_{n}$, and $E^{\mathrm{bgt}}_{n}$ were based on the resource limitations of each device $n$, whereas $\beta_{n}^{\mathrm{DE}}$, $\beta_{n}^{\mathrm{TE}}$, $\gamma_{n}^{\mathrm{DE}}$, $\gamma_{n}^{\mathrm{TE}}$, $P_{n}^{\mathrm{idle}}$, and $P_{n}^{\mathrm{max}}$ were empirically determined using the profiling and power monitoring tools mentioned in \cref{subsubsec:realOverview}. 
The communication channels and their respective parameters ($\sigma_{kl}$, $\tau_{kl}$, and $\rho_{kl}$), which were based on real-world measurements \cite{Huang2012, Shi2018, 5GAmericas, Vladan2021}, are included in \cref{table:commChannels}.
As we employ an offline approach, all model parameters were treated as fixed inputs during optimization, instantiated with conservative values derived from real-world measurements and profiling to ensure feasibility under adverse operating conditions. 
In line with the critical nature of the targeted use cases, the edge, hub, and cloud devices were reserved exclusively for the considered workflow application, ensuring that their computational resources were dedicated to its execution. 
It is noted that the framework can be re-executed with different parameter sets to enable design-time evaluation of alternative network or workload conditions.
For the weights $w_{\mathrm{rel}}$ and $w_{\mathrm{lat}}$ of the considered objectives we used values in the interval $[0, 1]$. We increased/decreased each weight by $0.05$ increments/decrements to investigate a wide range of cases. 
We considered applications with a low ($\mathit{\Lambda}=1$), moderate ($\mathit{\Lambda}=2$), and high ($\mathit{\Lambda}=\mathit{\Lambda}_{\mathrm{max}}=3$) criticality level. For each $\mathit{\Lambda}$, we experimented with the vulnerability thresholds $VT_{\mathrm{DE}}$ and $VT_{\mathrm{TE}}$ shown in \cref{table:thresholds}, and the adjustment coefficients $\kappa = 0.06$ and $\lambda = 3$. We selected the particular values so that in each case the vulnerability thresholds would be realistic and aligned with prior studies \cite{Kherraf2019, Kouloumpris2020, Xie2020}, but still challenging.
We implemented the proposed framework, including the resulting BILP formulation, in C++.
The resulting BILP model was solved using Gurobi Optimizer 10 \cite{Gurobi} with default parameter settings, on a server equipped with an Intel Xeon Gold 6240 processor @ 2.6\,GHz and 400\,GiB of RAM.
No parameter tuning strategies were required, as our framework uses fixed model parameters and the solver was run with default settings.

\subsection{Real-world workflow application}
\label{subsec:real}

\subsubsection{Overview}
\label{subsubsec:realOverview}
\begin{figure}[t]
    \centering
    \begin{minipage}[b]{0.55\columnwidth}
        \begin{table}[H]
            \setlength{\tabcolsep}{1.6pt}
            \centering
            \caption{Real-world workflow tasks.}
            \resizebox{\columnwidth}{!}{
                \begin{tabular}{ll}
                \toprule
                Task & Description\\
                \hline
                \multicolumn{2}{c}{Image Acquisition}\\
                \midrule
                $N_1$ & Capture image\\
                \midrule
                \multicolumn{2}{c}{Image Preprocessing}\\
                \midrule
                $N_2$ & Grayscale processing\\
                $N_3$ & Gamma correction\\
                $N_4$ & Range filter\\
                $N_5$ & Edge detection\\
                
                \midrule
                \multicolumn{2}{c}{Detection of Power Transmission Lines}\\
                \midrule
                $N_6$ & Map edge points to Hough space\\
                $N_7$ & Yield infinite lines\\
                $N_8$ & Convert infinite lines to finite lines\\
                $N_9$ & Reject non-parallel finite lines\\

                \midrule
                \multicolumn{2}{c}{Detection of Power Transmission Towers}\\
                \midrule
                $N_{10}$ & Attention mechanism\\
                $N_{11}$ & Image segmentation\\
                $N_{12}$ & Convolutional neural network inference\\
                $N_{13}$ & Acquire and post-process predictions\\
                $N_{14}$ & Draw detection boxes\\

                \midrule
                \multicolumn{2}{c}{Display Results}\\
                \midrule
                $N_{15}$ & Visualize inspection findings\\
                \bottomrule
                \end{tabular}
            }
        \label{table:realTasks}
        \end{table}
    \end{minipage}
    \hspace{4pt}
    \begin{minipage}[b]{0.38\columnwidth}
          \centering
          \includegraphics[width=0.42\columnwidth]{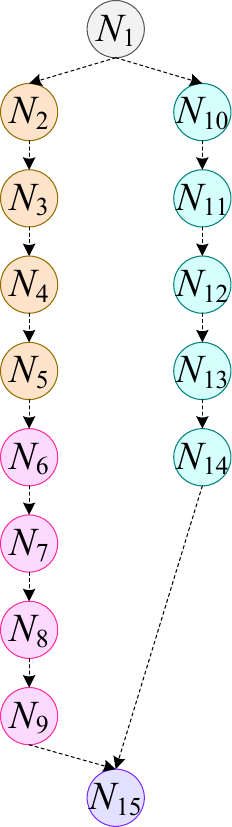}
            \caption{Real-world workflow TG.}
            \label{fig:realWorldApp}

            \begin{table}[H]
            \setlength{\tabcolsep}{4.8pt}
            \centering
            \caption{Node/Arc counts for real-world workflow.}
            \scriptsize
            \begin{tabular}{rrr}
            \toprule
            \multicolumn{1}{c}{Graph} & \multicolumn{1}{c}{\#Nodes} & \multicolumn{1}{c}{\#Arcs}\\
            \hline
            TG $G$ & 15 & 15\\
            EG $\dot{G}$ & 41 & 111 \\
            REG $\ddot{G}$ & 153 & 111 \\
            \bottomrule
            \end{tabular}
            \label{table:realWorldApp}
            \end{table}
    \end{minipage}   
\end{figure}

The considered real-world application is used for the autonomous UAV-based aerial inspection of power transmission towers and lines \cite{Savva2021}. Its tasks are described in \cref{table:realTasks}. Its TG $G$ is depicted in \cref{fig:realWorldApp}, where tasks are shown in different colors based on their grouping in \cref{table:realTasks}. 
Task $N_1$ is responsible for capturing an image through a camera mounted on the UAV. 
Tasks $N_2$--$N_5$ perform the initial processing of the captured image, while tasks $N_6$--$N_9$ apply a Hough transform-based line detector to identify power transmission lines for navigation purposes and to detect issues such as vegetation encroachment. 
On the other hand, tasks $N_{10}$--$N_{14}$ utilize a convolutional neural network-based detector to identify power transmission towers and to detect problems such as faulty insulators, while task $N_{15}$ is responsible for displaying the inspection results. 
Based on their specific functions, tasks $N_1$ and $N_{15}$ can only be allocated on devices $\mathrm{e}$ and $\mathrm{h}$, respectively. In contrast, tasks $N_2$--$N_{14}$ can be allocated on any device ($\mathrm{e}$, $\mathrm{h}$, or $\mathrm{c}$).

As the examined application concerns the inspection of a critical infrastructure, we considered that its criticality level is $\mathit{\Lambda} = \mathit{\Lambda}_{\mathrm{max}} = 3$.
For each node $N_{ik} \in \dot{G}$ we calculated $\epsilon_{ik}$ and $V_{ik}$ so that, using \eqref{eq:execMode} and the vulnerability thresholds for $\mathit{\Lambda} = 3$ in \cref{table:thresholds}, the percentage of tasks that required SE, DE, and TE on each device was the one shown in \cref{table:devices}. We selected the particular percentages based on the criticality level of the application, and considering that device $\mathrm{e}$ is less reliable than device $\mathrm{h}$, which in turn is less reliable than device $\mathrm{c}$.
We calculated $M_{i}$, $S_{i}$, $D_{i}$, $L_{ik}$, and $P_{ik}$ by profiling the execution of task $i$ on device $k$, using PowerTOP, perf, Sysprof, and a digital power monitoring tool, with measurements selected to reflect conservative operating conditions \cite{Powertop, Sysprof}.
$E_{ik}$ and $R_{ik}$ were determined using \eqref{eq:compEnergy} and \eqref{eq:reliability}, respectively, whereas $\nu_{i}$ was derived from the structure of $G$.
For each arc $A_{ik \rightarrow jl} \in \dot{G}$, we determined $\zeta_{ik \rightarrow jl}^{o}$ based on the structure of $\dot{G}$, and we calculated $CL_{ik \rightarrow jl}$ and $CE_{ik \rightarrow jl}$ using \eqref{eq:commLatency} and \eqref{eq:commEnergy}, respectively.
The primary task and its replicas $r_{in}^z \in N_{i\hat{k}}$ in $\ddot{G}$ inherited their parameters from the respective nodes $N_{in} \in \dot{G}$. For each candidate node $N_{i\hat{k}} \in \ddot{G}$, we calculated the aggregate parameters $L_{i\hat{k}}$, $E_{in}^{z, i\hat{k}}$, $V_{i\hat{k}}$, and $R_{i\hat{k}}$, using \eqref{eq:totalLatency}--\eqref{eq:totalReliability}. 
Each arc $A_{ik \rightarrow jl} \in \ddot{G}$ had the same parameters as $A_{ik \rightarrow jl} \in \dot{G}$. The number of nodes and arcs in $G$, $\dot{G}$, and $\ddot{G}$ for the real-world workflow are shown in \cref{table:realWorldApp}.

\begin{figure}[t]
    \centering
    \begin{minipage}[b]{.56\linewidth}
    \begin{table}[H]
        \centering
        \caption{Communication channels.}
        \scriptsize
        \setlength{\tabcolsep}{3pt}
        \begin{tabular}{crcc}
        \toprule
        Comm. Channel &  \multicolumn{1}{c}{$\sigma_{kl}$} &  $\tau_{kl}$ &  $\rho_{kl}$\\
        Between $k \rightarrow l$ &  \multicolumn{1}{c}{(Mbit/s)} &  (\SI{}{\micro \joule}/bit) & (\SI{}{\micro \joule}/bit)\\
        \hline
        $\mathrm{e} \rightarrow \mathrm{h}$  & $11.0$ \hspace{3pt} & $1.0$ & $0.70$\\
        $\mathrm{h} \rightarrow \mathrm{e}$  & $8.5$  \hspace{3pt} & $1.0$ & $0.70$\\
        $\mathrm{h} \rightarrow \mathrm{c}$  & $12.5$ \hspace{3pt} & $2.5$ & $1.25$\\
        $\mathrm{c} \rightarrow \mathrm{h}$  & $20.0$ \hspace{3pt} & $2.5$ & $1.25$\\
        \bottomrule
        \end{tabular}
        \label{table:commChannels}
    \end{table}
    \end{minipage}
    \hfill
    \begin{minipage}[b]{.32\linewidth}
    \begin{table}[H]
        \centering
        \caption{Vulnerability thresholds.}
        \scriptsize
        \begin{tabular}{ccc}
        \toprule
        $\mathit{\Lambda}$ & $VT_{\mathrm{DE}}$ & $VT_{\mathrm{TE}}$\\
        \hline
        1 & 0.06 & 0.18\\
        2 & 0.03 & 0.09\\
        3 & 0.02 & 0.06\\
        
        \bottomrule
        \end{tabular}
        \label{table:thresholds}
    \end{table}
    \end{minipage}
\end{figure}

\subsubsection{Experimental results}
\label{subsubsec:realResults}
\begin{figure}[t]
    \centering
    \includegraphics[width= 0.92 \columnwidth]{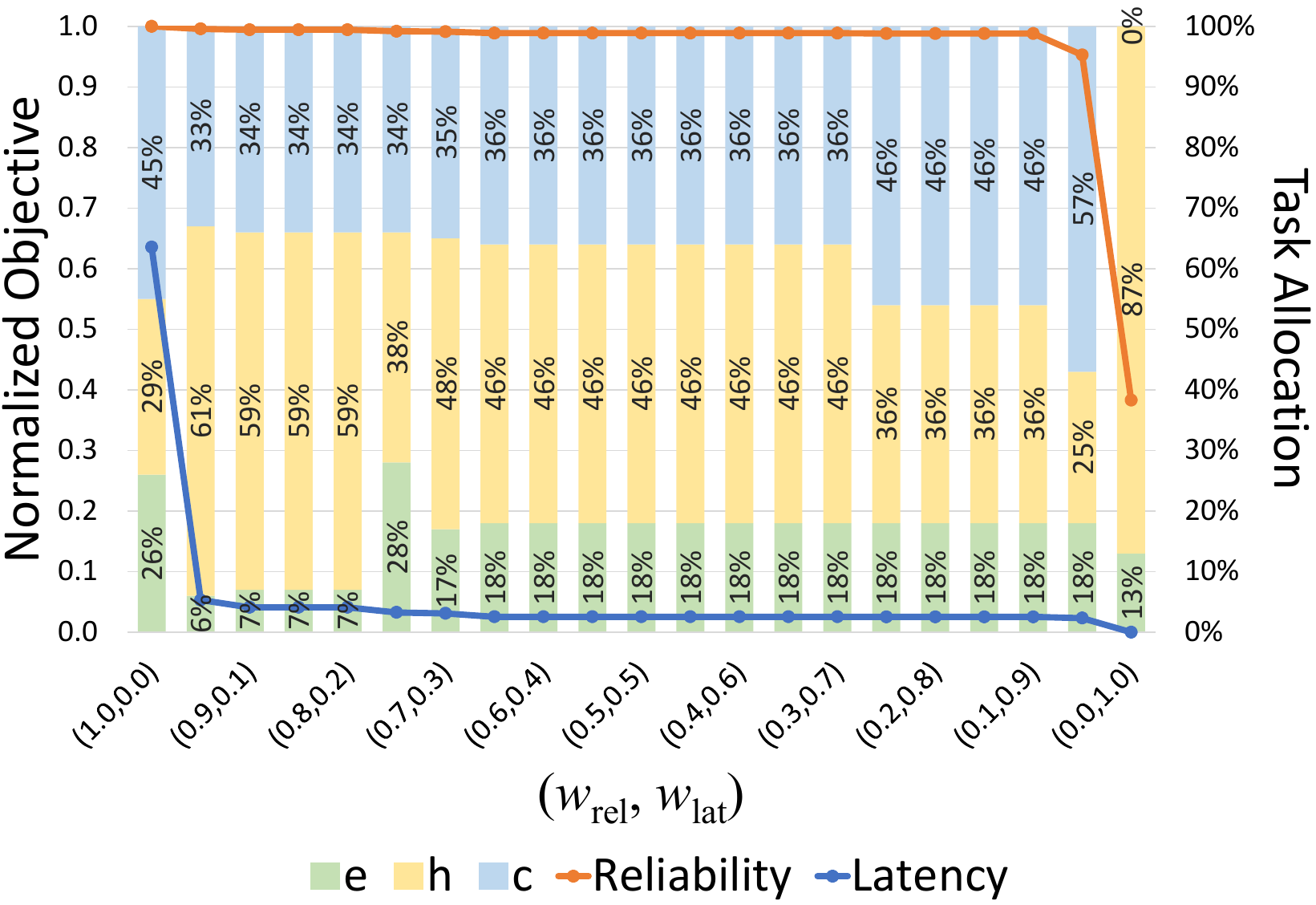}
    \caption{Normalized overall reliability and latency, and percentage of allocated tasks (primary and replicas) per device, with respect to $w_{\mathrm{rel}}$ and $w_{\mathrm{lat}}$, for the real-world workflow.}
    \label{fig:real}
\end{figure}

\cref{fig:real} shows the evaluation results for the real-world application, regarding the normalized overall reliability and latency, with respect to the utilized weights, $w_{\mathrm{rel}}$ and $w_{\mathrm{lat}}$. 
To facilitate the interpretation of the results, the negated normalized latency $f_{\mathrm{lat}}^{\prime \prime}$ is reported as positive (i.e., as if latency were minimized in the combined objective).   
The allocation of the tasks (primary and replicas) on each device ($\mathrm{e}$, $\mathrm{h}$, and $\mathrm{c}$) is also presented in percentage form for each weight pair.
When the objective was set to maximize overall reliability (i.e., $w_{\mathrm{rel}}=1$), our approach yielded the worst performance in terms of latency.  
However, although a higher percentage of tasks was expected on device $\mathrm{c}$ due to its higher reliability, the tasks were distributed across the three devices, with $26\%$ on device $\mathrm{e}$, $29\%$ on $\mathrm{h}$, and $45\%$ on $\mathrm{c}$.
On the other hand, when the objective was to minimize the overall latency (i.e., $w_{\mathrm{lat}}=1$), our method provided the minimum reliability. In this case, the tasks were allocated only on devices $\mathrm{e}$ and $\mathrm{h}$, with $13\%$ on $\mathrm{e}$ and $87\%$ on $\mathrm{h}$.
No tasks were allocated on device $\mathrm{c}$, due to the higher communication latency involved.

When we took into account both objectives (i.e., $0 < w_{\mathrm{rel}}, w_{\mathrm{lat}} < 1$), the majority of tasks were allocated on devices $\mathrm{h}$ and $\mathrm{c}$, which generally provided higher reliability and computational capacity, compared to device $\mathrm{e}$. It can also be observed that the overall reliability deteriorated less compared to the overall latency, as we decreased their corresponding weights (and thus their relative importance in the combined objective). 
When we considered both objectives as equally important (i.e., $w_{\mathrm{rel}} = w_{\mathrm{lat}} = 0.5$), near-optimal results were achieved for both reliability and latency, as their normalized values (0.99 and 0.02, respectively) were close to the optimal ones (1 and 0, respectively).

\begin{figure}[t]
    \centering
    \includegraphics[width= \columnwidth]{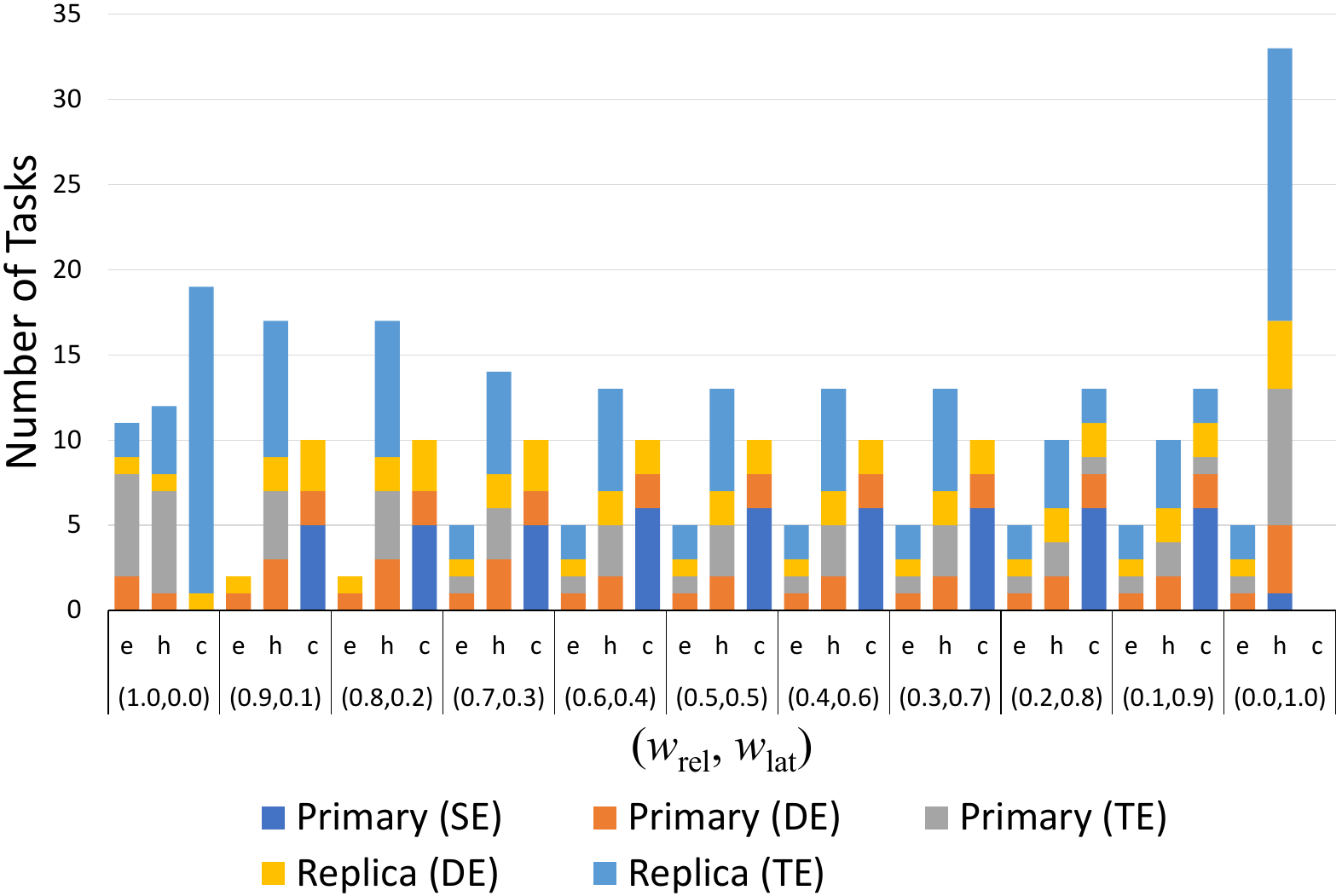}
    \caption{Number of tasks (primary and replicas) per device, with respect to $w_{\mathrm{rel}}$ and $w_{\mathrm{lat}}$, for the real-world workflow.}
    \label{fig:replicas}
\end{figure}

\begin{figure}[t]
    \centering
    \includegraphics[width= \columnwidth]{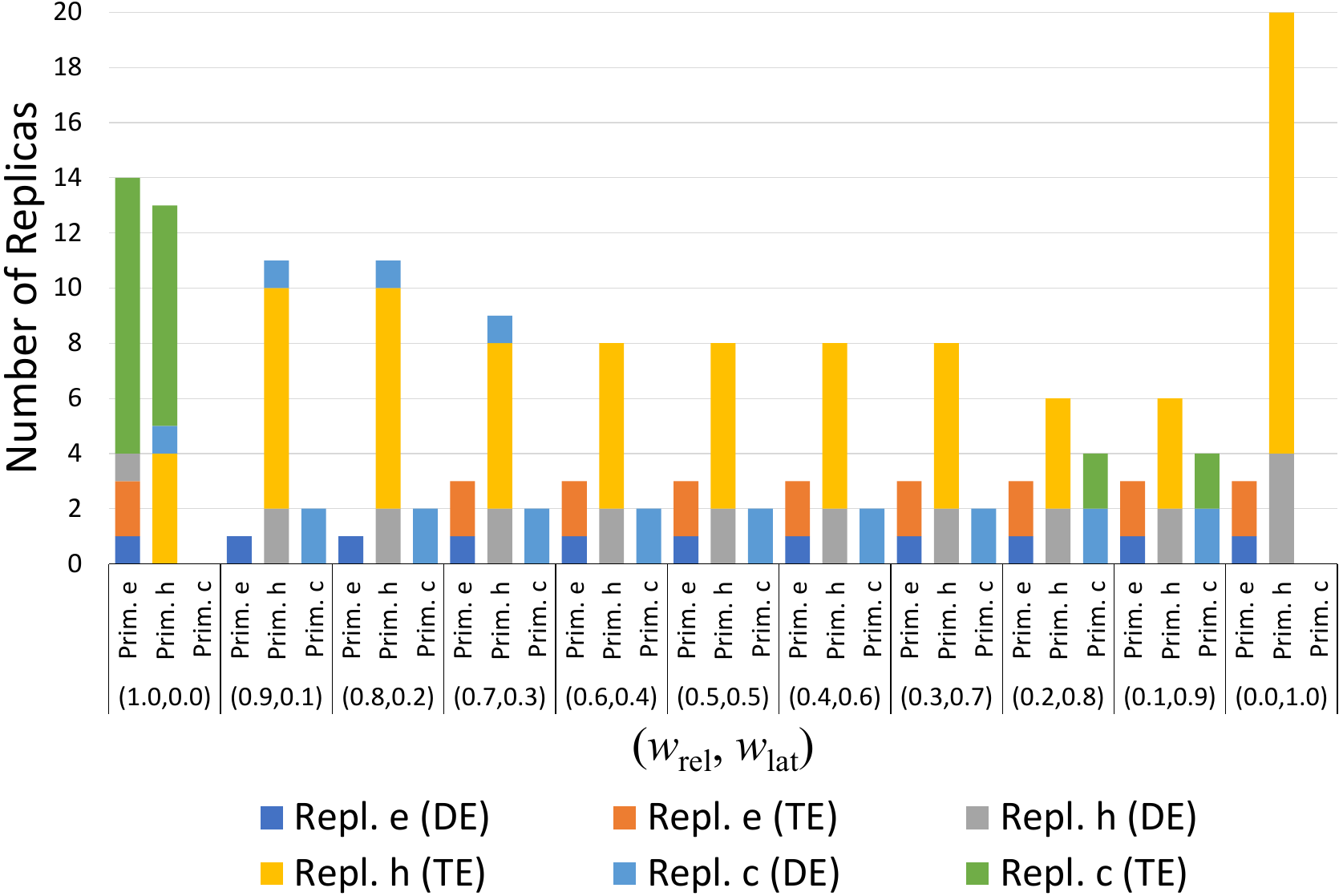}
    \caption{Number of replicas per device, with respect to the allocation of their primary tasks, for $w_{\mathrm{rel}}$ and $w_{\mathrm{lat}}$, in the case of the real-world workflow.}
    \label{fig:secReplicas}
\end{figure}

As some allocation cases in \cref{fig:real} were not intuitive (e.g., for $w_{\mathrm{rel}}=1$), a more detailed analysis is presented in \cref{fig:replicas,fig:secReplicas}. This analysis aims to provide deeper insights into how our approach allocated the primary tasks and their replicas across the three devices, as well as the utilized task replication technique in each case.
Specifically, \cref{fig:replicas} illustrates the number of tasks (primary and replicas) per device, with respect to $w_{\mathrm{rel}}$ and $w_{\mathrm{lat}}$. \cref{fig:secReplicas} depicts the number of replicas per device in relation to the allocation of their primary counterparts, for each weight pair.  
It is revealed that when $w_{\mathrm{rel}}=1$, no primary tasks were allocated on device $\mathrm{c}$, even though it provided the highest reliability. In contrast, all of the primary tasks were allocated on devices $\mathrm{e}$ and $\mathrm{h}$. However, their replicas (the majority of which were due to TE) were allocated on device $\mathrm{c}$. Evidently, this provided the highest total reliability for the tasks. 
On the other hand, as we increased the relative importance of latency, the majority of replicas (which were due to TE) were allocated on device $\mathrm{h}$, where their primary tasks were also allocated, as this provided lower latency. This trend was more prominent in the case where $w_{\mathrm{lat}}=1$.

\begin{figure}[t]
    \centering
    \includegraphics[width=0.82\columnwidth]{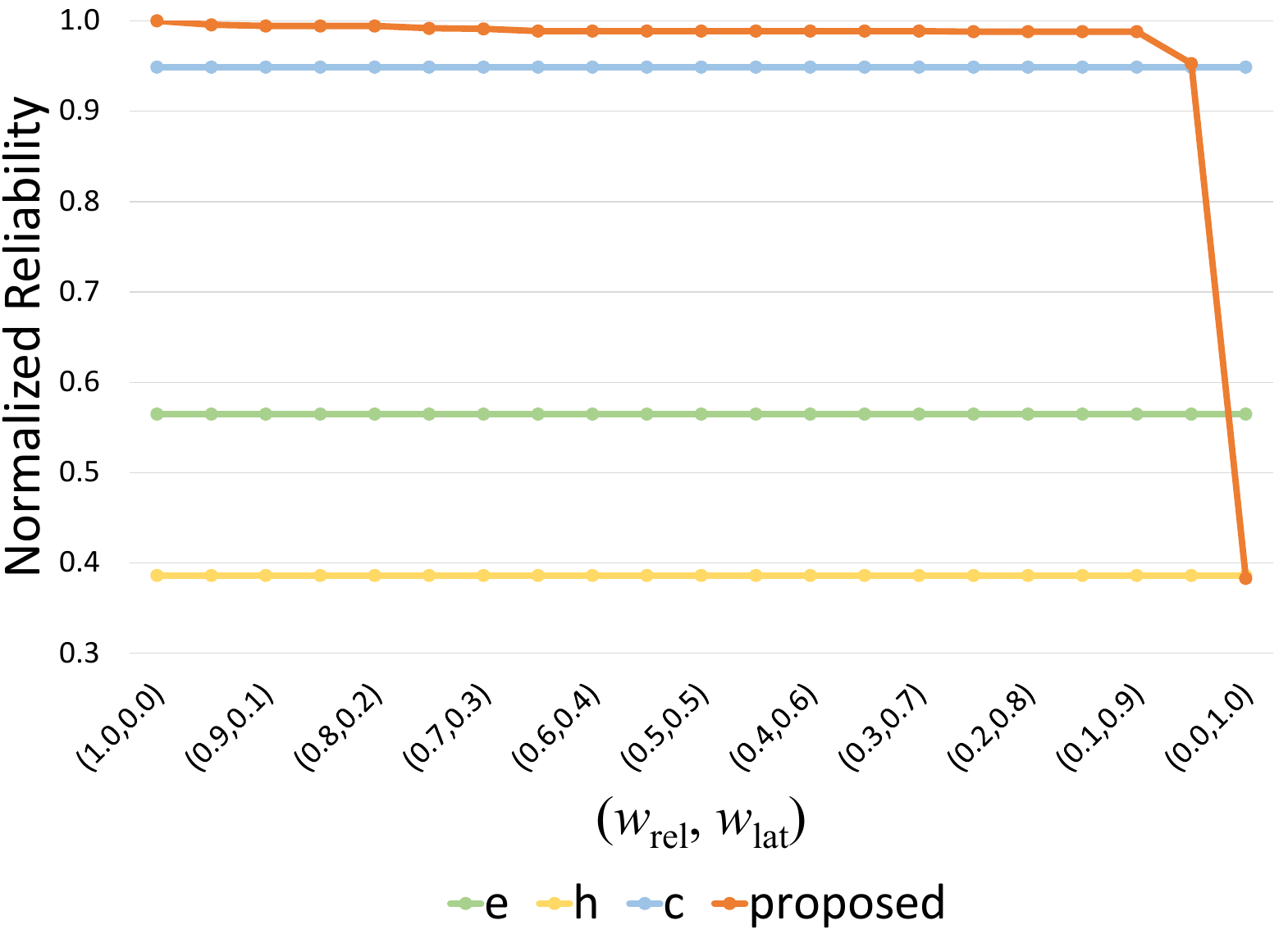}
    \caption{Comparison of proposed approach with baseline strategies (each corresponding to a specific device: $\mathrm{e}$, $\mathrm{h}$, or $\mathrm{c}$) with respect to normalized overall reliability for the real-world workflow, under varying weights $w_{\mathrm{rel}}$ and $w_{\mathrm{lat}}$.}
    \label{fig:extremeCasesReliability}
\end{figure}

\begin{figure}[t]
    \centering
    \includegraphics[width=0.82\columnwidth]{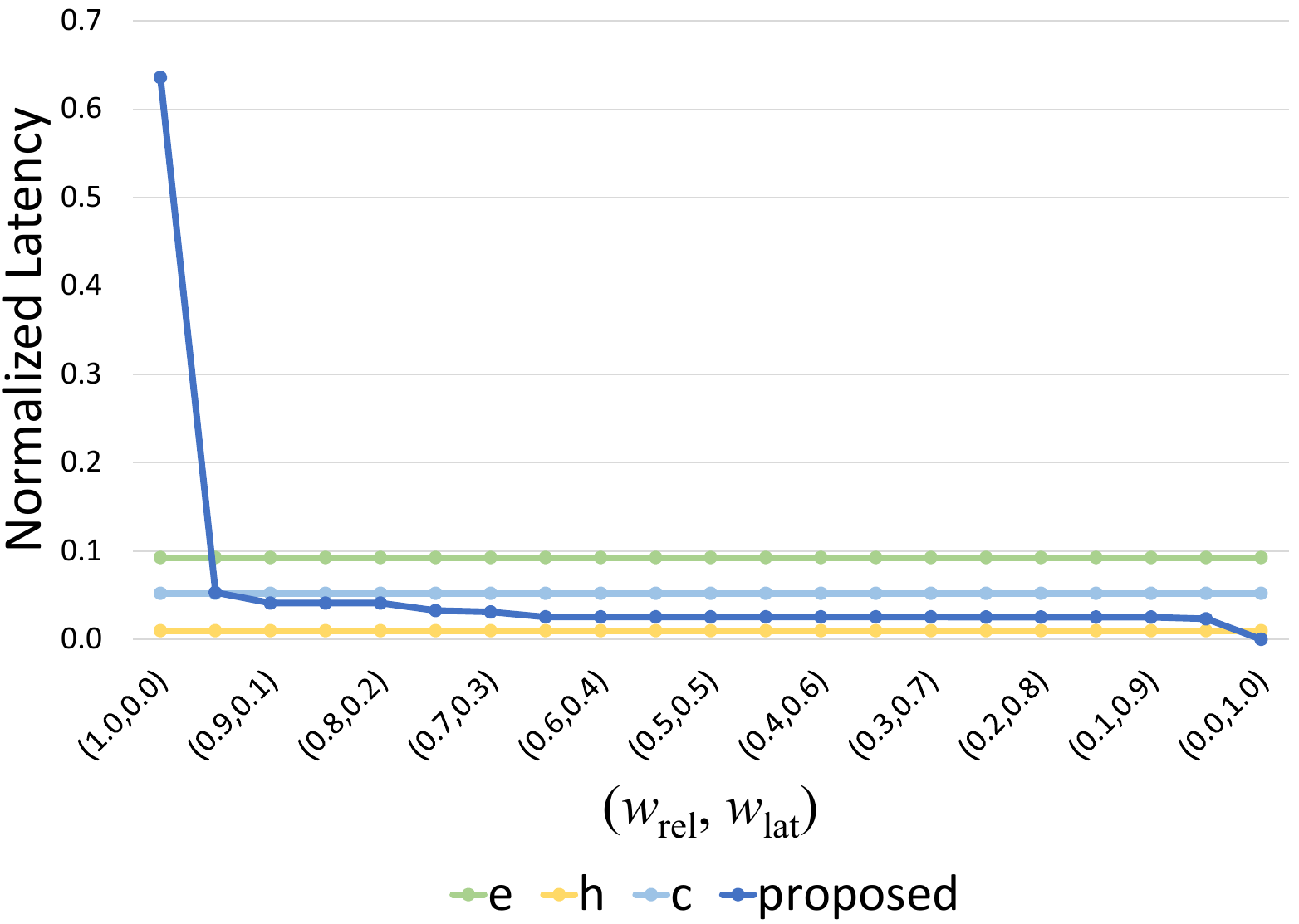}
    \caption{Comparison of proposed approach with baseline strategies (each corresponding to a specific device: $\mathrm{e}$, $\mathrm{h}$, or $\mathrm{c}$) with respect to normalized overall latency for the real-world workflow, under varying weights $w_{\mathrm{rel}}$ and $w_{\mathrm{lat}}$.}
    \label{fig:extremeCasesLatency}
\end{figure}

\paragraph{Comparison with baseline strategies}
\label{para:baselines}
To compare the performance of our approach against more restrictive task allocation scenarios, we evaluated it against three baseline strategies where all tasks (primary and replicas) were allocated on device $\mathrm{e}$, $\mathrm{h}$, or $\mathrm{c}$ (except for tasks $N_1$ and $N_{15}$, which required fixed allocation on devices $\mathrm{e}$ and $\mathrm{h}$, respectively). 
The comparative results regarding the normalized overall reliability and latency, with respect to $w_{\mathrm{rel}}$ and $w_{\mathrm{lat}}$, are illustrated in \cref{fig:extremeCasesReliability,fig:extremeCasesLatency}, respectively. Each baseline strategy is indicated by the device it concerns ($\mathrm{e}$, $\mathrm{h}$, or $\mathrm{c}$).
It can be observed that with regard to reliability, the proposed approach outperformed the baseline strategies across all examined weights (except for $w_{\mathrm{rel}}=0$), with an average absolute increase in normalized reliability of 0.36 (on a scale from 0 to 1).
In actual (non-normalized) values, the average reliability improvement was 84.19\%.
For $w_{\mathrm{rel}}=0$, only latency was optimized, resulting in a significant decrease in the reliability yielded by our method.
On the other hand, with respect to latency, the proposed approach outperformed the baseline strategies for $w_{\mathrm{lat}}=1$, with an average absolute decrease in normalized latency of 0.05 (on a scale from 0 to 1).
In actual (non-normalized) values, the average latency reduction was 49.81\%.
For $ 0 < w_{\mathrm{lat}} < 1$, our method allocated the majority of tasks (primary and replicas) on device $\mathrm{h}$, achieving similar performance to baseline strategy $\mathrm{h}$, which yielded the lowest latency among the three baselines.
For $w_{\mathrm{lat}}=0$, the latency provided by the proposed approach increased significantly, as only reliability was optimized in this case.
Overall, considering both reliability and latency, the results demonstrate that our method attained an optimized balance between the two objectives, compared to the baseline strategies.

\paragraph{Practical applicability}
\label{para:practicality}
The evaluation results regarding the real-world application show that the proposed approach highly depended on the relative importance of the two objectives. As reported by the Gurobi solver, our framework required 1653 variables, 1538 constraints, and an average time of 0.04 seconds to return an optimal solution in each case of the examined reliability and latency weights.
Consequently, considering the short execution time of the solver, as well as the fact that the experimental results were not always intuitive nor straightforward, our method can facilitate design space exploration with respect to the desired trade-off between reliability and latency.

\subsection{Synthetic workflow applications}
\label{subsec:synthetic}

\subsubsection{Overview}
\label{subsubsec:syntheticOverview}
\begin{figure}[!t]
    \centering
    \includegraphics[width=0.72\columnwidth]{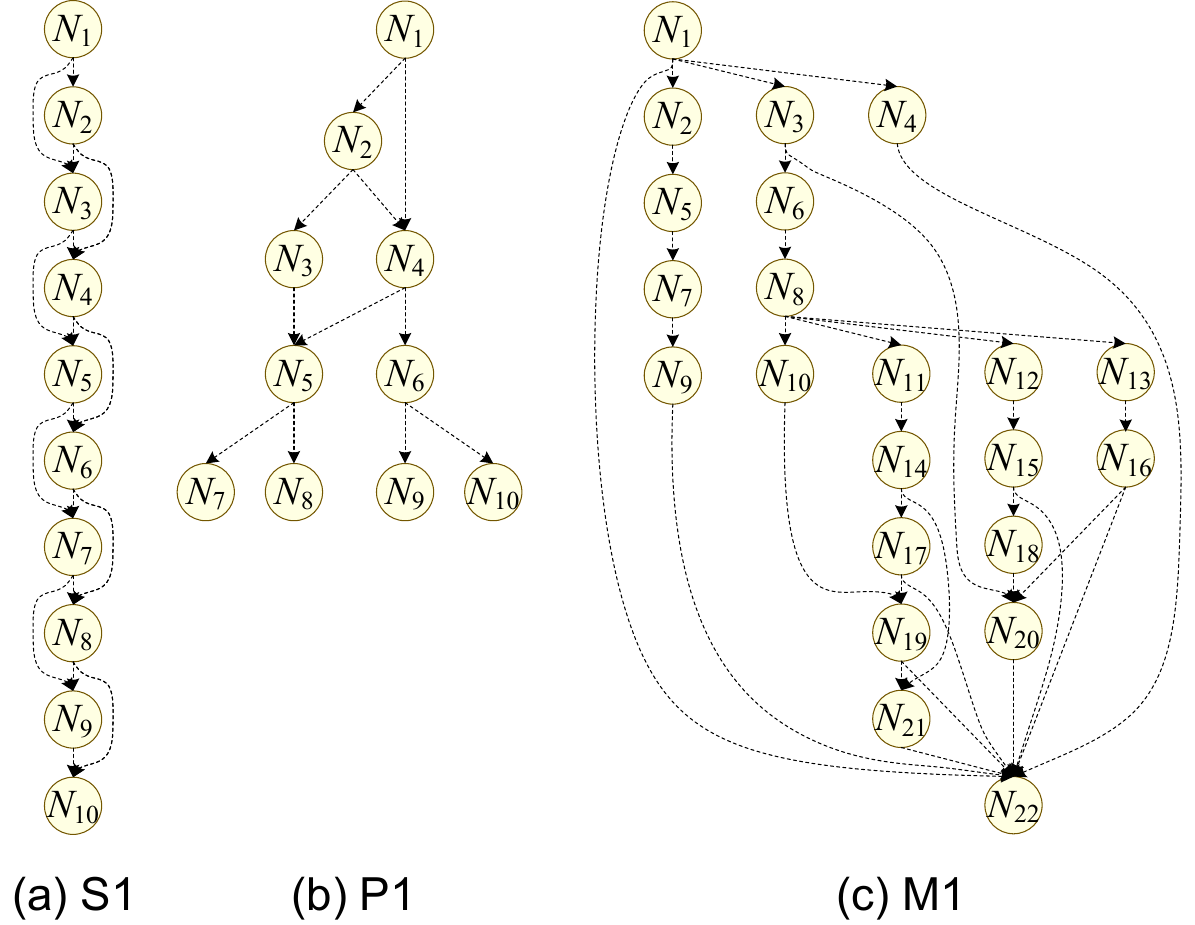}
    \caption{Sample TGs of synthetic workflows with 10 tasks of (a) serial, (b) parallel, and (c) mixed structure. Larger workflows with 100 and 1000 tasks were also generated and examined in our evaluation, as shown in \cref{table:syntheticTGs}.}
    \label{fig:syntheticTGs}
\end{figure}

We generated synthetic workflows using the random TG generator in \cite{Dick1998}. 
The datasets of our synthetically generated workflows are publicly available\footnote{\url{https://doi.org/10.5281/zenodo.10357101}}.
To examine representative workflow structures and evaluate the broad applicability and scalability of our framework, we generated TGs with serial (S), parallel (P), and mixed (M) structure types (with M being a combination of S and P), comprising 10, 100, and 1000 tasks. The input for the generation of each TG is shown in \cref{table:syntheticTGs}. 
\cref{fig:syntheticTGs} shows sample TGs generated for each structure type (S1, P1, and M1 in \cref{table:syntheticTGs}).
Due to the nature of the targeted applications, for each TG we considered that a percentage of randomly selected tasks required fixed allocation on devices $\mathrm{e}$ and $\mathrm{h}$, as shown in \cref{table:syntheticTGs}. Moreover, we considered three criticality levels, $\mathit{\Lambda} \in \{ 1, \allowbreak 2, \allowbreak 3 \}$, to investigate TGs with low, moderate, and high criticality, respectively.

\begin{table*}[t]
\setlength{\tabcolsep}{4pt}
\centering
\caption{Synthetic workflow TGs with their corresponding EGs \& REGs per $\mathit{\Lambda}$.}
\scriptsize
\resizebox{\textwidth}{!}{
\begin{tabular}{@{\extracolsep{2pt}}ccccccccccccrrr@{}}
\toprule

\multicolumn{3}{c}{TGFF Generator Input} & \multicolumn{2}{c}{Generated TG} & \multicolumn{1}{c}{Intermediate EG} & \multicolumn{9}{c}{Final REG}\\
\cline{1-3}
\cline{4-5}
\cline{6-6}
\cline{7-15}
\multirow{3}{*}{SN} & \multirow{3}{*}{\#Nodes} & \multirow{2}{*}{In/Out} & \multirow{2}{*}{\#Nodes/} & Fixed & \multirow{2}{*}{\#Nodes/} & \multicolumn{3}{c}{\#Candidate Nodes/Arcs} & \multicolumn{3}{c}{\#Variables/Constraints} & \multicolumn{3}{c}{Avg. Execution Time (s)}\\
\cline{7-9}
\cline{10-12}
\cline{13-15}

& & \multirow{2}{*}{Degree} & \multirow{2}{*}{Arcs} & Alloc. \% & \multirow{2}{*}{Arcs} & \multirow{2}{*}{$\mathit{\Lambda} = 1$} & \multirow{2}{*}{$\mathit{\Lambda} = 2$} & \multirow{2}{*}{$\mathit{\Lambda} = 3$} & \multirow{2}{*}{$\mathit{\Lambda} = 1$} & \multirow{2}{*}{$\mathit{\Lambda} = 2$} & \multirow{2}{*}{$\mathit{\Lambda} = 3$} & \multirow{2}{*}{$\mathit{\Lambda} = 1$} & \multirow{2}{*}{$\mathit{\Lambda} = 2$} & \multirow{2}{*}{$\mathit{\Lambda} = 3$}\\
& & & & ($\mathrm{e}$/$\mathrm{h}$) & & & & & & & & & & \\

\hline
S1 & 10 & 2\,/\,2   & 10\,/\,17 & 4\,/\,2  & 30\,/\,153            & 61\,/\,153 & 94\,/\,153 & 94\,/\,153               & 659\,/\,626 & 1542\,/\,1476 & 1542\,/\,1476 & 0.06  & 0.10  & 0.10\\ 

S2 & 100 & 2\,/\,2  & 100\,/\,197 & 4\,/\,3 & 286\,/\,1605         & 924\,/\,1605 & 1091\,/\,1605 & 1145\,/\,1605       & 17498\,/\,16782  & 24409\,/\,23526  & 26827\,/\,25890  & 1.43  & 2.16  & 2.10\\ 
   
S3 & 1000 & 2\,/\,2 & 1000\,/\,1997 & 5\,/\,3 & 2840\,/\,16089     & 8775\,/\,16089 & 10563\,/\,16089 & 11181\,/\,16089 & 162142\,/\,155375 & 232841\,/\,224286 & 260637\,/\,251464 & 26.78 & 41.19 & 50.94\vspace{3pt}\\ 
   
P1 & 10 & 2\,/\,2   & 10\,/\,11 & 4\,/\,2  & 28\,/\,93           & 87\,/\,93 & 118\,/\,93 & 123\,/\,93                & 871\,/\,809 & 1552\,/\,1459   & 1794\,/\,1696   & 0.03  & 0.05  & 0.06 \\ 
   
P2 & 100 & 2\,/\,3  & 100\,/\,129 & 4\,/\,2 & 288\,/\,1077       & 860\,/\,1077 & 1039\,/\,1077 & 1103\,/\,1077       & 9925\,/\,9249  & 14580\,/\,13725  & 16664\,/\,15745  & 0.62  & 0.76  & 0.90\\ 
   
P3 & 1000 & 2\,/\,3 & 999\,/\,1232 & 5\,/\,2 & 2857\,/\,10078    & 8630\,/\,10078 & 10486\,/\,10078 & 11122\,/\,10078 & 99665\,/\,92769 & 144674\,/\,135922 & 162657\,/\,153269 & 15.23 & 23.46 & 26.41\vspace{3pt}\\ 
   
M1 & 10 & 9\,/\,4    & 22\,/\,33 & 4\,/\,1 & 64\,/\,261             & 184\,/\,261 & 218\,/\,261 & 231\,/\,261            & 2284\,/\,2152  & 3204\,/\,3038   & 3559\,/\,3380   & 0.08 & 0.12  & 0.14\\ 
   
M2 & 100 & 10\,/\,3  & 109\,/\,141 & 1\,/\,3 & 319\,/\,1210         & 927\,/\,1210 & 1149\,/\,1210 & 1214\,/\,1210       & 11079\,/\,10378 & 17020\,/\,16097  & 18814\,/\,17826  & 0.40 & 0.72  & 0.75\\ 
   
M3 & 1000 & 12\,/\,4 & 1000\,/\,1224 & 5\,/\,2 & 2864\,/\,10055     & 8572\,/\,10055 & 10411\,/\,10055 & 11046\,/\,10055 & 99339\,/\,92775 & 144911\,/\,136508 & 162719\,/\,153681 & 6.50 & 10.97 & 13.68\\ 
\bottomrule
\end{tabular}
}
\label{table:syntheticTGs}
\end{table*}

In the case where the criticality level of a TG $G$ was $\mathit{\Lambda}=3$, we calculated $\epsilon_{ik}$ and $V_{ik}$ in $\dot{G}$ the same way as in the real-world application (\cref{subsubsec:realOverview}). In the cases where $\mathit{\Lambda}=1$ and $\mathit{\Lambda}=2$, we retained the same $V_{ik}$ as in the case of $\mathit{\Lambda}=3$. However, we adjusted $\epsilon_{ik}$ based on \eqref{eq:execMode} and the corresponding vulnerability thresholds in \cref{table:thresholds}.  
To assign random, yet realistic, values to the rest of the node and arc parameters in $\dot{G}$,
we first ran a number of relevant benchmarks \cite{Openbenchmarking, Geekbench5} on each device, as shown in \cref{table:benchmarks}.
For the benchmark scores reported in points (pts), higher values indicate better performance (in terms of latency), whereas for those reported in seconds (s), higher values indicate worse performance. We ranked the devices according to their benchmark score, from the slowest (rank 1) to the fastest (rank 3).
For each benchmark, we calculated the performance ratio $\theta_{k}$ of each device $k$. 
Specifically, for the benchmark scores in points, we calculated $\theta_{k}$ as the ratio of the score of device $k$ to the score of the previous device in the ranking, and for the scores in seconds, as the ratio of the score of the previous device to the score of device $k$.

Subsequently, we calculated the computational latency $L_{ik}$ for each node $N_{ik} \in \dot{G}$, by first assigning a random value to $L_{i \mathrm{e}}$, selected from those measured on device $\mathrm{e}$ in the case of the real-world application. 
We then set $L_{i \mathrm{h}}$ and $L_{i \mathrm{c}}$ as $L_{i \mathrm{h}} = L_{i \mathrm{e}} / \theta_{\mathrm{h}}$ and $L_{i \mathrm{c}} = L_{i \mathrm{h}} / \theta_{\mathrm{c}}$, respectively, where $\theta_{\mathrm{h}}$ and $\theta_{\mathrm{c}}$ were randomly selected from the respective values in \cref{table:benchmarks}.
Similarly, to determine the power consumption $P_{ik}$, we first assigned a value to $P_{i \mathrm{e}}$, randomly selected from those measured on device $\mathrm{e}$ for the real-world workflow. 
We then calculated $P_{i \mathrm{h}}$ and $P_{i \mathrm{c}}$ as $P_{i \mathrm{h}} = P_{i \mathrm{e}} \, \theta_{\mathrm{h}}$ and $P_{i \mathrm{c}} = P_{i \mathrm{h}} \, \theta_{\mathrm{c}}$, respectively. 
If the derived $P_{ik}$ was not in the interval $( P_{k}^{\mathrm{idle}}, P_{k}^{\mathrm{max}} ]$, we adjusted it accordingly. Specifically, if $P_{ik} \leq P_{k}^{\mathrm{idle}}$, we adjusted it to $P_{k}^{\mathrm{idle}} (1 + \omega)$, whereas if $P_{ik} > P_{k}^{\mathrm{max}}$, we adjusted it to $P_{k}^{\mathrm{max}} (1 - \omega)$. We randomly selected the adjustment factor $\omega$ in $[ 0.1\%, 0.5\% ]$, so that the adjusted value of $P_{ik}$ did not deviate significantly from its initially calculated value.  

Furthermore, we randomly selected $M_{i}$, $S_{i}$, and $D_{i}$ from the respective values measured in the case of the real-world application. We calculated the rest of the node and arc parameters in $\dot{G}$, and subsequently the parameters in $\ddot{G}$, as described in \cref{subsubsec:realOverview}. 
The number of nodes and arcs in each synthetic TG, as well as in the corresponding EG and REG (for each criticality level), are shown in \cref{table:syntheticTGs}.

\begin{table}[t]
\centering
\caption{Performance ranking of computational devices.}
\scriptsize
\resizebox{0.93\columnwidth}{!}{
\begin{tabular}{c|r|rr|rr} 
\toprule
 & \multicolumn{5}{c}{Device} \\
\hline

$\mathrm{e}$/$\mathrm{h}$/$\mathrm{c}$   & \multicolumn{1}{c|}{$\mathrm{e}$}    & \multicolumn{2}{c|}{$\mathrm{h}$}    & \multicolumn{2}{c}{$\mathrm{c}$} \\

\hline 

  & \multicolumn{1}{c|}{Raspberry Pi 3}  & \multicolumn{2}{c|}{Mi Notebook}  & \multicolumn{2}{c}{HPE ProLiant}\\ 
Benchmark   & \multicolumn{1}{c|}{Model B} &  \multicolumn{2}{c|}{Pro} &  \multicolumn{2}{c}{DL580 Gen10}		\\
\cline{2-6}

 & \multicolumn{1}{c|}{Score} & \multicolumn{1}{c}{Score} & \multicolumn{1}{c|}{$\theta_{\mathrm{h}}$}  & \multicolumn{1}{c}{Score} & \multicolumn{1}{c}{$\theta_{\mathrm{c}}$} \\
 
\hline

Geekbench 5  & 398.00\,pts \hspace{5pt}  & 3612.00\,pts & 9.08  & 33323.00\,pts       & 9.23 \\

NumPy  & 10.63\,pts \hspace{5pt}    & 284.68\,pts   & 26.78 & 405.71\,pts    & 1.43	\\

TensorFlow  & 12648.38\,s \hspace{5pt}  & 169.70\,s & 74.53	& 59.82\,s      & 2.84	\\

Scikit-Learn & 1373.54\,s \hspace{5pt}    & 10.11\,s  & 135.86    & 7.64\,s  & 1.32	\\

R Benchmark  & 3.80\,s \hspace{5pt}  & 0.50\,s   & 7.60  & 0.26\,s          & 1.92	\\

\hline

Rank & \multicolumn{1}{c|}{1} & \multicolumn{2}{c|}{2} & \multicolumn{2}{c}{3} \\

\bottomrule

\end{tabular}
}
\label{table:benchmarks}
\end{table}

\subsubsection{Experimental results}
\label{subsubsec:syntheticResults}

\begin{figure*}[t]
    \centering
    \subfloat[S1 (serial TG with 10 tasks)]{\includegraphics[width=.31\textwidth]{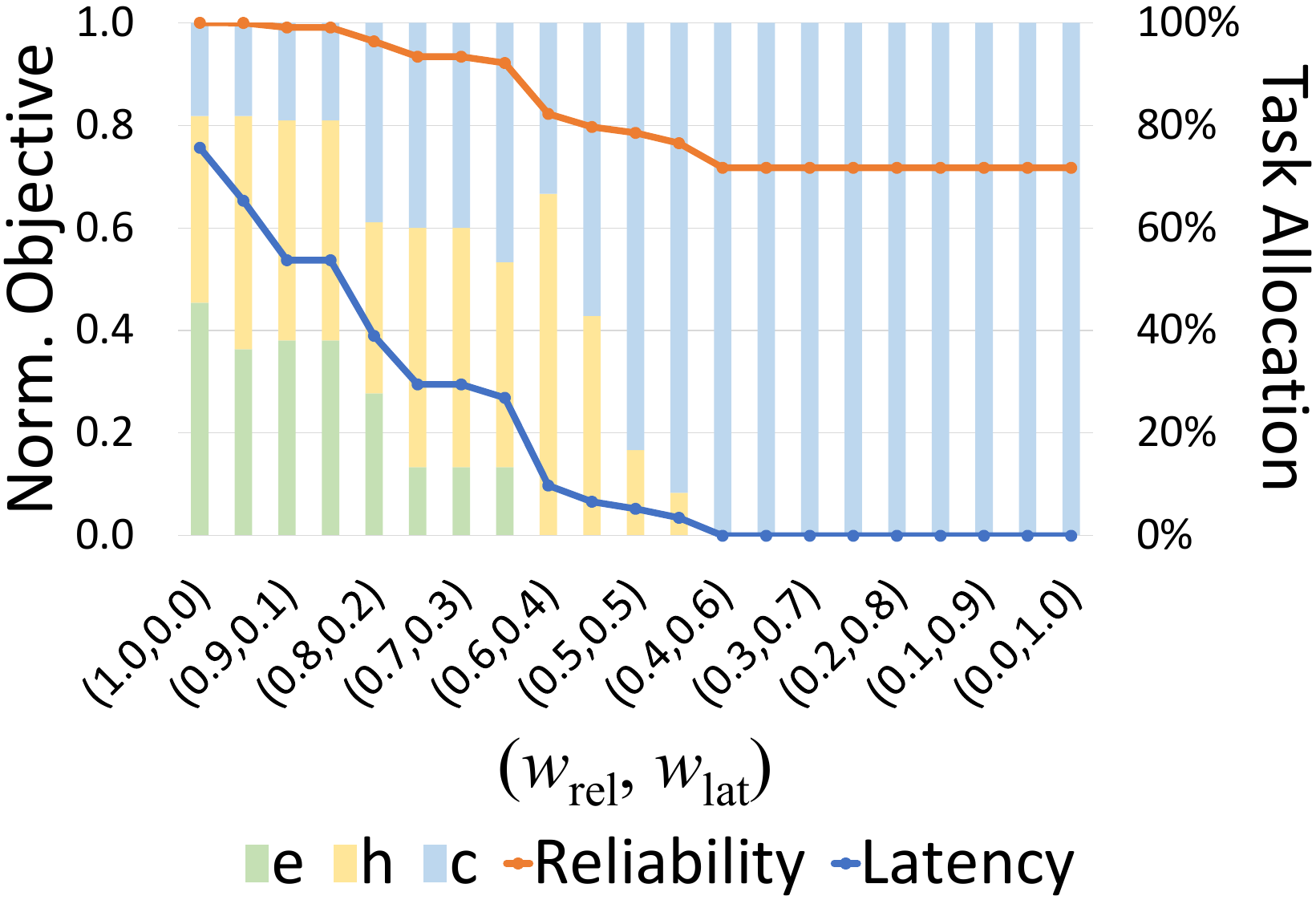}
        \label{fig:cl1_serial10}}
    \hfill
    \subfloat[S2 (serial TG with 100 tasks)]{\includegraphics[width=.31\textwidth]{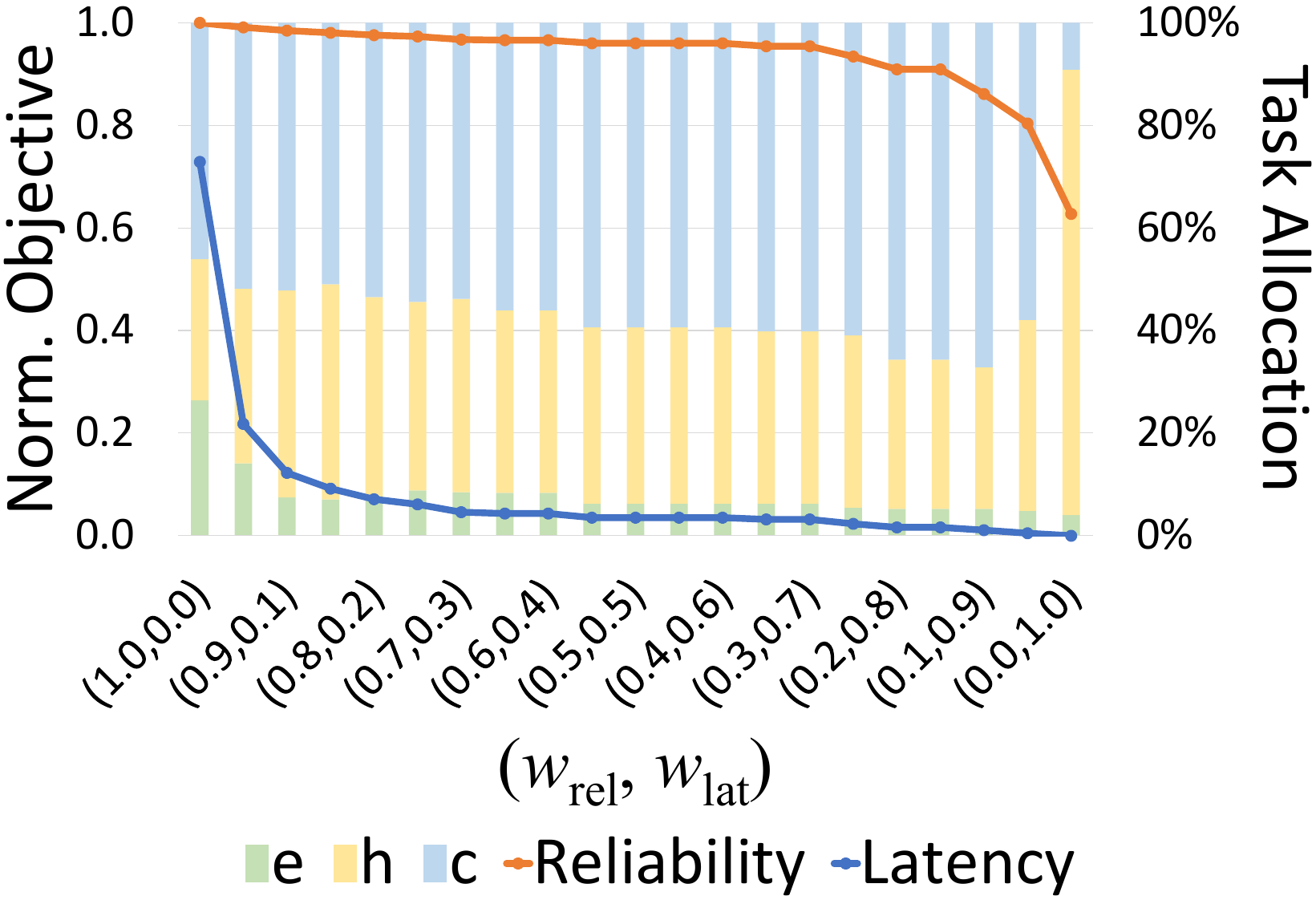}
        \label{fig:cl1_serial100}}
    \hfill
    \subfloat[S3 (serial TG with 1000 tasks)]{\includegraphics[width=.31\textwidth]{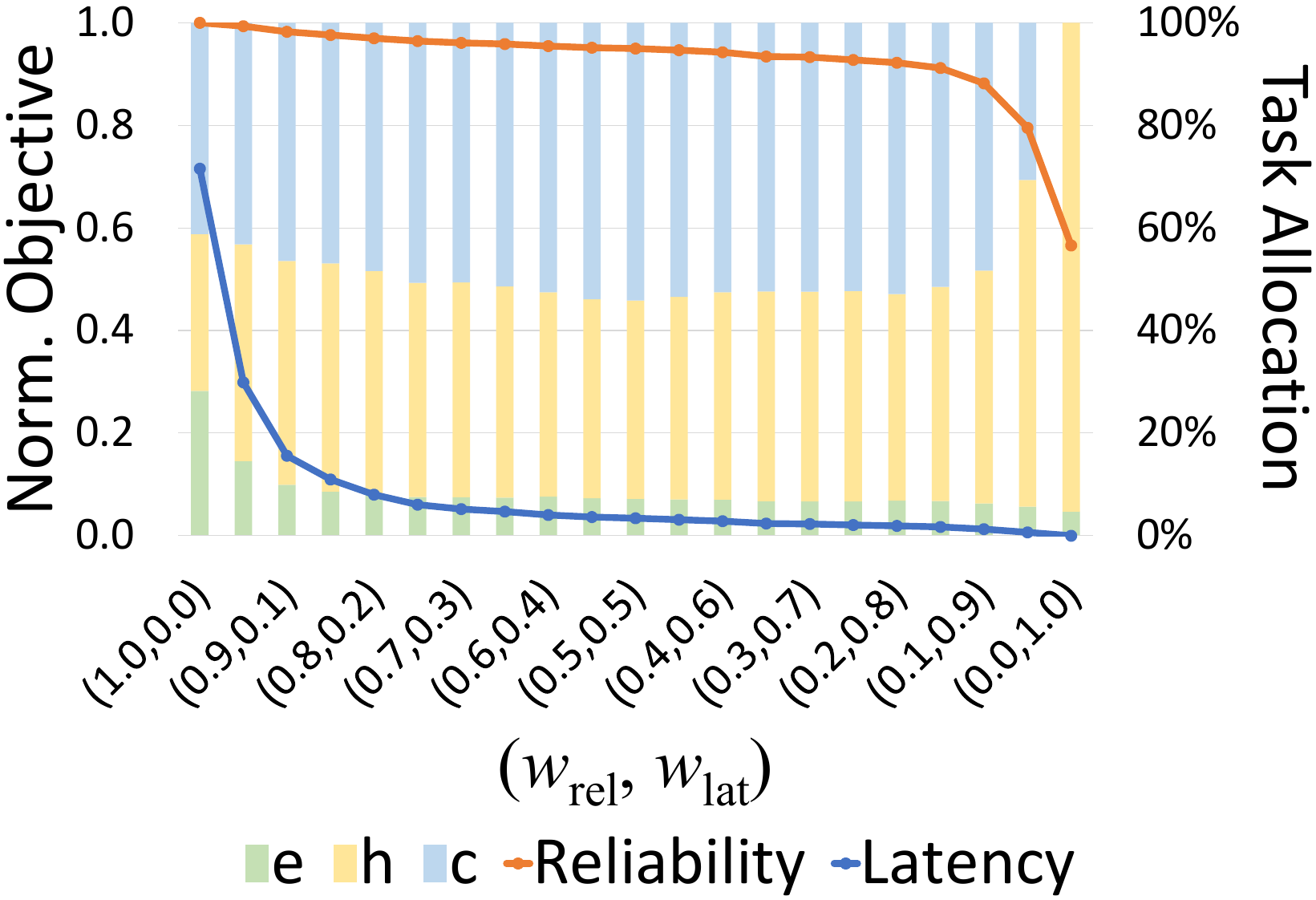}
        \label{fig:cl1_serial1000}}
    \\
    \subfloat[P1 (parallel TG with 10 tasks)]{\includegraphics[width=.31\textwidth]{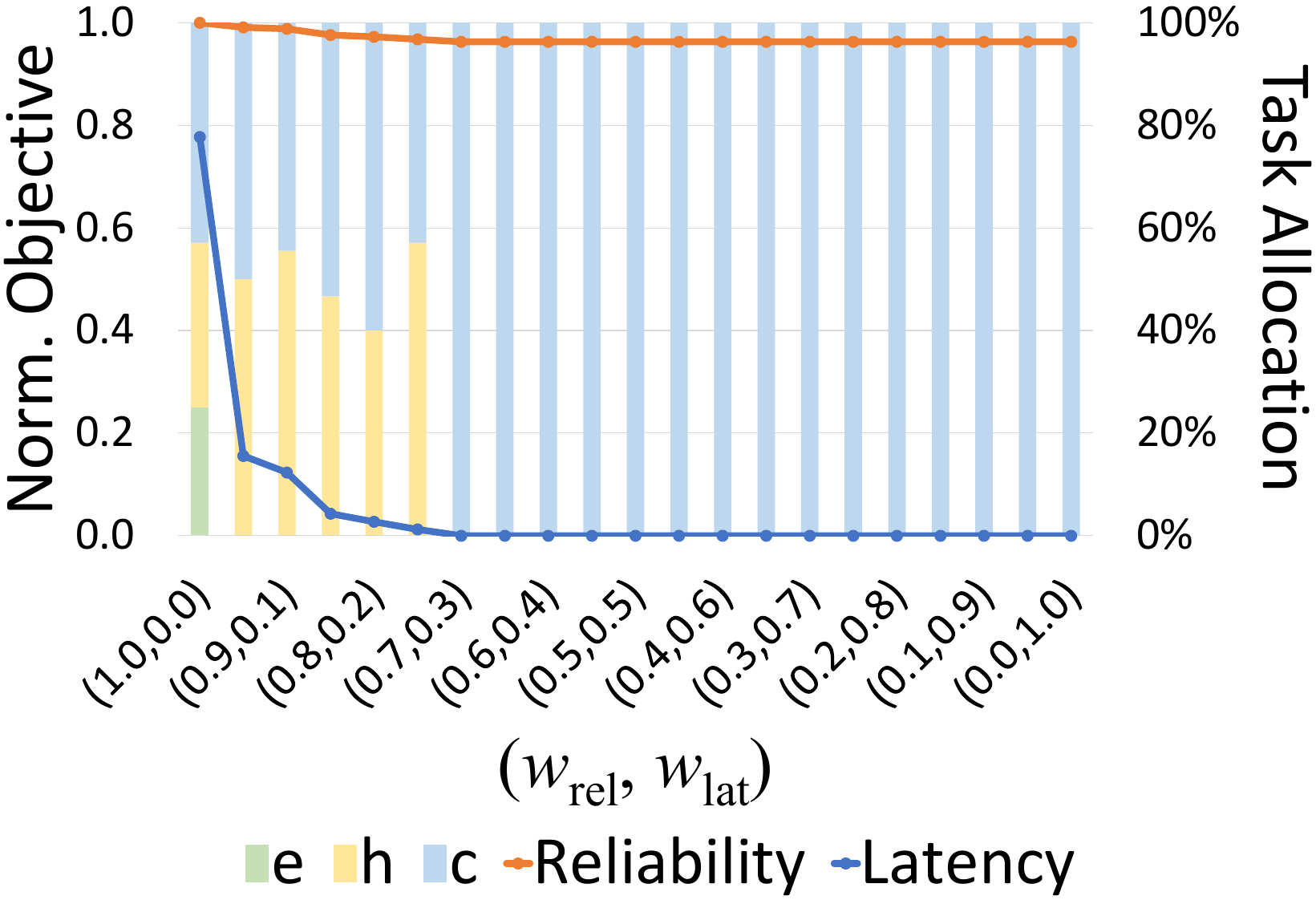}
        \label{fig:cl1_parallel10}}
    \hfill
    \subfloat[P2 (parallel TG with 100 tasks)]{\includegraphics[width=.31\textwidth]{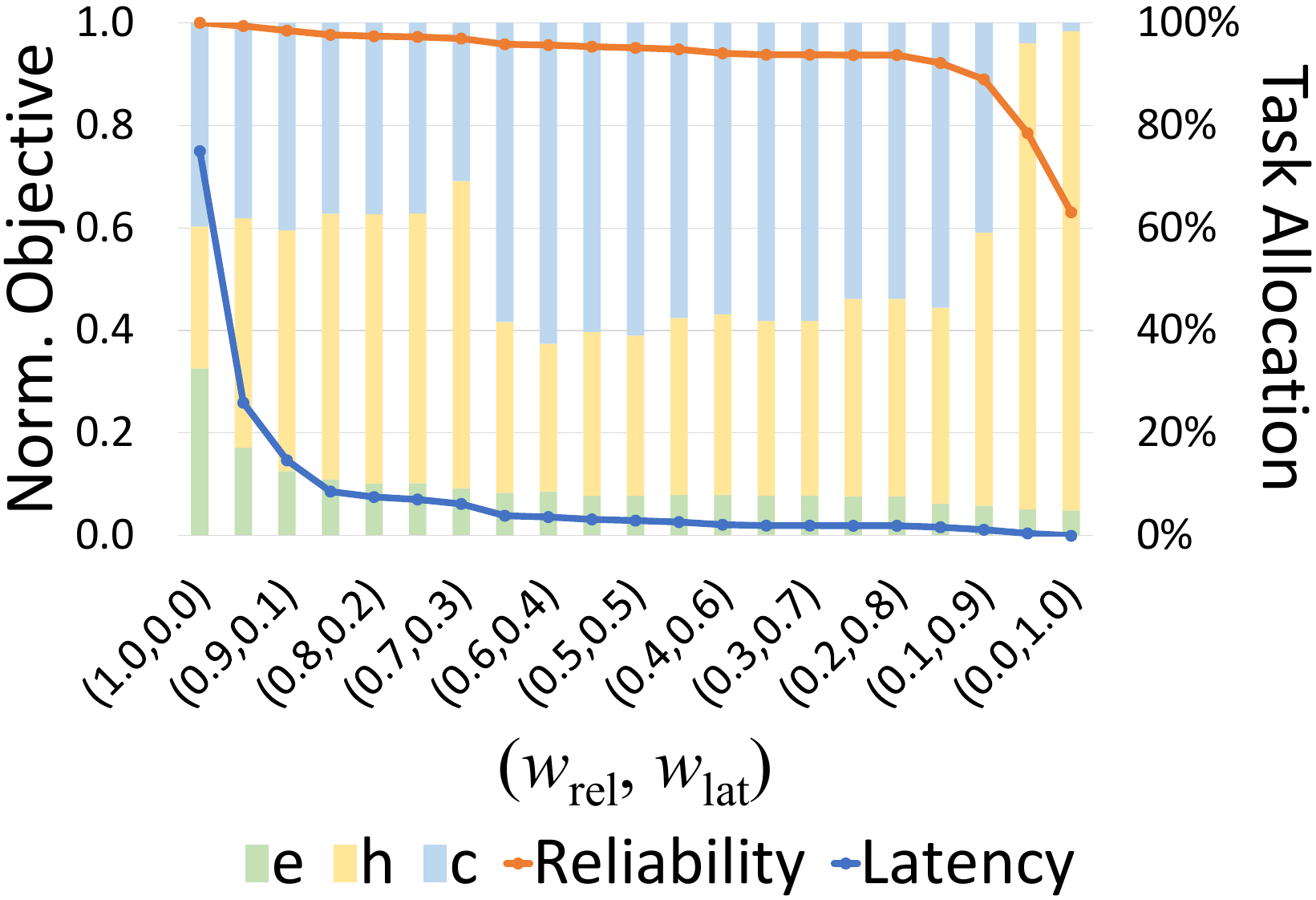}
        \label{fig:cl1_parallel100}}
    \hfill
    \subfloat[P3 (parallel TG with 1000 tasks)]{\includegraphics[width=.31\textwidth]{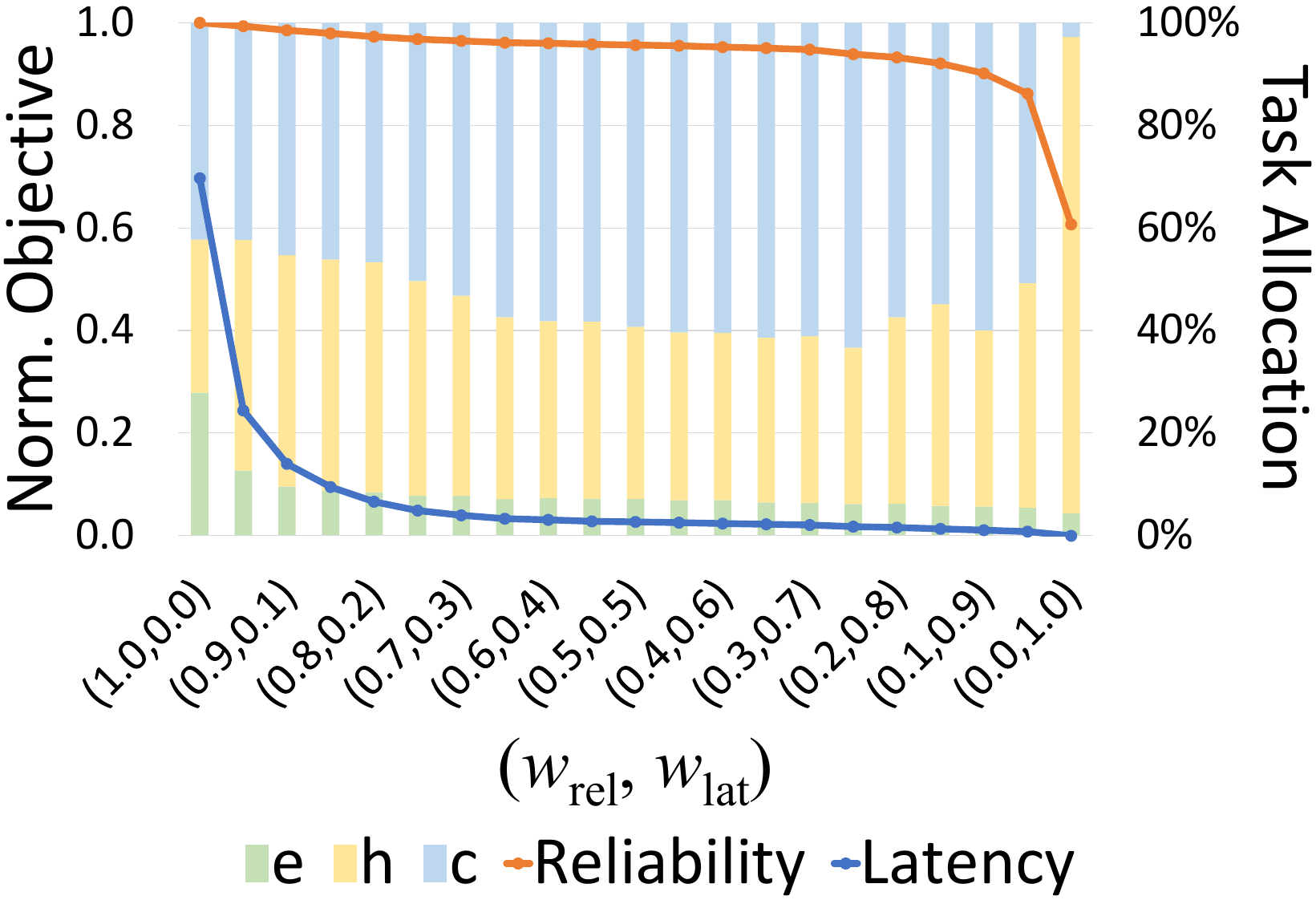}
        \label{fig:cl1_parallel1000}}
    \\
    \subfloat[M1 (mixed TG with 10 tasks)]{\includegraphics[width=.31\textwidth]{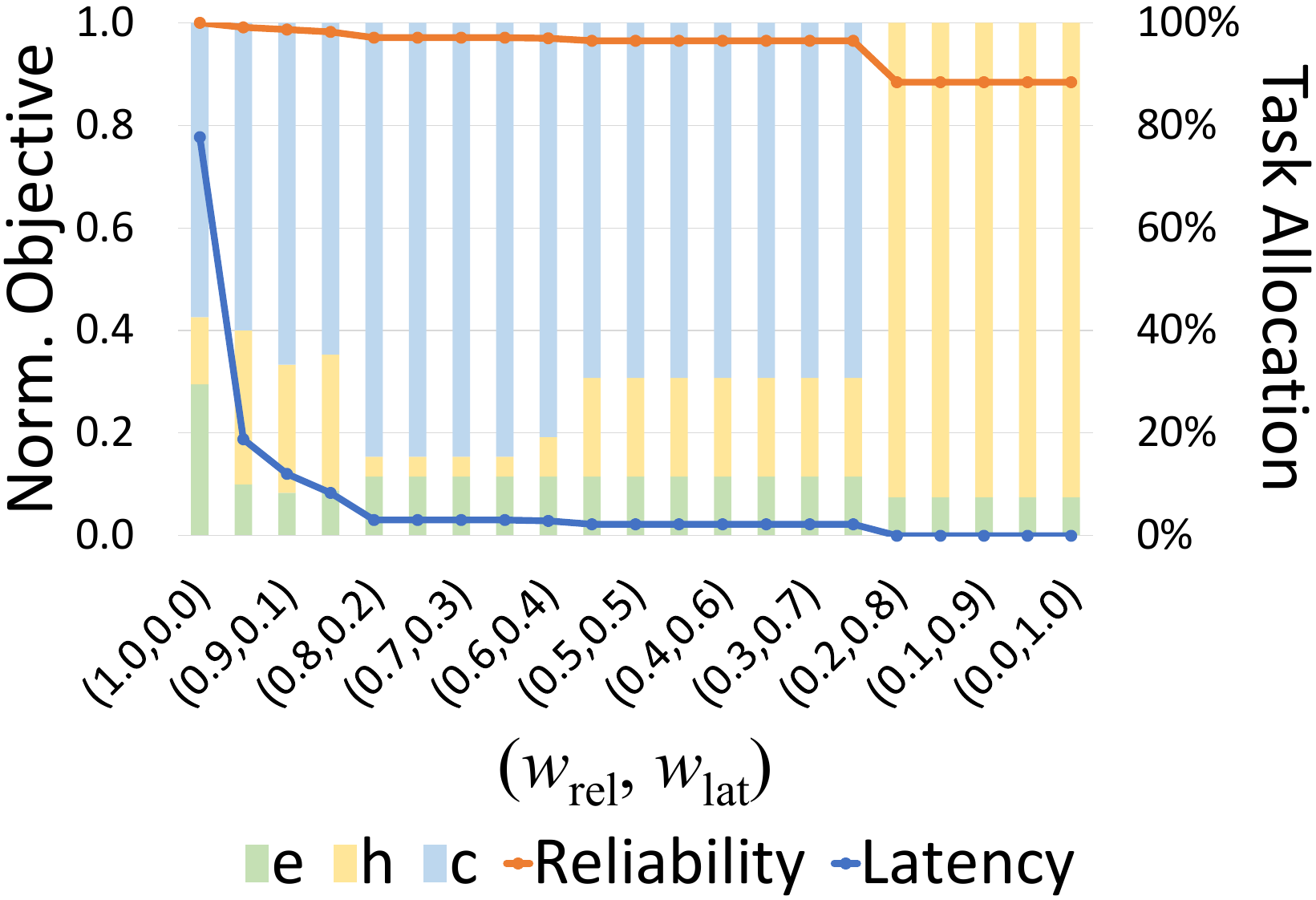}
        \label{fig:cl1_mixed10}}
    \hfill
    \subfloat[M2 (mixed TG with 100 tasks)]{\includegraphics[width=.31\textwidth]{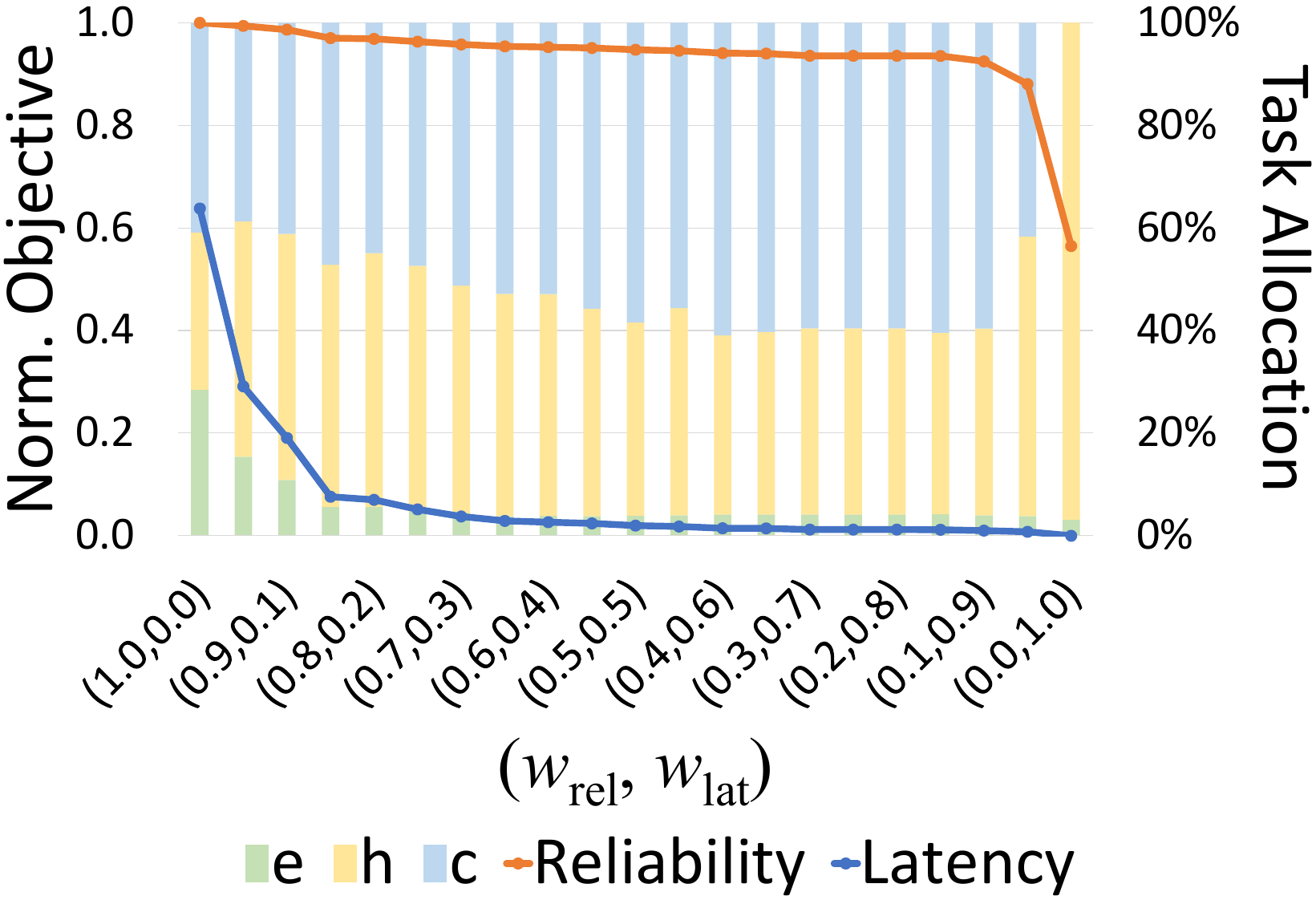}
        \label{fig:cl1_mixed100}}
    \hfill
    \subfloat[M3 (mixed TG with 1000 tasks)]{\includegraphics[width=.31\textwidth]{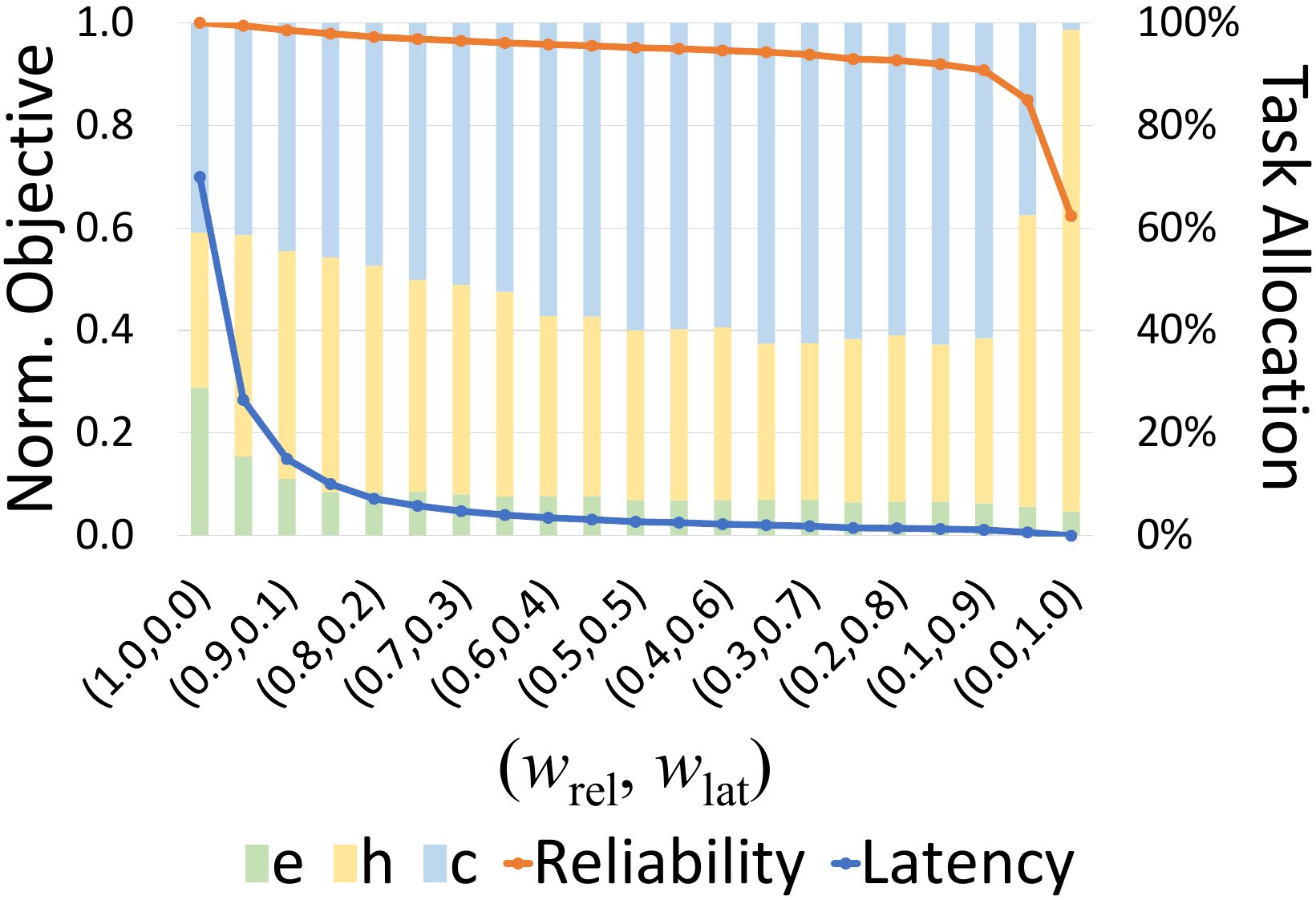}
        \label{fig:cl1_mixed1000}}
    \caption{Normalized overall reliability and latency, and percentage of allocated tasks (primary and replicas) per device, with respect to $w_{\mathrm{rel}}$ and $w_{\mathrm{lat}}$, for synthetic workflows with $\mathit{\Lambda} = 1$.}
    \label{fig:syntheticCL1}
\end{figure*}

\begin{figure*}[t]
    \centering
    \subfloat[S1 (serial TG with 10 tasks)]{\includegraphics[width=.31\textwidth]{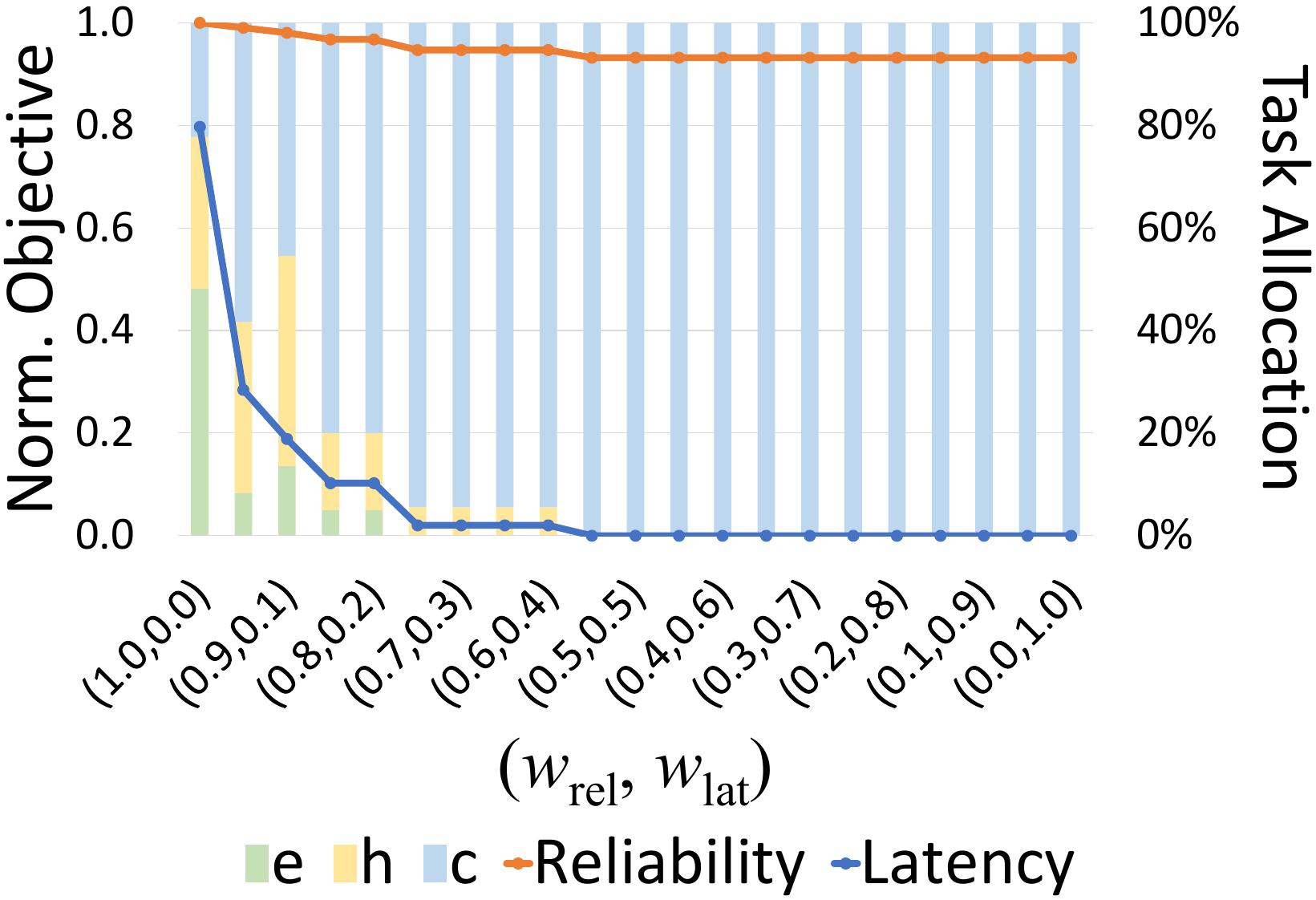}
        \label{fig:cl2_serial10}}
    \hfill
    \subfloat[S2 (serial TG with 100 tasks)]{\includegraphics[width=.31\textwidth]{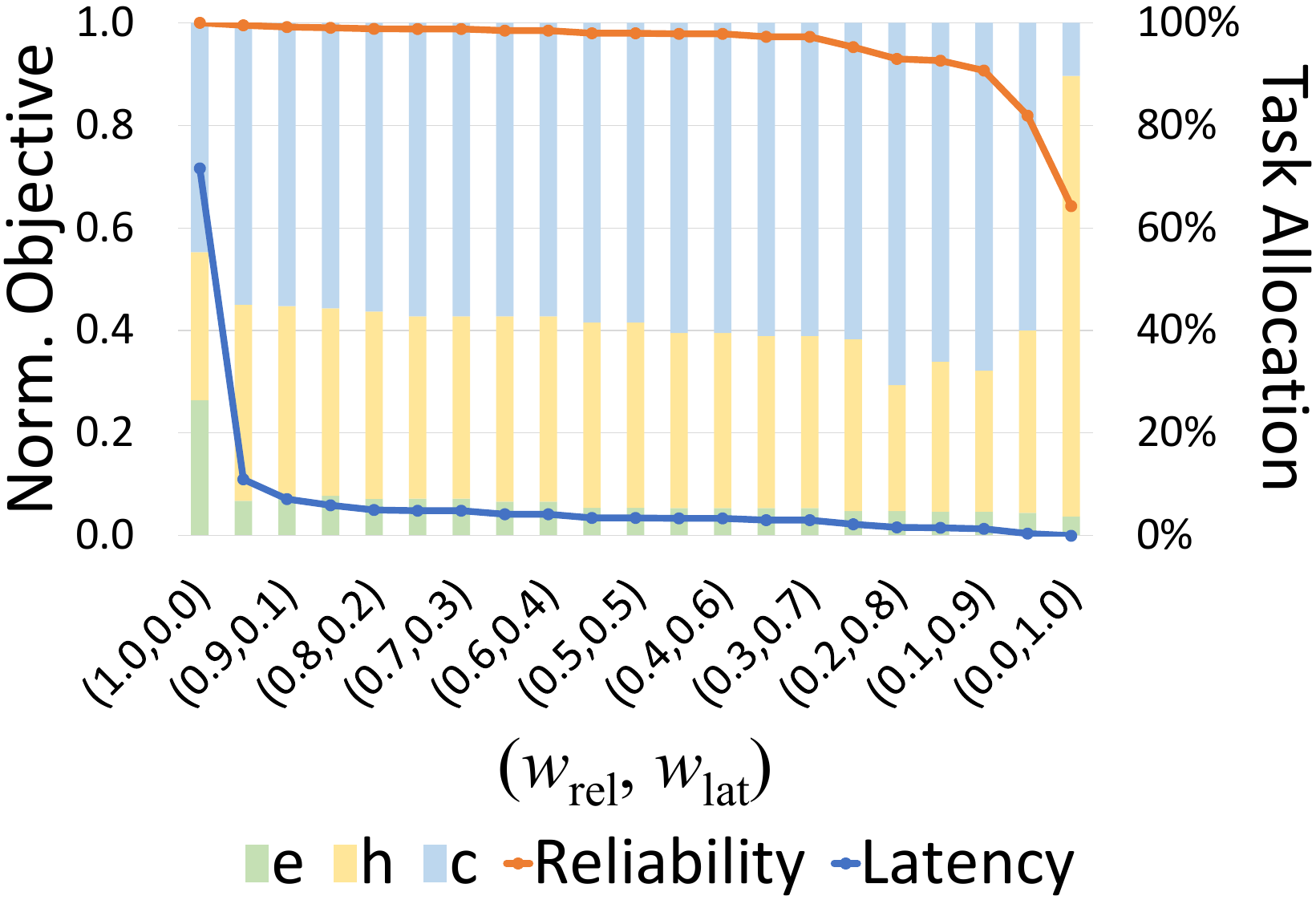}
        \label{fig:cl2_serial100}}
    \hfill
    \subfloat[S3 (serial TG with 1000 tasks)]{\includegraphics[width=.31\textwidth]{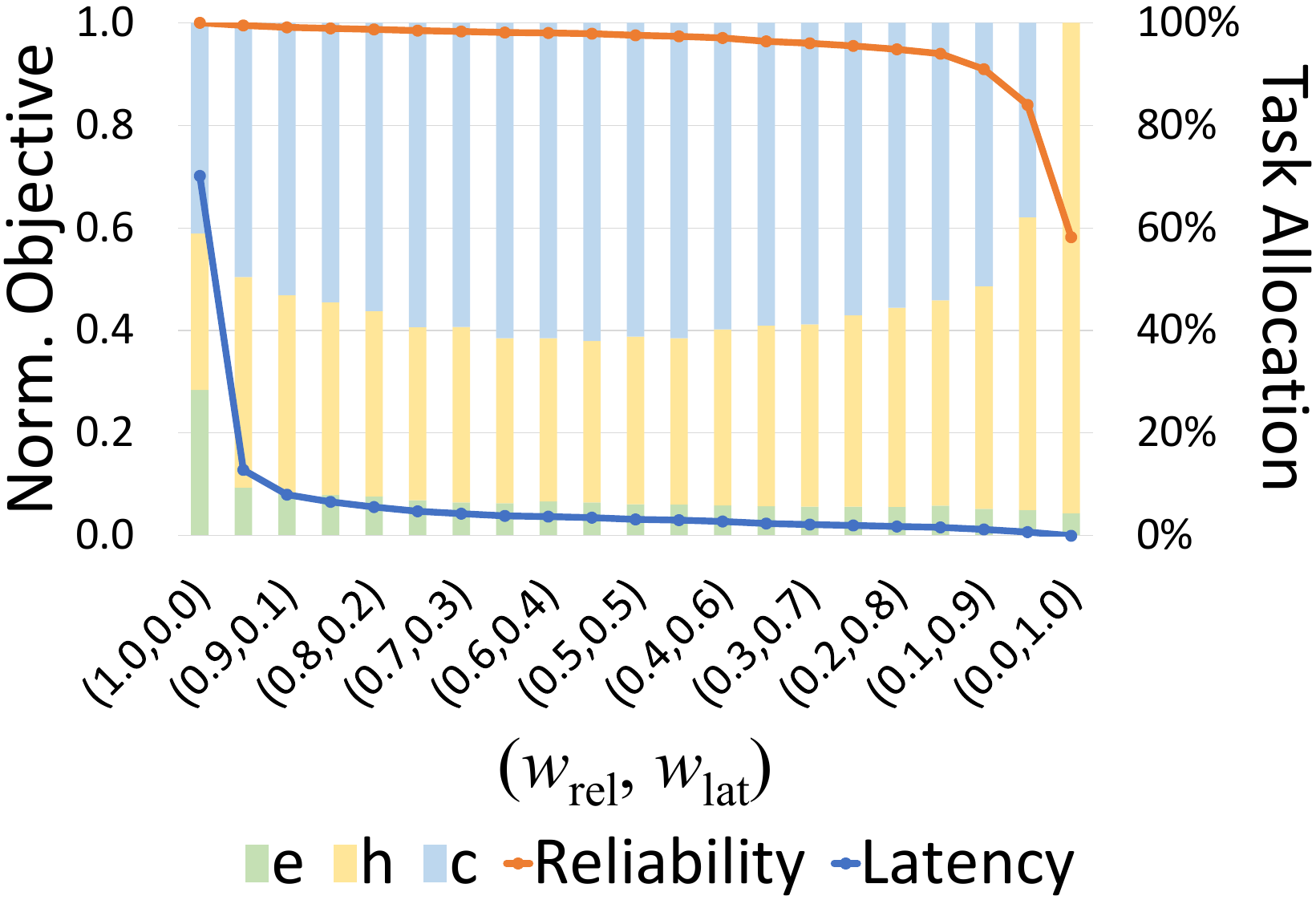}
        \label{fig:cl2_serial1000}}
    \\
    \subfloat[P1 (parallel TG with 10 tasks)]{\includegraphics[width=.31\textwidth]{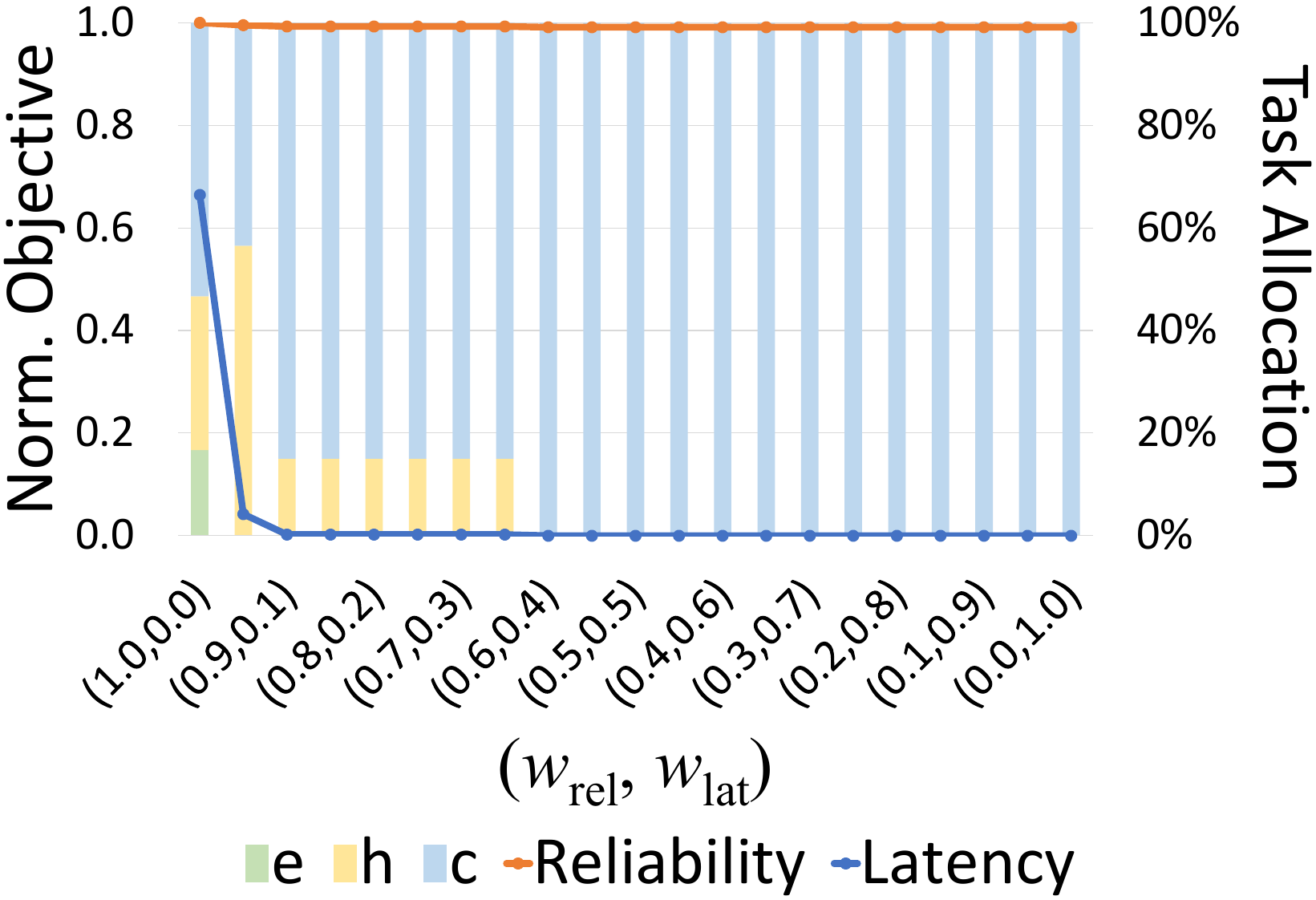}
        \label{fig:cl2_parallel10}}
    \hfill
    \subfloat[P2 (parallel TG with 100 tasks)]{\includegraphics[width=.31\textwidth]{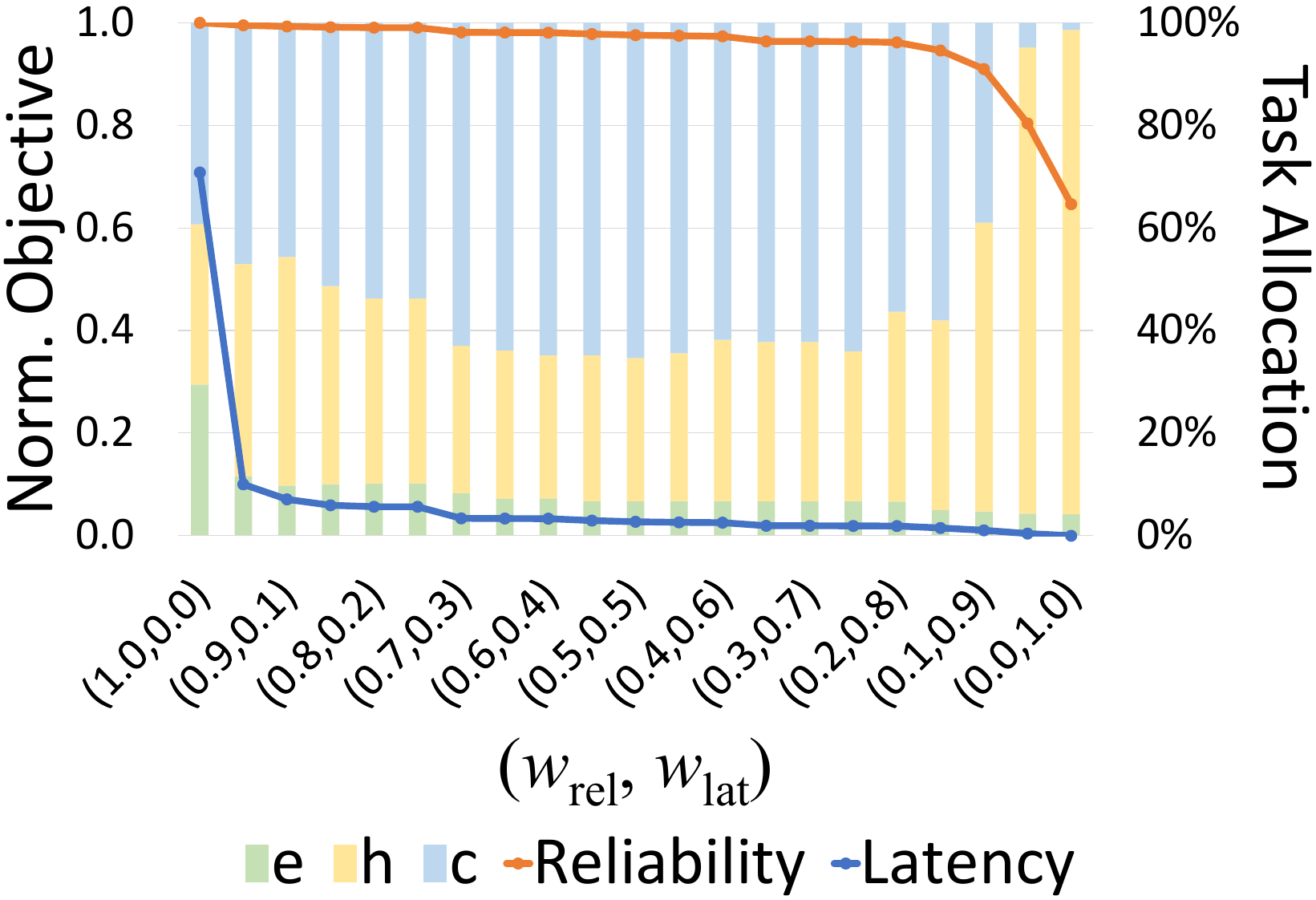}
        \label{fig:cl2_parallel100}}
    \hfill
    \subfloat[P3 (parallel TG with 1000 tasks)]{\includegraphics[width=.31\textwidth]{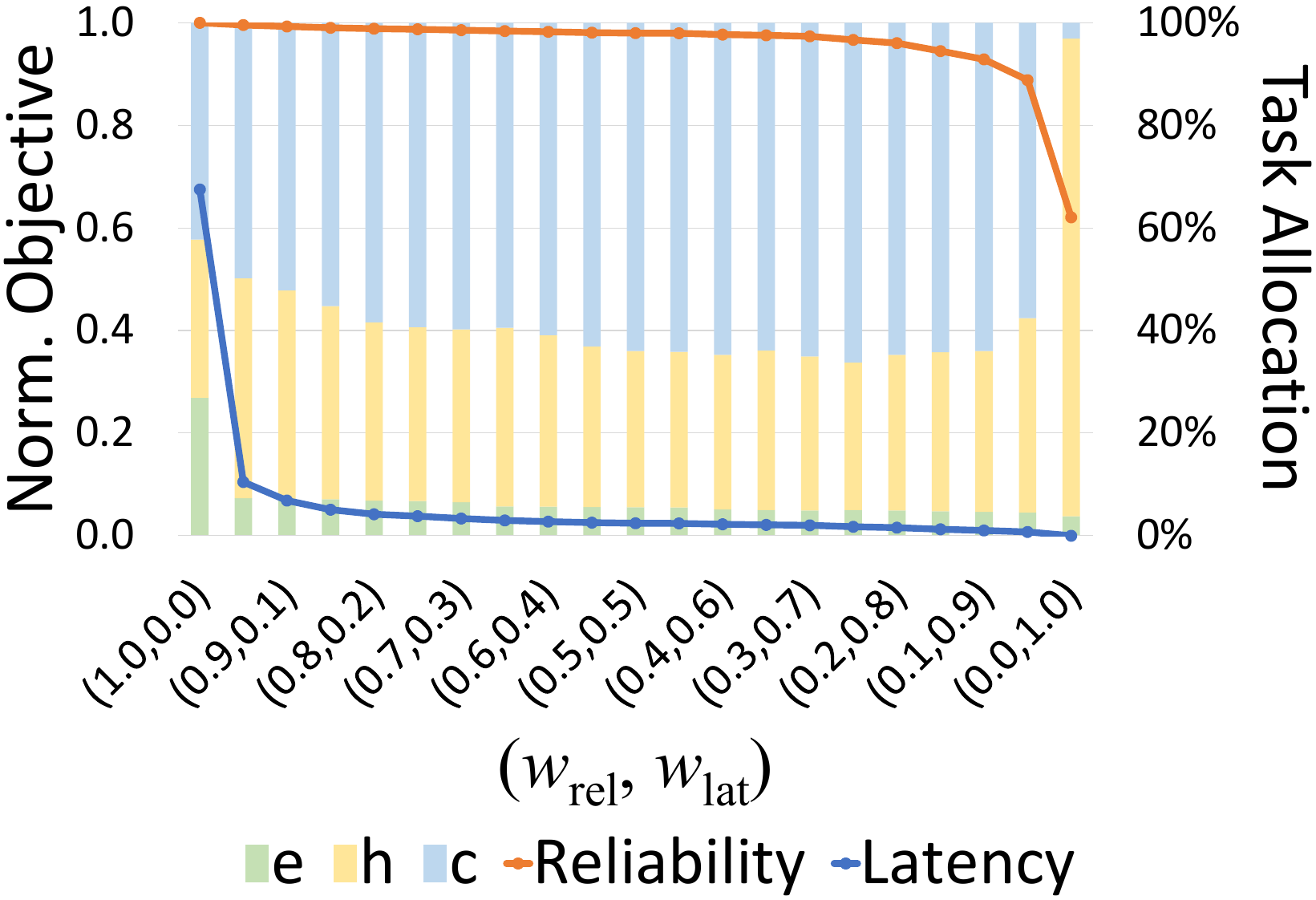}
        \label{fig:cl2_parallel1000}}
    \\
    \subfloat[M1 (mixed TG with 10 tasks)]{\includegraphics[width=.31\textwidth]{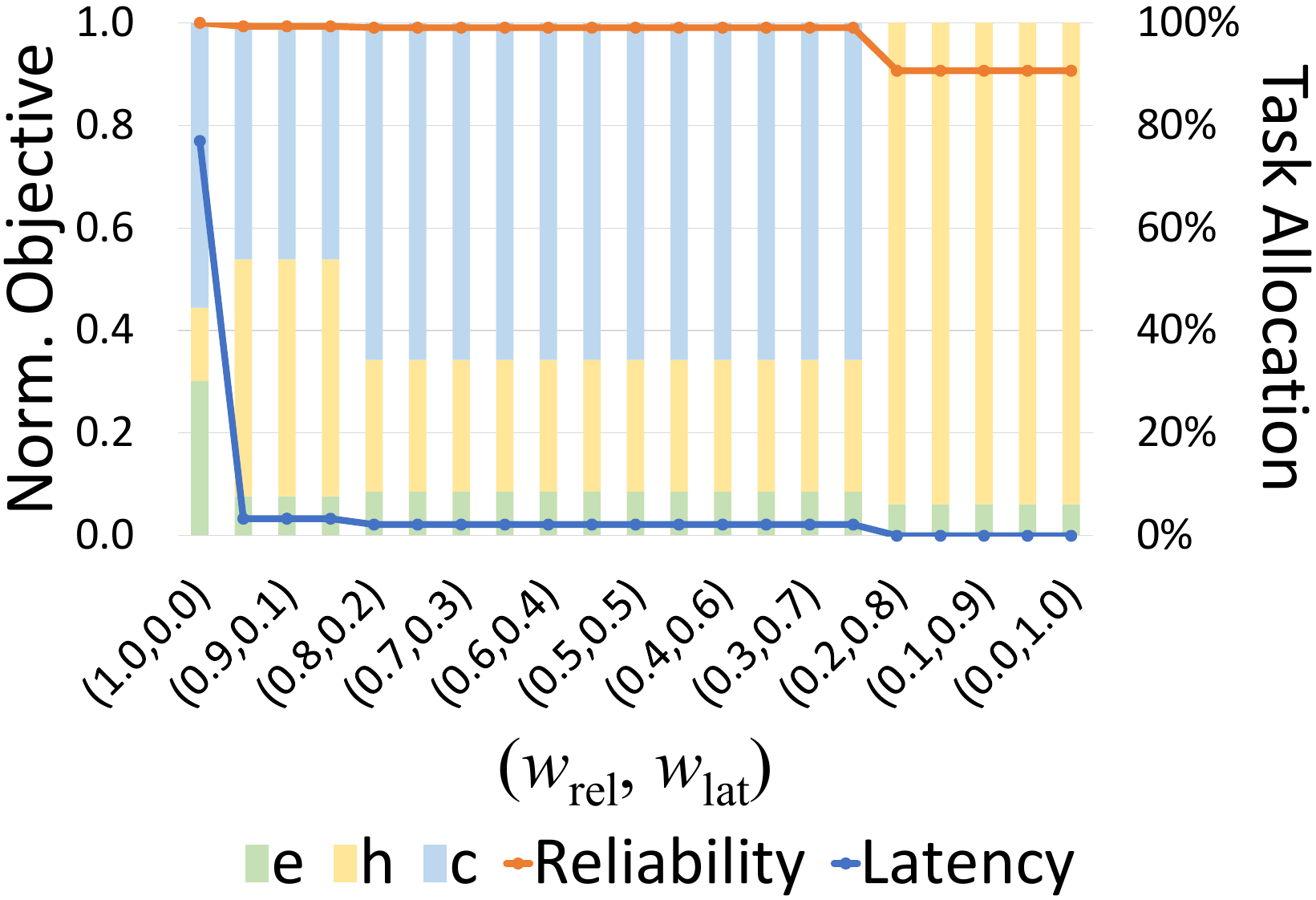}
        \label{fig:cl2_mixed10}}
    \hfill
    \subfloat[M2 (mixed TG with 100 tasks)]{\includegraphics[width=.31\textwidth]{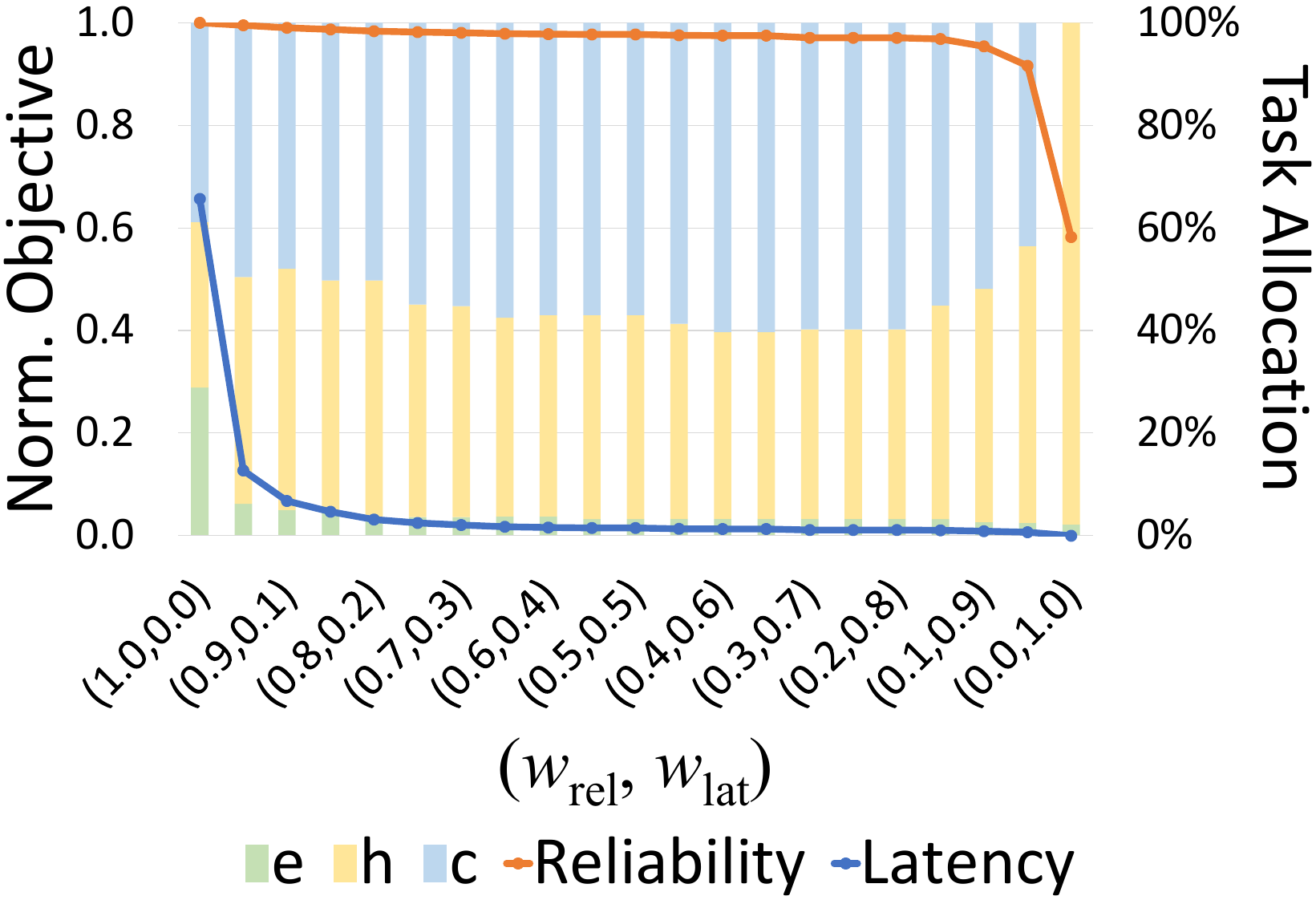}
        \label{fig:cl2_mixed100}}
    \hfill
    \subfloat[M3 (mixed TG with 1000 tasks)]{\includegraphics[width=.31\textwidth]{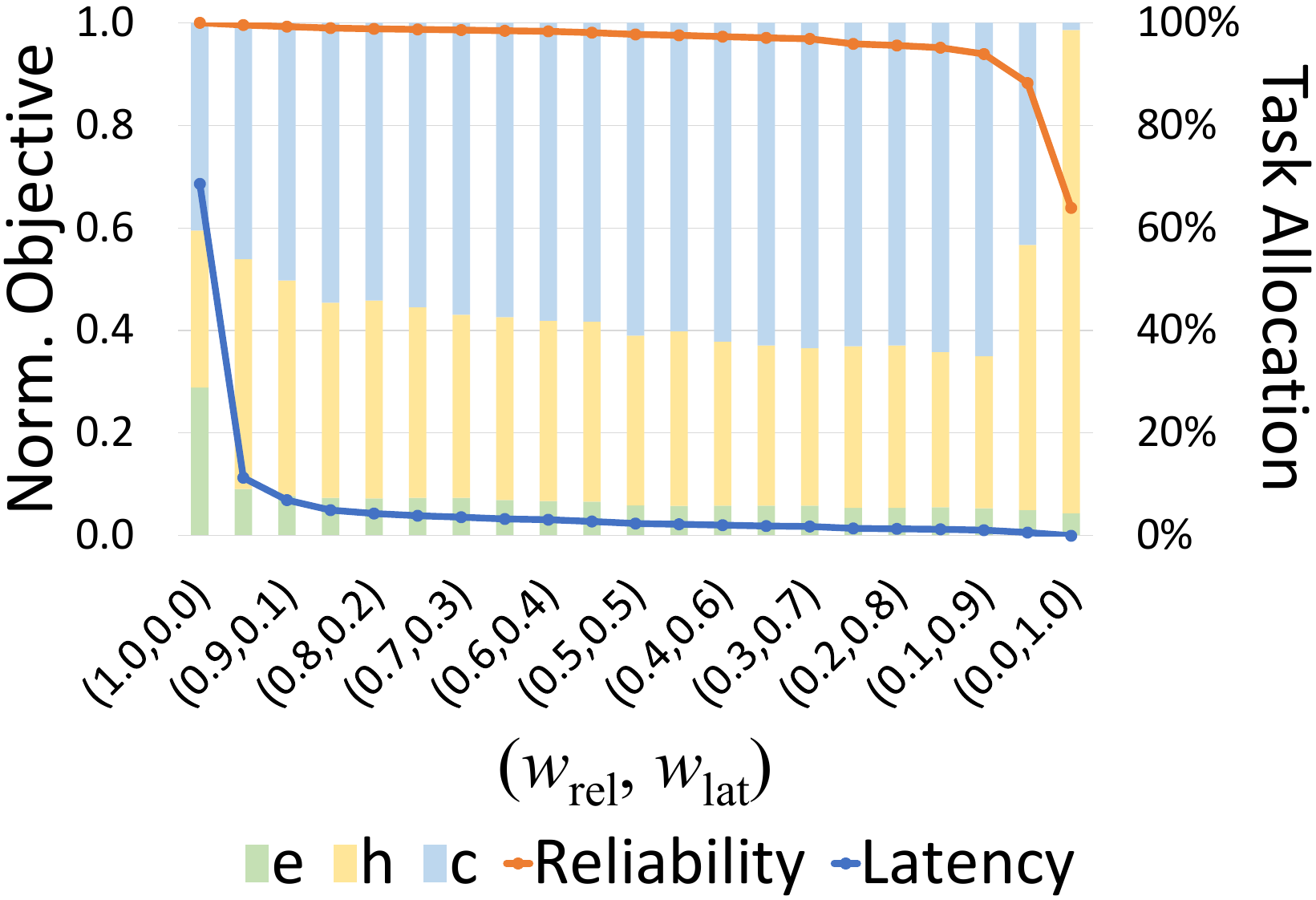}
        \label{fig:cl2_mixed1000}}
    \caption{Normalized overall reliability and latency, and percentage of allocated tasks (primary and replicas) per device, with respect to $w_{\mathrm{rel}}$ and $w_{\mathrm{lat}}$, for synthetic workflows with $\mathit{\Lambda} = 2$.}
    \label{fig:syntheticCL2}
\end{figure*}

\begin{figure*}[t]
    \centering
    \subfloat[S1 (serial TG with 10 tasks)]{\includegraphics[width=.31\textwidth]{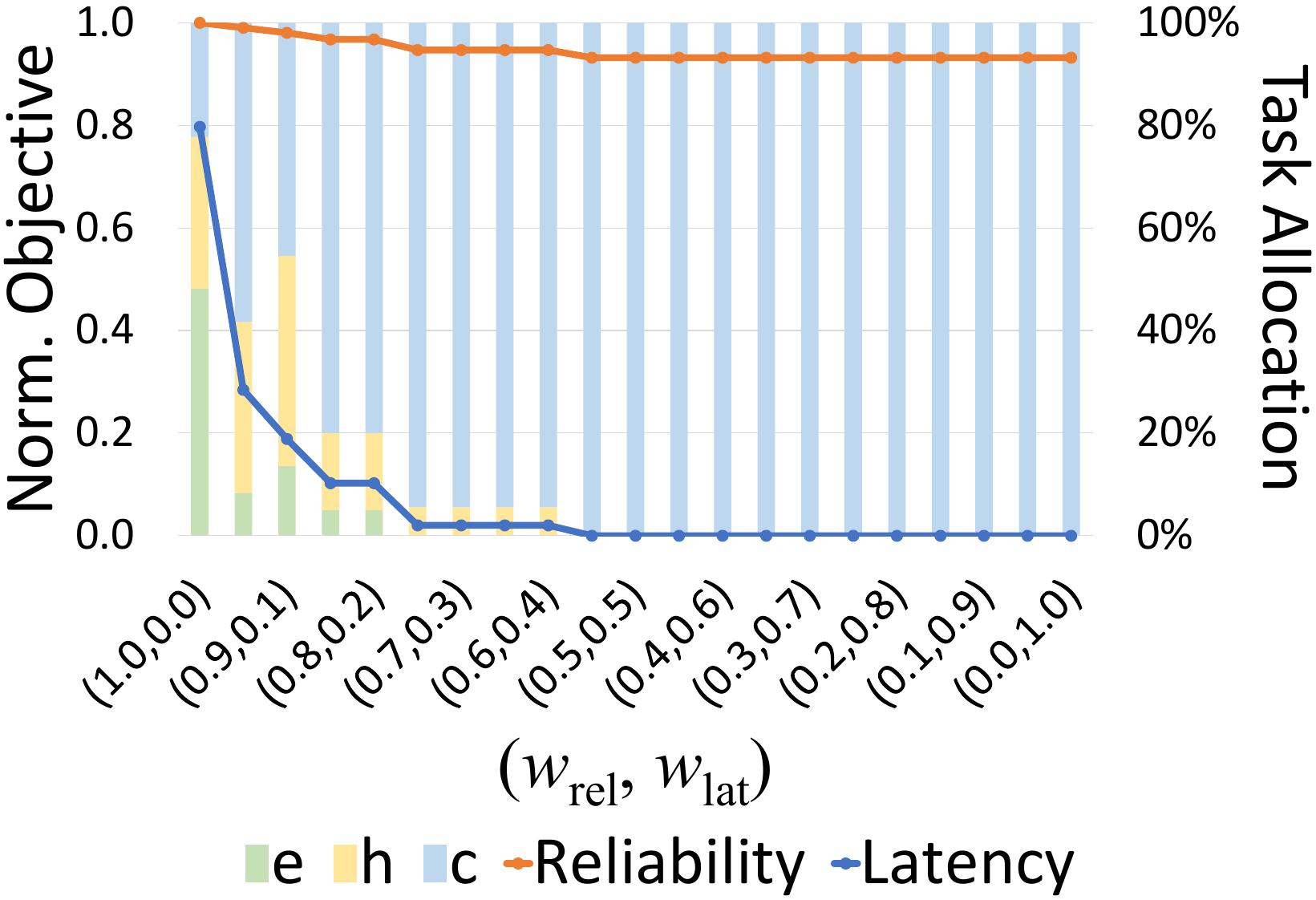}
        \label{fig:cl3_serial10}}
    \hfill
    \subfloat[S2 (serial TG with 100 tasks)]{\includegraphics[width=.31\textwidth]{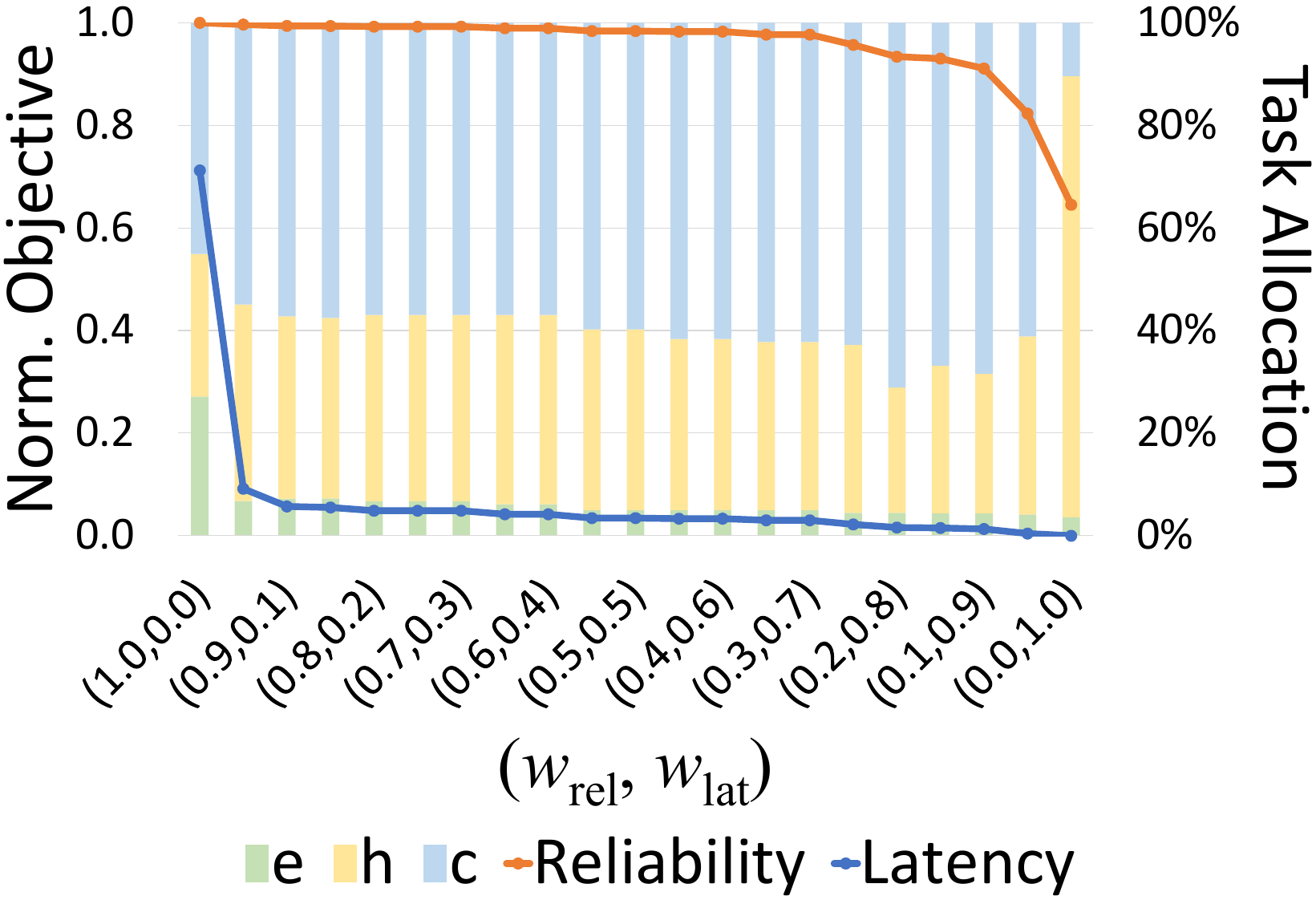}
        \label{fig:cl3_serial100}}
    \hfill
    \subfloat[S3 (serial TG with 1000 tasks)]{\includegraphics[width=.31\textwidth]{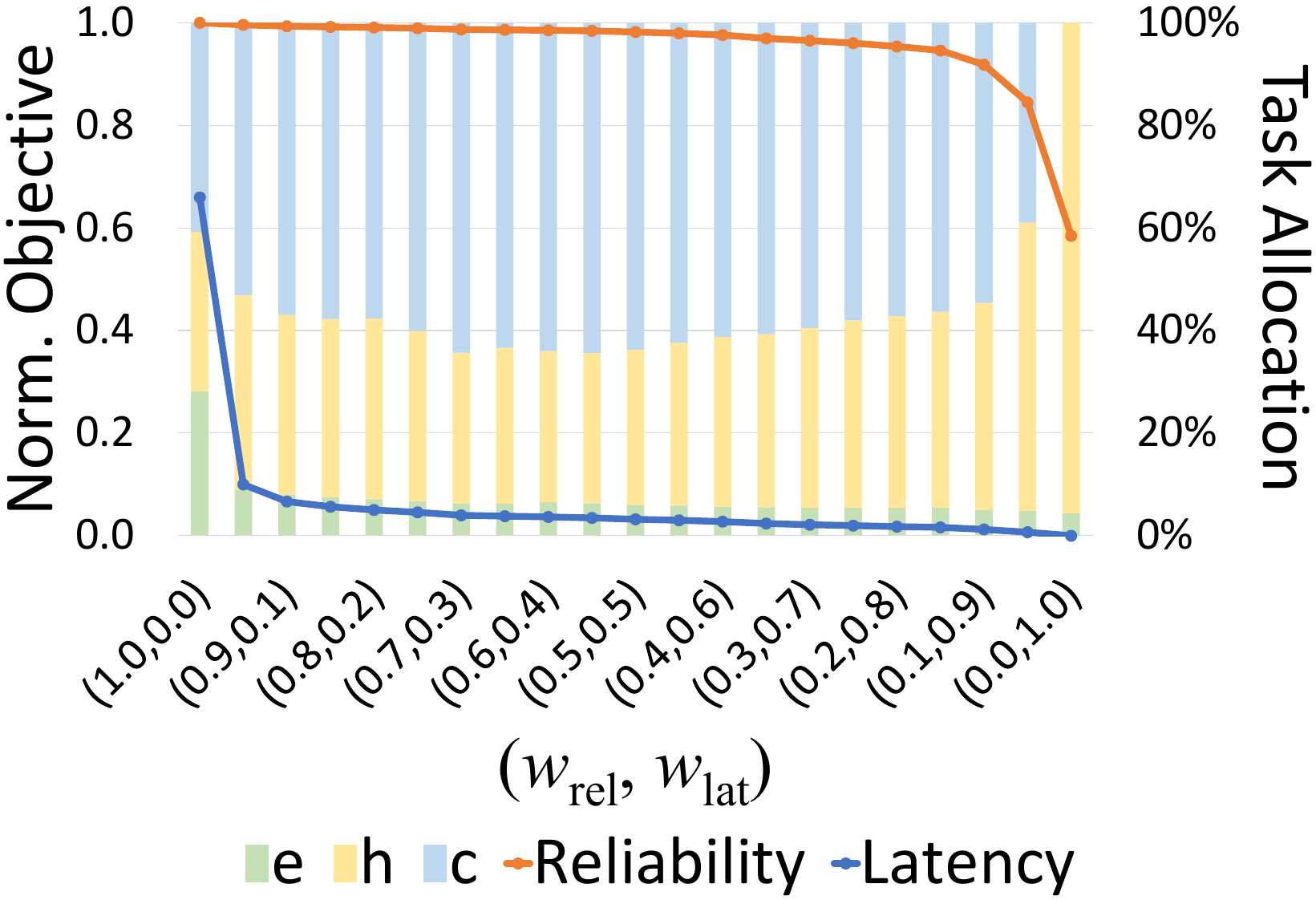}
        \label{fig:cl3_serial1000}}
    \\
    \subfloat[P1 (parallel TG with 10 tasks)]{\includegraphics[width=.31\textwidth]{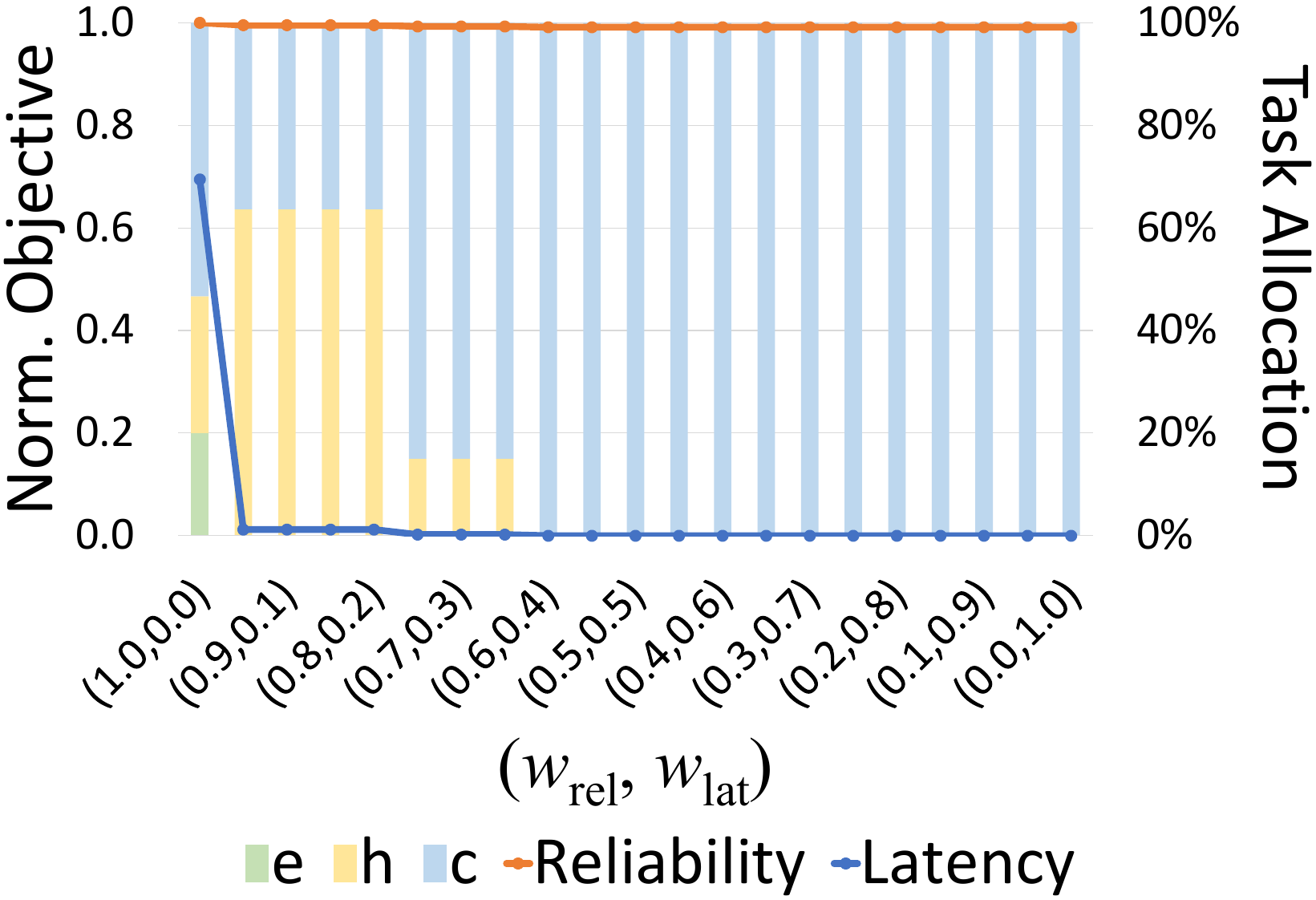}
        \label{fig:cl3_parallel10}}
    \hfill
    \subfloat[P2 (parallel TG with 100 tasks)]{\includegraphics[width=.31\textwidth]{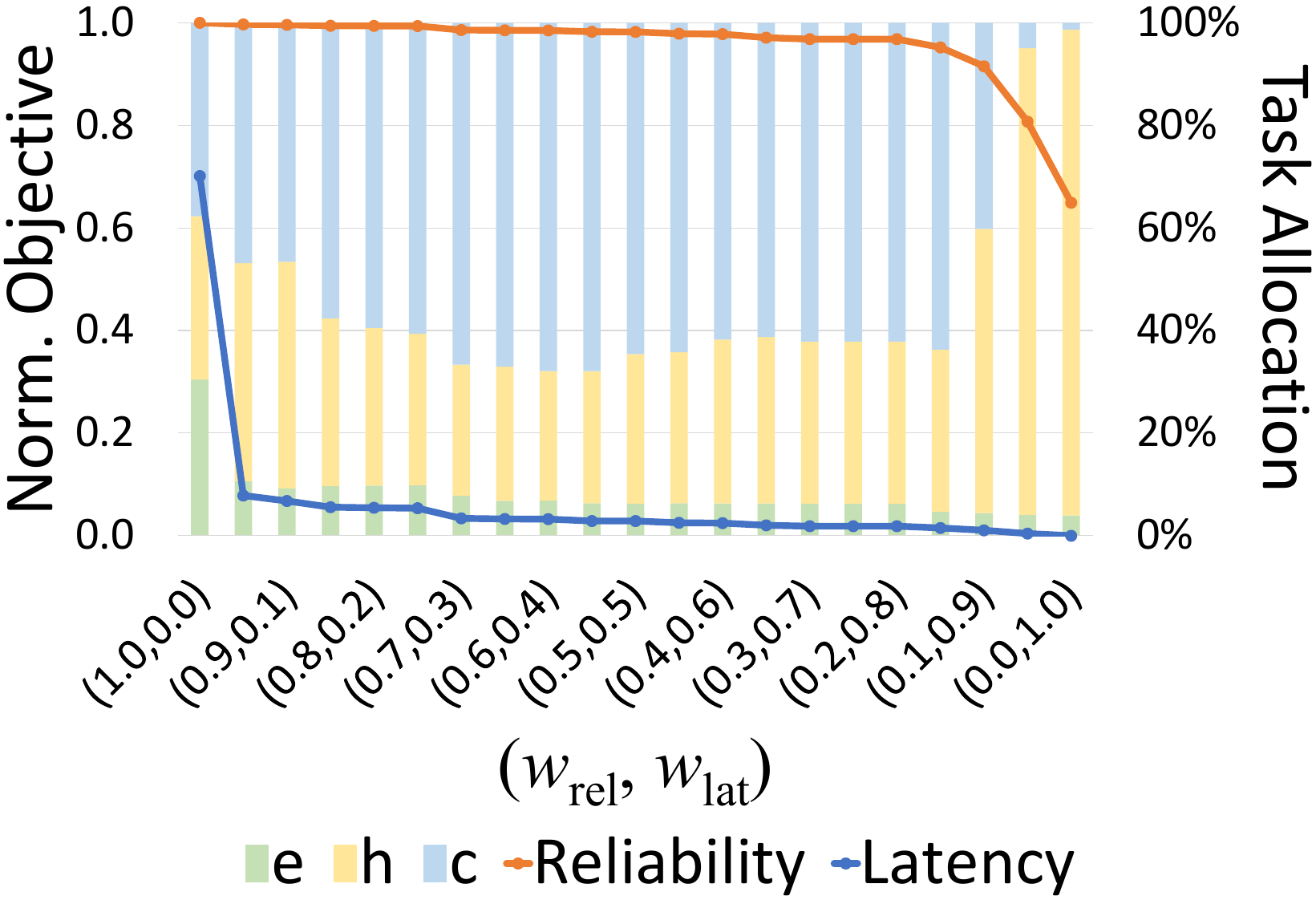}
        \label{fig:cl3_parallel100}}
    \hfill
    \subfloat[P3 (parallel TG with 1000 tasks)]{\includegraphics[width=.31\textwidth]{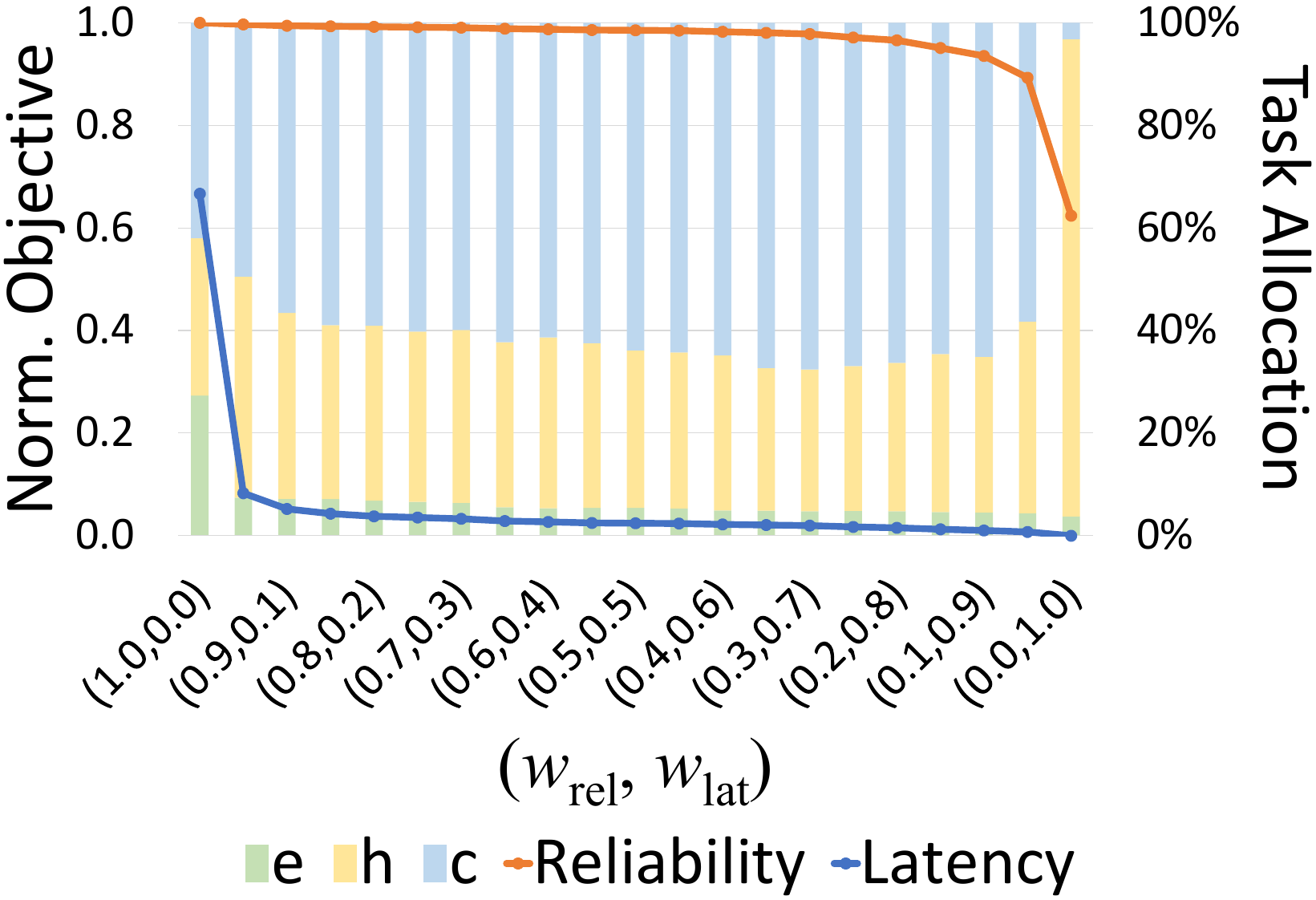}
        \label{fig:cl3_parallel1000}}
    \\
    \subfloat[M1 (mixed TG with 10 tasks)]{\includegraphics[width=.31\textwidth]{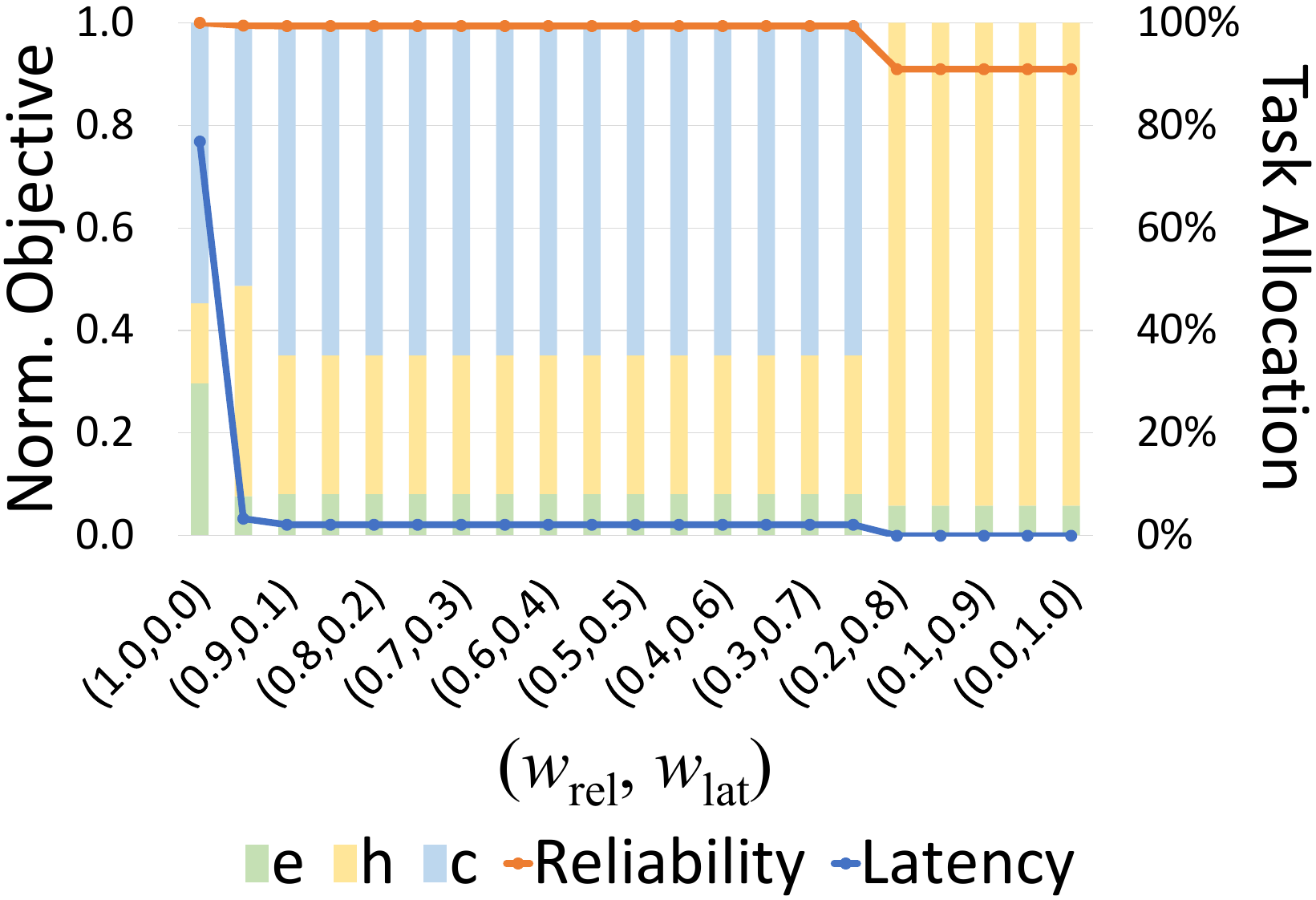}
        \label{fig:cl3_mixed10}}
    \hfill
    \subfloat[M2 (mixed TG with 100 tasks)]{\includegraphics[width=.31\textwidth]{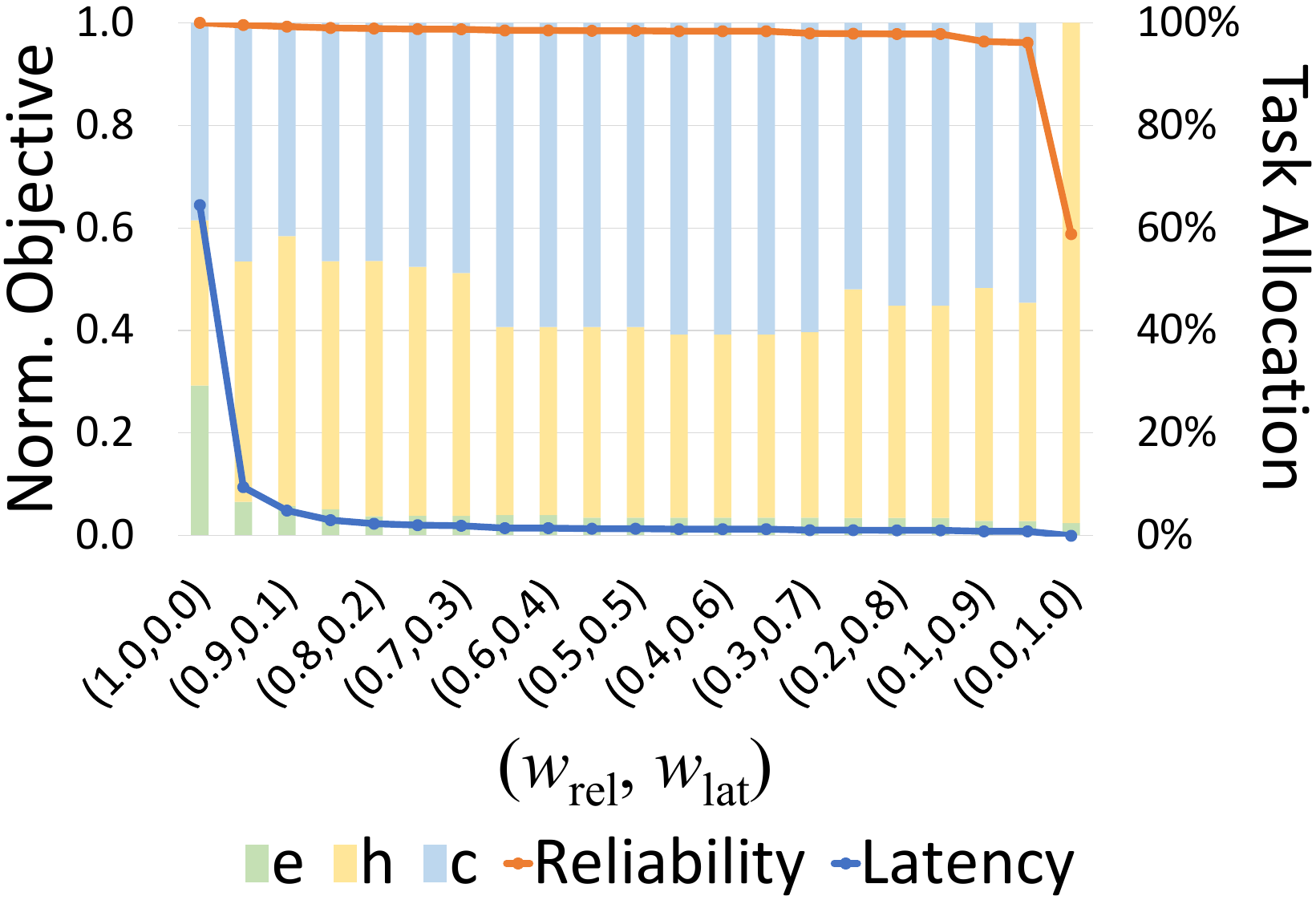}
        \label{fig:cl3_mixed100}}
    \hfill
    \subfloat[M3 (mixed TG with 1000 tasks)]{\includegraphics[width=.31\textwidth]{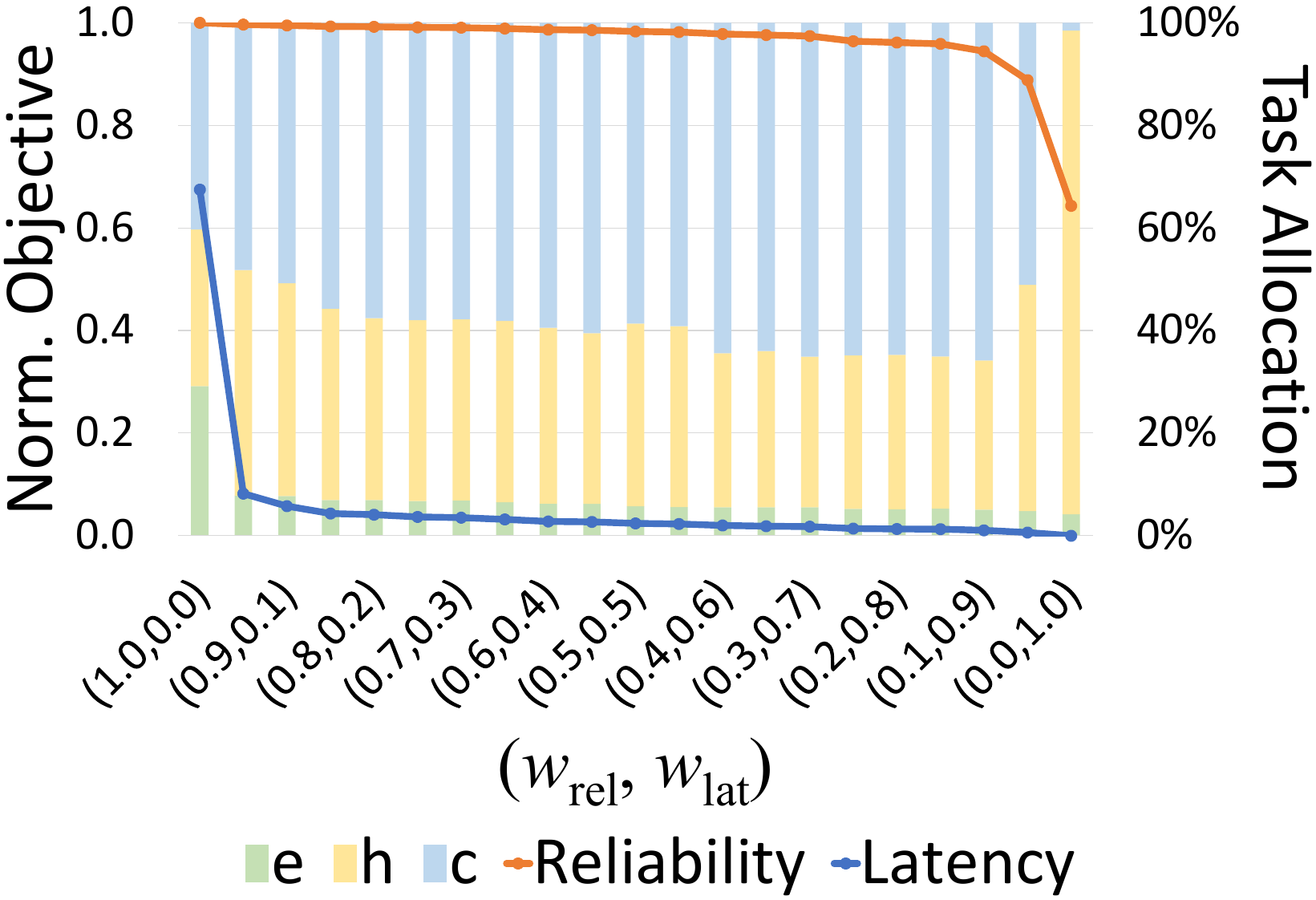}
        \label{fig:cl3_mixed1000}}
    \caption{Normalized overall reliability and latency, and percentage of allocated tasks (primary and replicas) per device, with respect to $w_{\mathrm{rel}}$ and $w_{\mathrm{lat}}$, for synthetic workflows with $\mathit{\Lambda} = 3$.}
    \label{fig:syntheticCL3}
\end{figure*}

The results regarding the normalized overall reliability and latency, with respect to $w_{\mathrm{rel}}$ and $w_{\mathrm{lat}}$, for synthetic workflows with criticality level $\mathit{\Lambda} \in \{1, \allowbreak 2, \allowbreak 3\}$ (each evaluated using TGs of serial, parallel, and mixed structure, and with 10, 100, and 1000 tasks), are illustrated in Figs. \ref{fig:syntheticCL1}, \ref{fig:syntheticCL2}, and \ref{fig:syntheticCL3}, respectively.
The results also show the percentage of tasks (primary and replicas) that were allocated on devices $\mathrm{e}$, $\mathrm{h}$, and $\mathrm{c}$, in each case. 
In general, the observations made for the real-world workflow also hold for the synthetic workflows.
S1 and P1 are notable exceptions, since as we increased $w_{\mathrm{lat}}$ from $0$ to $1$, more tasks were allocated on device $\mathrm{c}$, instead of device $\mathrm{h}$, in contrast to the other TGs. This occurred, as S1 and P1 had a smaller number of tasks compared to other TGs, as shown in \cref{table:syntheticTGs}. Consequently, for S1 and P1 the calculated percentage of tasks requiring fixed allocation on devices $\mathrm{e}$ and $\mathrm{h}$ was $0.4$, and thus rounded to $0$. Therefore, as tasks could be allocated on any device, they were all allocated on device $\mathrm{c}$, which provided the lowest computational latency.

For each of the examined workflow structures, it can be observed that the difference in the results was more prominent between cases with 10 and 100 tasks than between cases with 100 and 1000 tasks. 
This was due to the fact that TGs with 10 tasks led to significantly different task allocations, compared to those with 100 and 1000 tasks.
Moreover, as in the case of the real-world workflow, it can be observed that the overall latency deteriorated more compared to reliability, as we reduced their weights. Thus, the overall latency was more sensitive to the changes in its relative importance, compared to reliability.

\paragraph{Sensitivity to criticality level}
For $\mathit{\Lambda} \in \{2, \allowbreak 3 \}$, the deterioration in reliability and latency started to occur when their respective weights (and thus their relative importance) were closer to zero. In contrast, for $\mathit{\Lambda} = 1$, the deterioration started to occur at larger weights. 
The reason was that higher criticality levels resulted in a greater number of task replicas, due to the lower vulnerability thresholds, and thus the overall reliability was improved. 
In addition, the greater number of replicas also benefited the latency objective for larger $w_{\mathrm{lat}}$, as it forced more primary tasks to be allocated along with their replicas on devices $\mathrm{h}$ and $\mathrm{c}$, which provided lower computational latency.
Hence, our framework adapted predictably to different levels of application criticality.

\paragraph{Scalability analysis}
\label{para:scalability}
\cref{table:syntheticTGs} shows the BILP problem size (number of variables and constraints) and the average execution time required to obtain a solution, as reported by the Gurobi solver, for each examined TG.
For all TG structures (serial, parallel, and mixed), the solver execution time increased with workflow size. 
On average, serial TGs required the longest solver runtimes (0.06--50.94 seconds), followed by parallel TGs (0.03--26.41 seconds), while mixed TGs generally required the shortest runtimes (0.08--13.68 seconds). 
The shorter execution times for mixed TGs were due to their balanced structure, which avoided the long dependency chains of serial TGs and the excessive parallelism of parallel TGs that made finding the optimal task allocation more challenging in those cases. 
Furthermore, as the criticality level $\mathit{\Lambda}$ increased, the number of replicas grew, increasing the problem size and thus the solver runtime. However, even for TGs with 1000 tasks and high criticality, runtimes remained short and practical for all workflow structures.

Overall, the experimental results demonstrate the scalability and applicability of our approach to workflows of different structures (serial, parallel, and mixed), sizes (10, 100, and 1000 tasks), and criticality levels (low, moderate, and high). 
The framework consistently provided optimal solutions for the specified trade-offs between reliability and latency in a short time frame, ranging on average from 0.03 to 50.94 seconds. 
Given that this is an exact, design-time (i.e., offline) method, and considering the NP-hard nature of the problem \cite{Khalil2018}, the solver execution times are both reasonable and practical.

\section{Conclusions and future work}
\label{sec:conclusions}

We proposed an exact, reliability- and latency-driven multi-objective optimization framework to allocate the tasks of a workflow application in the edge-hub-cloud continuum. We examined a streamlined yet challenging architecture, comprising an edge device, a hub device, and a cloud server. Our approach utilizes a comprehensive BILP formulation to jointly optimize the overall reliability and latency of the application, while considering constraints often overlooked in related studies, such as the limited memory, storage, and energy capacities of the devices, the reliability requirements of the tasks, as well as the computational and communication latency and energy required for the execution of each task.
The proposed formulation is facilitated by a two-step task graph transformation technique. It takes into account the defined trade-off between the reliability and latency goals, as well as the criticality level of the application, while employing widely used time redundancy methods. 

We evaluated our framework using a real-world workflow application, as well as representative synthetic workflows that we generated for this purpose. 
Through extensive experimentation, we demonstrated the effectiveness, scalability, and applicability of our approach to workflows of different structures, sizes, and criticality levels. 
In the real-world workflow, the proposed method achieved average improvements of 84.19\% in reliability and 49.81\% in latency over baseline strategies, across relevant objective trade-offs.
The average solution time across all examined task graphs ranged from 0.03 to 50.94 seconds. 
The usefulness of our framework is further highlighted by experimental results that were sometimes non-intuitive, particularly in how tasks were allocated based on the relative importance of each objective.
While this work focuses on critical workflow applications where a single edge device suffices for reliable and efficient operation, in future work we plan to evaluate the proposed framework in scenarios requiring multiple edge devices to assess the additional challenges such architectures entail. Furthermore, we aim to incorporate heuristics for dynamic task allocation and explore machine learning techniques to support online allocation decisions through the prediction of reliability and latency trends.

\section*{Acknowledgments}
This work has been supported by the European Union’s Horizon 2020 research and innovation programme under grant agreement No. 739551 (KIOS CoE) and from the Government of the Republic of Cyprus through the Cyprus Deputy Ministry of Research, Innovation and Digital Policy.

\biboptions{numbers,sort&compress}
\bibliographystyle{elsarticle-num} 
\bibliography{6_references.bib}

\end{document}